\documentclass[vecphys]{svmult}

\usepackage{makeidx}         % allows index generation
\usepackage{graphicx}        % standard LaTeX graphics tool
\usepackage{multicol}        % used for the two-column index
\usepackage[bottom]{footmisc}% places footnotes at page bottom

\usepackage{bbm}
\usepackage{amssymb}
\usepackage{axodraw}
\usepackage{subfigure}

\newcommand{\eqref}[1]{(\ref{#1})}
\newcommand{\Eqref}[1]{Eq.~(\ref{#1})}
\newcommand{\pat}{\partial_t}
\newcommand{\Tr}{\mathrm{Tr}}
\newcommand{\tr}{\mathrm{tr}}
\newcommand{\ddk}{\frac{\D}{\D k}}
\newcommand{\Nc}{N_{\mathrm{c}}}
\newcommand{\Nf}{N_{\mathrm{f}}}
\newcommand{\text}[1]{\mathrm{#1}}  % just an alias a la amsmath
\newcommand{\mG}{\mathcal{G}}
\newcommand{\bG}{\bar{\mathcal{G}}}
\newcommand{\yb}{\bar{\psi}}
\newcommand{\fss}[1]{#1\!\!\!/}
\newcommand{\fsl}[1]{#1\!\!\!\!/}
\newcommand{\scb}[1]{\scalebox{1.5}[1.5]{#1}}

\newcommand{\lb}{\lambda}
\newcommand{\mb}{m}
\newcommand{\hb}{h}
\newcommand{\te}{\tilde{\epsilon}}

\makeindex             % used for the subject index
\begin{document}

\title*{Introduction to the Functional RG and Applications to Gauge
  Theories} 
\author{Holger Gies}
\institute{Institute for Theoretical Physics, Heidelberg University,
  Philosophenweg 16, D-69120 Heidelberg, Germany,
\texttt{h.gies@thphys.uni-heidelberg.de}
}
\maketitle

\section*{Prologue}
\label{sec:prologue}

This lecture course\footnote{Lectures held at the 2006 ECT* School
"Renormalization Group and Effective Field Theory Approaches to
Many-Body Systems", Trento, Italy.} is intended to fill the gap
between graduate courses on quantum field theory and specialized
reviews or forefront-research articles on functional renormalization
group approaches to quantum field theory and gauge theories.

These lecture notes are meant for advanced students who want to get
acquainted with modern renormalization group (RG) methods as well as
functional approaches to quantum gauge theories.  In the first
lecture, the functional renormalization group is introduced with a
focus on the flow equation for the effective average action.  The
second lecture is devoted to a discussion of flow equations and
symmetries in general, and flow equations and gauge symmetries in
particular. The third lecture deals with the flow equation in the
background formalism which is particularly convenient for analytical
computations of truncated flows. The fourth lecture concentrates on
the transition from microscopic to macroscopic degrees of freedom;
even though this is discussed here in the language and the context of
QCD, the developed formalism is much more general and will be useful
also for other systems. Sections which have an asterisk $\ast$ in the
section title contain more advanced material and may be skipped during
a first reading.

This is not a review.  I apologize for many omissions of further
interesting and important aspects of this field (and their
corresponding references). General reviews and more complete reference
lists can be found in
\cite{Fisher:1998kv,Morris:1998da,Litim:1998nf,Aoki:2000wm,%
Bagnuls:2000ae,Berges:2000ew,Polonyi:2001se,Pawlowski:2005xe}.  A
guide to more specialized literature is given in the Further-Reading
subsections at the end of some sections.

\addtolength{\textheight}{-1.5cm}

\section{Introduction}
\label{sec:intro}

Quantum and statistical field theory are two sides of the same medal,
representing the fundament on which modern physics is built. Both
branches have been molded by the concept of the renormalization
group. The renormalization group deals with the physics of scales. A
central theme is the understanding of the macroscopic physics at long
distances or low momenta in terms of the fundamental microscopic
interactions. Bridging this gap from micro to macro scales requires a
thorough understanding of quantum or statistical fluctuations on all
the scales in between.

Field theories with gauge symmetry are of central importance, in
particular, since all elementary particle-physics interactions are
described by gauge theories. Moreover, nonabelian gauge theories --
and most prominently quantum chromodynamics (QCD) -- are in many
respects paradigmatic, since they exhibit numerous features that are
encountered in various field theoretical systems both in particle
physics as well as condensed-matter systems. During the transition
from micro to macro scales, these theories turn from weak to strong
coupling, the relevant degrees of freedom are changed, the realization
of the fundamental symmetries is different on the various scales, and
the phase diagram is expected to have a rich structure, being formed
by different and competing collective phenomena.

A profound understanding of gauge theories thus requires not just one
but a whole toolbox of field theoretical methods. In addition to
analytical perturbative methods for weak coupling and numerical
lattice gauge theory for arbitrary couplings, \emph{functional
  methods} begin to bridge the gap, since they are not restricted to
weak coupling and can still largely be treated analytically. Functional
methods aim at the computation of generating functionals of
correlation functions, such as the effective action that governs the
dynamics of the macroscopic expectation values of the fields. These
generating functionals contain all relevant physical information about
a theory, once the microscopic fluctuations have been integrated out. 

The functional RG combines this functional approach with the RG idea
of treating the fluctuations not all at once but successively from
scale to scale \cite{Wilson:1971bg,Wilson:1973jj}.  Instead of
studying correlation functions after having averaged over all
fluctuations, only the \emph{change} of the correlation functions as
induced by an infinitesimal momentum shell of fluctuations is
considered. From a structural viewpoint, this allows to transform the
functional-integral structure of standard field theory formulations
into a functional differential structure
\cite{Wegner:1972ih,Nicoll:1977hi,Polchinski:1983gv,Wetterich:1992yh}.
This goes along not only with a better analytical and numerical
accessibility and stability, but also with a great flexibility of
devising approximations adapted to a specific physical system. In
addition, structural investigations of field theories from first
principles such as proofs of renormalizability can more elegantly and
efficiently be performed with this strategy
\cite{Polchinski:1983gv,Warr:1986we,Hurd:1989up,Keller:1990ej}.

The central tool of the functional RG is given by a flow equation.
This flow equation describes the evolution of correlation functions or
their generating functional under the influence of fluctuations. It
connects a well-defined initial quantity, e.g., the microscopic
correlation functions in a perturbative domain, in an exact manner
with the desired full correlation functions after having integrated
out the fluctuations. Hence, solving the flow equation corresponds to
solving the full theory. 

The complexity of quantum gauge theories and QCD are a serious
challenge for all field theoretical methods. The construction of flow
equations for gauge theories has to take special care of the gauge
symmetry, i.e., the invariance of the theory under \emph{local}
transformations of the fields in coordinate space \cite{Reuter:1992uk,%
  Reuter:1993kw,Bonini:1993kt,Ellwanger:1994iz,D'Attanasio:1996jd}.
The beauty of gauge symmetry is turned into the beast of complex
dynamical equations and nontrivial symmetry constraints both of which
have to be satisfied by the flow of the correlation functions.
Nevertheless, the success of the functional RG in many branches of
physics as described in other lectures of this volume makes it a
promising tool also for gauge theories. The recent rapid development
of functional methods and their application to gauge theories and QCD
also in the strong-coupling domain confirm this expectation. In
combination with and partly complementary to other field theoretical
methods, the functional RG has the potential to shed light on some of
the still hardly accessible parameter regions of quantum gauge
theories.

\section{Functional RG Approach to Quantum Field Theory}
\label{sec:funrg}

\subsection{Basics of QFT}
\label{ssec:basics}

In quantum field theory (QFT), all physical information is stored in
correlation functions. For instance, consider a collider experiment
with two incident beams and $(n-2)$ scattering products. All
information about this process can be obtained from the
$n$-\emph{point function}, a correlator of $n$ quantum fields. In QFT,
we obtain this correlator by definition from the product of $n$ field
operators at different spacetime points $\varphi(x_n)$ averaged over
all possible field configurations (quantum fluctuations).

In Euclidean QFT, the field configurations are weighted with an
exponential of the action $S[\varphi]$,
\begin{equation}
\langle \varphi(x_1) \dots \varphi(x_n)\rangle := \mathcal
N \int \mathcal D \varphi\, \varphi(x_1) \dots \varphi(x_n)\,
\E^{-S[\varphi]}, \label{1.1}
\end{equation}
where we fix the normalization $\mathcal N$ by demanding that $\langle
1\rangle =1$. We assume that Minkowski-valued correlators can be
defined from the Euclidean ones by analytic continuation.  We also
assume that a proper regularized definition of the measure can be
given (for instance, using a spacetime lattice discretization), which
we formally write as $\int \mathcal D \varphi \to \int_\Lambda
\mathcal D \varphi$; here, $\Lambda$ denotes an ultraviolet (UV)
cutoff. This regularized measure should also preserve the symmetries of
the theory: for a symmetry transformation $U$ which acts on the fields,
$\varphi\to \varphi^U$, and leaves the action invariant,
$S[\varphi]\to S[\varphi^U]\equiv S[\varphi]$, the invariance of the
measure implies
\begin{equation}
\int_\Lambda \mathcal D\varphi \to \int_\Lambda \mathcal D \varphi^U
\equiv \int_\Lambda \mathcal D\varphi. \label{1.2}
\end{equation}
For simplicity, let $\varphi$ denote a real scalar field; the
following discussion also holds for other fields such as fermions with
minor modifications.  All $n$-point correlators are summarized by the
generating functional $Z[J]$,
\begin{equation}
Z[J]\equiv\E^{W[J]} =\int\mathcal D \varphi\, \E^{-S[\varphi]+\int
  J\varphi}, \label{1.3}
\end{equation}
with source term $\int J \varphi = \int \D^D x\, J(x) \varphi(x)$. All
$n$-point functions are obtained by functional differentiation:
\begin{equation}
\langle \varphi(x_1) \dots \varphi(x_n)\rangle =\frac{1}{Z[0]} \left(
  \frac{\delta^n Z[J]}{\delta J(x_1) \dots \delta J(x_n)}
  \right)_{J=0}.\label{1.4}
\end{equation}
Once the generating functional is computed, the theory is
solved.%\footnote{Beyond summarizing the correlators, the generating
%  functional also has the physical meaning of a \emph{vacuum
%    persistence amplitude} in the presence of a source $J$, $Z[J]/Z[0]
%  \equiv \langle 0|0\rangle_J$.}

In Eq.~\eqref{1.3}, we have also introduced the generating functional
of \emph{connected correlators}\footnote{In this short introduction,
  we use but make no attempt at fully explaining the standard QFT
  nomenclature; for the latter, we refer the reader to any standard
  QFT textbook, such as \cite{QFTtextbook1,QFTtextbook2}.}, $W[J]=\ln
Z[J]$, which, loosely speaking, is a more efficient way to store the
physical information.  An even more efficient information storage is
obtained by a Legendre transform of $W[J]$: the \emph{effective
  action} $\Gamma$:
\begin{equation}
\Gamma[\phi]=\sup_J\left( \int J\phi -W[J]\right). \label{1.5}
\end{equation}
For any given $\phi$, a special $J\equiv J_\mathrm{sup}=J[\phi]$ is
singled out for which $\int J\phi -W[J]$ approaches its supremum. Note
that this definition of $\Gamma$ automatically guarantees that
$\Gamma$ is convex. At $J=J_{\sup}$, we get
\begin{eqnarray}
0&\stackrel{!}{=}& \frac{\delta}{\delta J(x)} \left( \int J\phi
  -W[J]\right) \nonumber\\
&\Rightarrow&\quad \phi=\frac{\delta W[J]}{\delta J} =\frac{1}{Z[J]}
\frac{\delta Z[J]}{\delta J} = \langle \varphi \rangle_J. \label{1.6}
\end{eqnarray}
This implies that $\phi$ corresponds to the expectation value of
$\varphi$ in the presence of the source $J$. The meaning of $\Gamma$
becomes clear by studying its derivative at $J=J_{\sup}$
\begin{eqnarray}
\frac{\delta \Gamma[\phi]}{\delta \phi(x)}
=-\int_y \frac{\delta W[J]}{\delta J(y)} \frac{\delta
  J(y)}{\delta\phi(x)} + \int_y \frac{\delta J(y)}{\delta\phi(x)}
\phi(y) + J(x) \stackrel{\eqref{1.6}}{=} J(x). \label{1.7}
\end{eqnarray}
This is the \emph{quantum equation of motion} by which the effective
action $\Gamma[\phi]$ governs the dynamics of the field expectation
value, taking the effects of all quantum fluctuations into account. 

From the definition of the generating functional, we can
straightforwardly derive an equation for the effective action:
\begin{equation}
\E^{-\Gamma[\phi]}=\int_\Lambda \mathcal D \varphi\,
\exp\left(-S[\phi+\varphi] +\int \frac{\delta
    \Gamma[\phi]}{\delta\phi}\,\varphi \right). \label{1.8}
\end{equation}
Here, we have performed a shift of the integration variable, $\varphi
\to \varphi+\phi$. We observe that the effective action is determined
by a nonlinear first-order functional differential equation, the
structure of which is itself a result of a functional integral. An
exact determination of $\Gamma[\phi]$ and thus an exact solution has
so far been found only for rare, special cases.

As a first example of a functional technique, a solution of
Eq.~\eqref{1.8} can be attempted by a \emph{vertex expansion} of
$\Gamma[\phi]$,
\begin{equation}
\Gamma[\phi]=\sum_{n=0}^\infty\, \frac{1}{n!} \int \D^D x_1 \dots \D^D
x_n\, \Gamma^{(n)}(x_1,\dots,x_n)\, \phi(x_1) \dots \phi(x_n), 
\label{1.9}
\end{equation}
where the expansion coefficients $\Gamma^{(n)}$ correspond to the
\emph{one-particle irreducible (1PI) proper vertices}. Inserting
\Eqref{1.9} into \Eqref{1.8} and comparing the coefficients of the
field monomials results in an infinite tower of coupled
integro-differential equations for the $\Gamma^{(n)}$: the
Dyson-Schwinger equations. This functional method of constructing
approximate solutions to the theory via truncated Dyson-Schwinger
equations, i.e., via a finite truncation of the series \Eqref{1.9} has
its own merits and advantages; their application to gauge theories is
well developed; see, e.g., \cite{Alkofer:2000wg,Roberts:2000aa,%
  Maris:2003vk,Fischer:2006ub}.  Here, we proceed by amending the RG
idea to functional techniques in QFT.

\subsection{RG Flow equation}
\label{ssec:RGflow}

A versatile approach to the computation of $\Gamma$ is based on RG
concepts \cite{Wetterich:1992yh}. Whereas a computation via
\Eqref{1.8} or via Dyson-Schwinger equations corresponds to
integrating-out all fluctuations at once, we can implement Wilson's
idea of integrating out modes momentum shell by momentum shell.

\begin{figure}[t]
\centering

\vspace{0.6cm} 

\includegraphics[height=5cm]{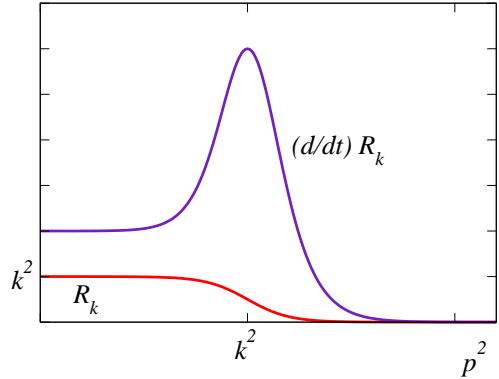}

\caption{Sketch of a regulator function $R_k(p^2)$ (lower curve) and
  its derivative $\pat R_k(p^2)$ (upper curve). Whereas the regulator
  provides for an IR regularization for all modes with $p^2\lesssim
  k^2$, its derivative implements the Wilsonian idea of integrating
  out fluctuations within a momentum shell near $p^2\simeq k^2$.  }
\label{fig:reg1}       % Give a unique label
\end{figure}

In terms of $\Gamma$, we are looking for an interpolating action
$\Gamma_k$, which is also called \emph{effective average action}, with
a momentum-shell parameter $k$, such that $\Gamma_k$ for $k\to\Lambda$
corresponds to the bare action to be quantized; the full
quantum action $\Gamma$ should be approached for $k\to 0$, 
\begin{equation}
\Gamma_{k\to\Lambda}\simeq S_{\mathrm{bare}}, \quad
\Gamma_{k\to0}=\Gamma. \label{1.10}
\end{equation}
This can indeed be constructed from the generating functional. For
this, let us define the IR regulated functional
\begin{eqnarray}
\E^{W_k[J]} &\equiv& Z_k[J]:=\exp \left(-\varDelta S_k
  \left[\frac{\delta}{\delta J}\right] \right)\, Z[J]\nonumber\\
&=& \int_\Lambda \mathcal D\varphi \, \E^{-S[\varphi] -\varDelta
  S_k[\varphi] +\int J\varphi }, \label{1.11}
\end{eqnarray}
where
\begin{equation}
\varDelta S_k[\varphi] =\frac{1}{2} \int \frac{\D^D q}{(2\pi)^{D}}\,
\varphi(-q) R_k(q) \varphi(q)\label{1.12}
\end{equation}
is a regulator term which is quadratic in $\varphi$ and can be viewed
as a momentum-dependent mass term. The regulator function $R_k(q)$
should satisfy
\begin{equation}
\lim_{q^2/k^2\to 0}\, R_k(q) >0,\label{1.13a}
\end{equation}
which implements an IR regularization. For instance, if $R_k\sim k^2$
for $q^2\ll k^2$, the regulator screens the IR modes in a mass-like
fashion, $m^2\sim k^2$. Furthermore,
\begin{equation}
\lim_{k^2/q^2\to 0}\, R_k(q) =0, \label{1.13b}
\end{equation}
which implies that the regulator vanishes for $k\to0$. As an immediate
consequence, we automatically recover the standard generating
functional as well as the full effective action in this limit:
$Z_{k\to0}[J] = Z[J]$ and $\Gamma_{k\to 0}=\Gamma$. The third
condition is
\begin{equation}
\lim_{k^2\to \Lambda\to \infty}\, R_k(q) \to \infty,
   \label{1.13c} 
\end{equation}
which induces that the functional integral is dominated by the
stationary point of the action in this limit. This justifies the use
of a saddle-point approximation which filters out the classical field
configuration and the bare action, $\Gamma_{k\to\Lambda} \to S+
\mathrm{const.}$. A sketch of a typical regulator that satisfies these
three requirements is shown in Fig.~\ref{fig:reg1}. Incidentally, the
regulator is frequently written as
\begin{equation}
R_k(p^2) = p^2\, r(p^2/k^2), \label{1.13d}
\end{equation}
where $r(y)$ is a dimensionless regulator shape function with a
dimensionless momentum argument. The requirements
\eqref{1.13a}-\eqref{1.13c} translate in a obvious manner into
corresponding requirements for $r(y)$. 

Since we already know that the interpolating functional $\Gamma_k$
exhibits the correct limits, let us now study the intermediate
trajectory. We start with the generating functional $W_k[J]$, using the
abbreviations
\begin{equation}
t=\ln \frac{k}{\Lambda}, \qquad \pat= k\frac{\D}{\D k}.\
\end{equation}
Keeping the source $J$ fixed, i.e., $k$ independent, we obtain
\begin{eqnarray}
\pat W_k[J]&=&-\frac{1}{2} \int \mathcal D \varphi\, \varphi(-q) \,
\pat R_k(q)\, \varphi(q) \E^{-S-\varDelta S+\int J\varphi} \nonumber\\
&=&-\frac{1}{2} \int \frac{\D^D q}{(2\pi)^D}  \,
\pat R_k(q)\, G_k(q) +\pat \varDelta S_k[\phi]. \label{1.14}
\end{eqnarray}
Here, we have defined the \emph{connected} propagator
\begin{equation}
G_k(p)=\left( \frac{\delta^2 W_k}{\delta J \delta J} \right)(p) =
\langle \varphi(-p) \varphi(p) \rangle -\langle \varphi(-p)  \rangle
\langle\varphi(p) \rangle. \label{1.15}
\end{equation}
(Note that we frequently change from coordinate to momentum space or
vice versa by Fourier transformation for reasons of convenience.)
Now, we are in a position to define the interpolating effective action
$\Gamma_k$ by a slightly modified Legendre transform,\footnote{Now,
  only the ``sup'' part of $\Gamma_k$ is convex. For finite $k$, any
  non-convexity of $\Gamma_k$ must be of the form of the last regulator
  term of \Eqref{1.16}.} 
\begin{equation}
\Gamma_k[\phi]=\sup_J\left( \int J\phi -W_k[J]\right)- \varDelta
S_k[\phi]. \label{1.16} 
\end{equation}
Since we later want to study $\Gamma_k$ as a functional of a
$k$-independent field $\phi$, it is clear from \Eqref{1.16} that the
source $J\equiv J_\mathrm{sup}=J[\phi]$ for which the supremum is
approached is necessarily $k$ dependent. As before, we get at
$J=J_{\sup}$:
\begin{equation}
\phi(x) = \langle \varphi(x) \rangle_J=\frac{\delta W_k[J]}{\delta
  J(x)}. \label{1.17}
\end{equation}
The quantum equation of motion receives a regulator
modification, 
\begin{equation}
J(x)=\frac{\delta \Gamma_k[\phi]}{\delta \phi(x)} + \big(R_k
\phi\big)(x). \label{1.18}
\end{equation}
From this, we deduce\footnote{In case of fermionic Grassmann-valued
  fields, the following $\phi$ derivative should act on \Eqref{1.18}
  from the right.\label{foot:grass}}:
\begin{equation}
\frac{\delta J(x)}{\delta\phi(y)} = \frac{\delta^2
  \Gamma_k[\phi]}{\delta \phi(x) \delta \phi(y)} +
  R_k(x,y). \label{1.19}
\end{equation}
On the other hand, we obtain from \Eqref{1.17}:
\begin{equation}
\frac{\delta \phi(y)}{\delta J(x')} =\frac{\delta^2 W_k[J]}{\delta
  J(x') \delta J(y)} \equiv G_k(y-x'). \label{1.20}
\end{equation}
This implies the important identity
\begin{eqnarray}
\delta(x-x')&=&\frac{\delta J(x)}{\delta J(x')} =\int \D^D y\,
\frac{\delta J(x)}{\delta \phi(y)} \frac{\delta \phi(y)}{\delta J(x')}
\nonumber\\
&=&\int \D^D y\,(\Gamma_k^{(2)}[\phi] + R_k)(x,y) G_k(y-x'), \label{1.21b}
\end{eqnarray}
or, in operator notation,
\begin{equation}
\mathbbm{1} = (\Gamma_k^{(2)} +R_k)\, G_k. \label{1.21}
\end{equation}
Here, we have introduced the short-hand notation
\begin{equation}
\Gamma_k^{(n)}[\phi] = \frac{\delta^n \Gamma_k[\phi]}{\delta\phi \dots
  \delta \phi}. 
\end{equation}
Collecting all ingredients, we can finally derive the flow equation
for $\Gamma_k$ for fixed $\phi$ and at $J=J_{\sup}$
\cite{Wetterich:1992yh}:
\begin{eqnarray}
\pat \Gamma_k[\phi]&=& -\pat W_k[J]|_\phi + \int (\pat J) \phi -\pat
\varDelta S_k[\phi] = -\pat W_k[J]|_J -\pat \varDelta S_k[\phi]
\nonumber\\ 
&\stackrel{\eqref{1.14}}{=}& \frac{1}{2} \int \frac{\D^D q}{(2\pi)^D}\,
\pat R_k(q) \, G_k(q)\nonumber\\
&\stackrel{\eqref{1.21}}{=}& \frac{1}{2}\, \Tr\, \left[ \pat R_k
  \left( \Gamma_k^{(2)}[\phi] +R_k \right)^{-1} \right]. \label{1.22}
\end{eqnarray}
This flow equation forms the starting point of all our further
investigations. Hence, let us carefully discuss a few of its
properties, as they are apparent already at this stage:

\begin{figure}[t]
\begin{displaymath}
  \pat\Gamma_k[\phi]=\frac{1}{2}\, \pat R_k\quad 
  \begin{minipage}{2cm}
    \scalebox{0.8}[0.8]{
      \begin{picture}(70,60)(5,0) 
%        \SetOffset(3,10)
        \CArc(40,30)(29,0,360) 
        \CArc(40,30)(31,0,360)
        \CBoxc(10,30)(10,10){Red}{Red} 
      \end{picture}}
  \end{minipage}
\end{displaymath}

\caption{Diagrammatic representation of the flow equation
  \eqref{1.22}: the flow of $\Gamma_k$ is given by a one-loop form,
  involving the full propagator $G_k=(\Gamma_k^{(2)}+R_k)^{-1}$
  (double line) and an operator insertion in the form of $\pat R_k$
  (filled box).}
\label{fig:oneloop}
\end{figure}

\begin{itemize}
\item The flow equation is a functional differential equation for
  $\Gamma_k$. In contrast to \Eqref{1.8}, no functional integral has
  to be performed to reveal the full structure of the equation. 
  
\item We have \emph{derived} the flow equation from the standard
  starting point of QFT: the generating functional. But a different --
  if not inverse -- perspective is also legitimate. We may
  \emph{define} QFT based on the flow equation. For given suitable
  initial conditions, for instance, by defining the bare action at a
  high UV cutoff scale $k=\Lambda$, the flow equation defines a
  trajectory to the full quantum theory described by the full
  effective action $\Gamma$. In the case of additional symmetries, the
  QFT-defining flow equation may be supplemented by symmetry
  constraints to the effective action.
  
\item The purpose of the regulator is actually twofold: by
  construction, the occurrence of $R_k$ in the denominator of
  \Eqref{1.22} guarantees the IR regularization by construction. In
  addition to this and thanks to the conditions \eqref{1.13a} and
  \eqref{1.13b}, the derivative $\pat R_k$ occurring in the numerator
  of \Eqref{1.22} ensures also UV regularization, since its
  predominant support lies on a smeared momentum shell near $p^2\sim
  k^2$. A typical shape of the regulator and its derivative is
  depicted in Fig.~\ref{fig:reg1}. The peaked structure of $\pat R_k$
  implements nothing but the Wilsonian idea of integrating over
  momentum shells and implies that the flow is localized in momentum
  space.
  
\item The solution to the flow equation \eqref{1.22} corresponds
  to an RG trajectory in \emph{theory space}. The latter is a space of
  all action functionals spanned by all possible invariant operators
  of the field. The two ends of the trajectory are given by the
  initial condition $\Gamma_{k\to
  \Lambda}=S_{\mathrm{bare}}$, and the full effective action
  $\Gamma_{k\to 0}=\Gamma$.

\item Apart from the conditions \eqref{1.13a}-\eqref{1.13c}, the
  regulator can be chosen arbitrarily. Of course, the precise form of
  the trajectory depends on the regulator $R_k$. The variation of the
  trajectory with respect to $R_k$ reflects the RG scheme dependence
  of a non-universal quantity; see Fig.~\ref{fig:theoryspace}.
  Nevertheless, the final point on the trajectory is independent of
  $R_k$ as is guaranteed by Eqs.~\eqref{1.13a}-\eqref{1.13c}.
  
{\unitlength=1mm
\begin{figure}[t]
\centering

\begin{picture}(70,50) 
\put(0,0){
\includegraphics[height=5cm]{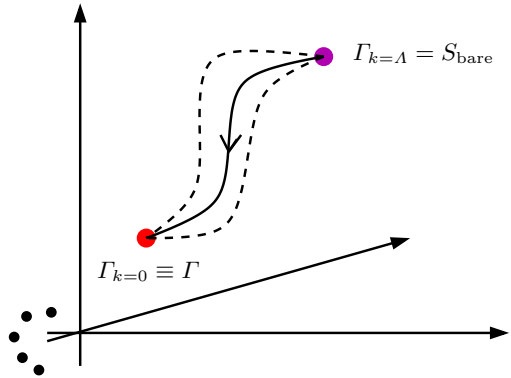}}
\put(47,42){$\Gamma_{k=\Lambda}=S_{\text{bare}}$}
\put(13,13){$\Gamma_{k=0}\equiv \Gamma$} 
\end{picture}
\caption{Sketch of the RG flow in theory space. Each axis labels a
  different operator which spans the effective action, e.g., $\phi^2$,
  $(\partial\phi)^2$, etc. Once the initial conditions in terms of the
  bare action $\Gamma_{k=\Lambda}=S_{\text{bare}}$ are given, the
  solution to the flow equation \eqref{1.22} is a trajectory (solid
  line) in this space of action functionals, ending at the full
  quantum effective action $\Gamma\equiv \Gamma_{k=0}$. A change of
  the regulator $R_k$ can modify the trajectory (dashed lines), but
  the end point $\Gamma$ stays always the same.}
\label{fig:theoryspace}       % Give a unique label
\end{figure}
}

\item The flow equation has a one-loop structure, but is nevertheless
  an exact equation, as is signaled by the occurrence of the exact
  propagator in the loop; see Fig.~\ref{fig:oneloop}. The one-loop
  structure is a direct consequence of $\varDelta S_k$ being quadratic
  in the field operator $\varphi$ which is coupled to the source
  \cite{Litim:2002xm}.

\item Perturbation theory can immediately be re-derived from the
  flow equation. For instance, imposing the loop expansion on
  $\Gamma_k$, $\Gamma_k=S+\hbar\Gamma_k^{\mathrm{1-loop}}+ \mathcal
  O(\hbar^2)$, it becomes obvious that, to one-loop order,
  $\Gamma^{(2)}_k$ can be replaced by $S^{(2)}$ on the right-hand side
  of \Eqref{1.22}. From this, we infer:
  \begin{eqnarray}
    \pat \Gamma_k^{\mathrm{1-loop}}&=& \frac{1}{2}\, \Tr\, \left[\pat
    R_k \left( S^{(2)}+R_k\right)^{-1} \right]= \frac{1}{2}\, \pat\,
    \Tr\, \ln (S^{(2)}+R_k) \nonumber\\
    &\Rightarrow& \Gamma^{\mathrm{1-loop}}=S+\frac{1}{2}\, \Tr\, \ln
    S^{(2)} + \mathrm{const.} \nonumber
  \end{eqnarray}
  The last formula corresponds to the standard one-loop effective
  action, as it should. 

\end{itemize}

In practice, the versatility of the flow equation is its most
important strength: beyond perturbation theory, various systematic
approximation schemes exist which can be summarized under the
\emph{method of truncations}.

A first example for such an approximation scheme has already been
given above in \Eqref{1.9}, the vertex expansion, which now reads
\begin{equation}
\Gamma_k[\phi]=\sum_{n=0}^\infty\, \frac{1}{n!} \int \D^D x_1 \dots \D^D
x_n\, \Gamma_k^{(n)}(x_1,\dots,x_n)\, \phi(x_1) \dots \phi(x_n). 
\label{1.}
\end{equation}
Upon inserting this expansion into the flow equation \eqref{1.22}, we
obtain flow equations for the vertex functions $\Gamma_k^{(n)}$ which
interpolate between the bare and the fully dressed vertices. These
equations are similar but not identical to Dyson-Schwinger equations,
as will be discussed in more detail below. 

As a second example, let us introduce the \emph{operator expansion}
which constructs the effective action from operators of increasing
mass dimension. Focusing in particular on derivative operators, we
arrive at the gradient expansion. For instance, for a theory with one
real scalar field, we obtain
\begin{equation}
\Gamma_k= \int \D^D x\left[ V_k(\phi) + \frac{1}{2} Z_k(\phi)\,
  (\partial_\mu \phi)^2 + \mathcal{O}(\partial^4) \right],
\end{equation}
where, for instance, $V_k(\phi)$ corresponds to the effective
potential. 

Further examples of this type can easily be constructed by combining
these two expansions in various ways. Formally, expansions of the
effective action should be \emph{systematic}; this implies that a
classification scheme exists which classifies all possible building
blocks of the effective action and relates them to a definite order in
the expansion. Truncations based on such expansions should also be
\emph{consistent} in the sense that, once the maximal order of the
truncated expansion is chosen, all terms up to this order are kept in
the flow equation. 

Systematics and consistency are, of course, only a necessary condition
for the construction of a reliable truncation -- they are not
necessarily sufficient. As a word of caution, let us stress that these
two conditions do not guarantee a rapid convergence of the truncated
effective action towards the true result. As for any expansion,
convergence properties have to be checked separately, in order to
estimate or even control the truncation errors.
 
Only one general recipe for the construction of a truncation exists:
let your truncation scheme be guided by physics by making sure that
the truncation includes the most relevant degrees of freedom of a
given problem. 

\subsection{Euclidean Anharmonic Oscillator}
\label{ssec:AnhOsc}

Let me illustrate the capabilities of the flow equation by a simple
example: a 0+1 dimensional real scalar field theory, or, in other
words, the Euclidean quantum mechanical anharmonic oscillator. This
system has first been studied by RG techniques in
\cite{Horikoshi:1998sw}. The bare action that we want to quantize is
\begin{equation}
S=\int d\tau \left( \frac{1}{2} \dot{x}^2+\frac{1}{2} \omega^2 x^2 +
  \frac{\lambda}{24} x^4 \right), \label{AO.1}
\end{equation}
with $\omega^2,\lambda>0$.\footnote{Also the double-well potential
  with $\omega^2<0$ can be studied with RG techniques, see
  \cite{Kapoyannis:2000sp,Zappala:2001nv}.}  We will mainly be
  interested in the determination of the ground state energy which we
  expect to be predominantly influenced by the effective
  potential. Hence, we consider the truncation
\begin{equation}
\Gamma_k[x]=\int d\tau \left( \frac{1}{2} \dot{x}^2 + V_k(x)
\right). \label{AO.2}
\end{equation}
For a concrete computation, we have to choose a regulator which
conforms to the conditions \eqref{1.13a}-\eqref{1.13c}. The following
choice is not only simple and convenient, it is also an \emph{optimal}
choice for the present problem, since it improves the stability
properties of our flow equation \cite{Litim:2000ci},
\begin{equation}
R_k(p)= \left(k^2 -p^2 \right)
\theta(k^2-p^2), \label{AO.3}
\end{equation}
which implies that $\pat R_k=2k^2\, \theta(k^2-p^2)$. On the
right-hand side of the flow equation \eqref{1.22}, we need
$\Gamma_k^{(2)}=(-\partial_\tau^2 +V_k''(x))\delta(\tau-\tau')$. In
order to project the flow equation onto the flow of the effective
potential, it suffices to consider the special case
$x=\mathrm{const.}$, for which the right-hand side can immediately be
Fourier transformed, e.g., $-\partial_\tau^2 \to p_\tau^2$. We obtain
the flow of the effective potential:
\begin{eqnarray}
\pat V_k(x) &=& \frac{1}{2} \int_{-\infty}^\infty \frac{dp_\tau}{2\pi}\,
\frac{2k^2 \theta(k^2-p_\tau^2)}{k^2 +V_k''(x)} \nonumber\\
\Rightarrow\quad \ddk  V_k(x)&=& \frac{1}{\pi}\,
\frac{k^2}{k^2+ V_k''(x)}. \label{AO.4}
\end{eqnarray}
This is a partial differential equation for the effective
potential. Aiming at the ground state energy, it suffices to study a
polynomial expansion of the potential,
\begin{equation}
V_k(x)= \frac{1}{2} \omega_k^2 x^2 + \frac{1}{24} \lambda_k x^4 +\dots
+ \tilde{E}_k, \label{AO.5}
\end{equation}
where the effects of the fluctuations are now encoded in a
scale-dependent frequency $\omega_k$ and coupling $\lambda_k$; their
initial conditions at $k=\Lambda$ are given by the parameters $\omega$
and $\lambda$ in the bare action \eqref{AO.1}. Unfortunately,
$\tilde{E}_{k}$ is not identical to the desired ground state energy
$E_{0,k}$, but differs by further $x$-independent contributions.  This
becomes already clear by looking at the UV limit $k\to\Lambda$, where
the regulator term becomes $\sim \frac{1}{2} \Lambda^2 x^2$,
contributing a harmonic-oscillator-like ground state energy $\sim
\frac{1}{2} \Lambda$ to $\tilde{E}$. In order to extract the true
ground state energy from the flow of $\tilde{E}_k$,
\begin{equation}
\ddk \tilde{E}_k= \frac{1}{\pi} \, \frac{k^2}{k^2+\omega_k^2},
\label{AO.6}
\end{equation}
we can perform a controlled subtraction to avoid the build-up of the
unphysical contributions: in the limit $\lambda=\omega=0$, the ground
state energy has to stay zero, $E_{0,k}=0$. This fixes the subtraction
term for \Eqref{AO.6}, which then reads:
\begin{equation}
\ddk {E}_{0,k}= \frac{1}{\pi} \left(
  \frac{k^2}{k^2+\omega_k^2}-1 \right),
\label{AO.7}
\end{equation}
Expanding \Eqref{AO.4} to higher orders, yields
\begin{eqnarray}
\ddk  \omega_k^2 &=& - \frac{2}{\pi} \,
\frac{k^2}{(k^2+\omega_k^2)^2}\, \frac{\lambda_k}{2}  \label{AO.8},\\ 
\ddk  \lambda_k &=& \frac{24}{\pi} \,
\frac{k^2}{(k^2+\omega_k^2)^3}\, \left(\frac{\lambda_k}{2}\right)^2+
\dots, \label{AO.9} 
\end{eqnarray}
where the ellipsis in the last equation denotes contributions from
higher-order terms $\sim x^6$, which we neglect here for simplicity.
We have boiled the flow equation down to a coupled set of first-order
ordinary differential equations, which can be viewed as the RG $\beta$
functions of the generalized couplings $E_{0,k},\omega_k, \lambda_k$.
These equations can diagrammatically be displayed as in
Fig.~\ref{fig:FEdiag}. The diagrams look very similar to one-loop
perturbative diagrams, but there are important differences: all
internal lines and vertices correspond to full propagators and full
vertices (in our simple truncation here, the vertex represents a full
running $\lambda_k$ and the propagators contain the running
$\omega_k^2$). Furthermore, one propagator in each loop carries a
regulator insertion, implying the replacement $G\to G_k \pat R_k G_k$
in comparison with perturbative diagrams.

{\unitlength=1pt
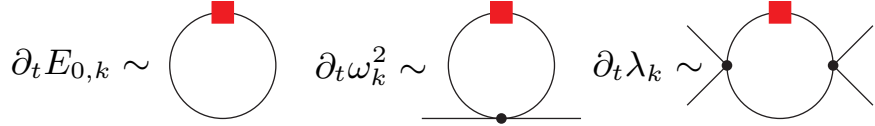
\begin{figure}[t]
  \centering
    \scalebox{1}[1]{
      \begin{picture}(300,80)(0,0)
        \Text(-10,40)[l]{\scb{$\pat E_{0,k}\sim$}}
        \CArc(70,40)(20,0,360)
        \CBoxc(70,60)(8,8){Red}{Red}
        \Text(105,40)[l]{\scb{$\pat \omega_{k}^2\sim$}}
        \CArc(175,40)(20,0,360) 
        \CBoxc(175,60)(8,8){Red}{Red}
        \Line(145,20)(205,20)
        \Vertex(175,20){2}
        \Text(210,40)[l]{\scb{$\pat \lambda_{k}\sim$}}
        \CArc(280,40)(20,0,360) 
        \CBoxc(280,60)(8,8){Red}{Red}
        \Line(245,55)(260,40)
        \Line(245,25)(260,40) 
        \Line(315,55)(300,40)
        \Line(315,25)(300,40)
        \Vertex(260,40){2}
        \Vertex(300,40){2}
      \end{picture}
    }
\caption{Diagrammatic representation of
  Eqs.~\eqref{AO.7}-\eqref{AO.9}. The diagrams look similar to
  one-loop perturbative diagrams with all internal propagators and
  vertices being fully dressed quantities. One internal line always
  carries the regulator insertion $\pat R_k$ (filled box). (One
  further diagram for $\pat\lambda_k$ involving a 6-point vertex is
  dropped, as in \Eqref{AO.9}.)}
\label{fig:FEdiag}
\end{figure}
}

It is instructive, to solve these three equations
\eqref{AO.7}-\eqref{AO.9} with various approximations. We begin with
dropping the anharmonic coupling $\lambda=0$, implying that
$\omega_k=\omega_{k=\Lambda}\equiv\omega$; integrating the remaining
flow of the ground state energy yields
\begin{equation}
E_0\equiv E_{0,k=0} =\int_0^\infty \D k\, \ddk E_{0,k}
\stackrel{\eqref{AO.7}}{=} \frac{1}{2}\, \omega, \label{AO.10}
\end{equation}
which corresponds to the ground state energy of the harmonic
oscillator, as it should ($\hbar=1$).

Now, let us try to do a bit better with the minimal nontrivial
approximation: we drop the running of the anharmonic coupling
$\lambda_k\to\lambda$, integrate the flow of the frequency
\eqref{AO.8}, insert the resulting $\omega_k$ into Eq.~\eqref{AO.7}
and integrate the energy. Expanding the result perturbatively in
$\lambda$, we find
\begin{equation}
E_0= \frac{1}{2} \omega + \frac{3}{4}\, \omega \left(
    \frac{\lambda}{24 \omega^3} \right) 
  - \frac{82}{40}\, \omega \left(
    \frac{\lambda}{24 \omega^3} \right)^2 + \dots,\label{AO.11}
\end{equation}
which can immediately be compared with direct 2nd-order perturbation
theory \cite{Bender:1969si},
\begin{equation}
E_0^{\mathrm{PT}}= \frac{1}{2} \omega + \frac{3}{4}\, \omega \left(
    \frac{\lambda}{24 \omega^3} \right) 
  - \frac{105}{40}\, \omega \left(
    \frac{\lambda}{24 \omega^3} \right)^2 + \dots.\label{AO.12}
\end{equation}
Our first-order ``one-loop'' result agrees exactly with perturbation
theory, whereas the second-order ``two-loop'' coefficient has the
right sign and order of magnitude but comes out too small with a $\sim
20$\% error. However, it should be kept in mind that we have obtained
this two-loop estimate from a cheap calculation which involved only
a one-loop integral with an RG-improved propagator.

Of course, we do not have to stop with perturbation theory. We can
look at the full integrated result based on Eqs. \eqref{AO.7} and
\eqref{AO.8} for any value of $\lambda$.  For instance, let us boldly
study the strong-coupling limit, where the asymptotics is known to be
of the form,
\begin{equation}
E_0=\left(\frac{\lambda}{24} \right)^{1/3} \left[ \alpha_0 + \mathcal
 O \left( \lambda^{-2/3} \right) \right]. \label{AO.13}
\end{equation}
The constant $\alpha_0$ has been determined in \cite{Janke:1995wt} to
a high precision by means of large-order variational perturbation
theory: $\alpha_0=0.66798\dots$.

With the simplest nontrivial approximation based on Eqs. \eqref{AO.7}
and \eqref{AO.8}, we obtain
$\alpha_0^{\mathrm{\eqref{AO.7},\eqref{AO.8}}}=0.6920\dots$, which
differs from the full solution by merely 4\%.  Now, solving all three
equations \eqref{AO.7}-\eqref{AO.9} simultaneously, we find
$\alpha_0^{\mathrm{\eqref{AO.7}-\eqref{AO.9}}}=0.6620\dots$, which
corresponds to a 1\% error.

\begin{figure}[t]
\begin{center}
\subfigure{}{\scalebox{1}[1]{
\includegraphics[width=5.5cm]{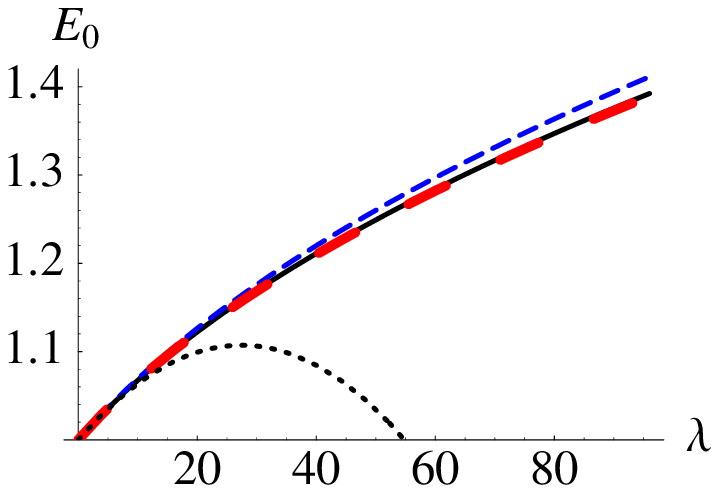}
}}
%\hspace{0.4cm}
\subfigure{}{\scalebox{1}[1]{
\includegraphics[width=5.5cm]{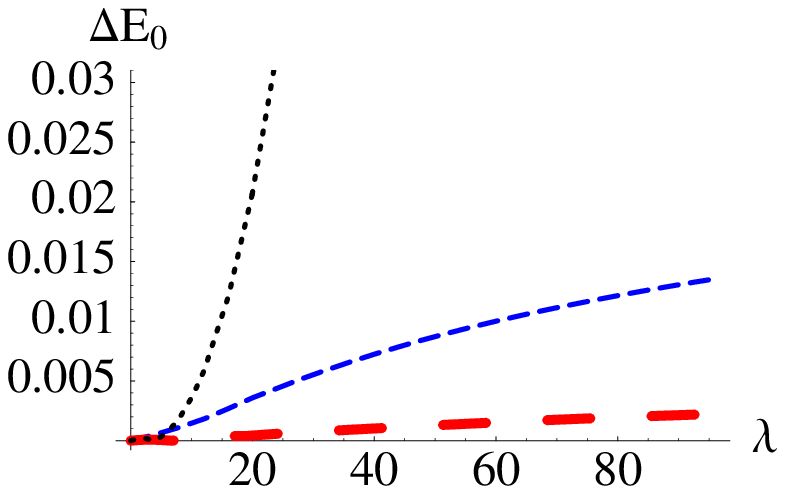}
}}
\end{center}
%\vspace{-1.0cm}
\caption{Left panel: ground state energy of the anharmonic oscillator
  for $\omega=2$ versus the anharmonic coupling $\lambda$: exact
  result (solid/black line), 2-loop perturbation theory (dotted/black
  line), flow-equation estimate based on
  Eqs.~\eqref{AO.7},\eqref{AO.8} (short-dashed/blue line) or on
  Eqs.~\eqref{AO.7}-\eqref{AO.9} (long-dashed/red line). Right panel:
  Relative error of the different estimates.}
\label{fig:AHO}
\end{figure}

A plot of the ground state energy is depicted in Fig.~\ref{fig:AHO}
(left panel) for $\omega=2$ and $\lambda=0\dots 100$. Whereas the
perturbative estimate (dotted line) is a good approximation for small
$\lambda$, it becomes useless for larger coupling. Already the
simplest approximation based on Eqs. \eqref{AO.7}
and \eqref{AO.8} (short-dashed/blue line) is a reasonable estimate
of the exact result (solid/black line), obtained from the exact
integration of the Schr\"odinger equation as quoted in
\cite{Galindo}. The flow equation estimate based on
Eqs.~\eqref{AO.7}-\eqref{AO.9} (long-dashed/red line) is hardly
distinguishable from the exact result. For better visibility, the
relative errors of these estimates are shown in Fig.~\ref{fig:AHO}
(right panel). Over the whole range of $\lambda$, the error of our
estimate based on Eqs.~\eqref{AO.7}-\eqref{AO.9} (long-dashed/red
line) does not exceed 0.3\%.

The quality of the result is remarkable in view of the extremely
simple approximation of the full flow. In particular, the truncation
to a low-order polynomial potential does not seem to be justified at
large coupling. In fact, there is no reason, why the dropped
higher-order terms, e.g., $\sim x^6$, should be small compared to the
terms kept.
 
The lesson to be learned is the following: it does not really matter
whether the terms dropped are small compared to the terms kept. It
only matters whether their influence on the terms that belong to the
truncation is small or large. 

\subsection{Further reading: regulator dependence and optimization${}^*$}
\label{ssec:FRO}

The reliability of the solution of a truncated RG flow is an important
question that needs to be addressed in detail for each application. In
absence of an obvious small expansion parameter, the convergence of
any systematic and consistent expansion of the effective action can be
checked by studying the quantitative influence of higher-order
terms. Whereas such computations can become rather extensive, an
immediate check can be performed by regulator studies. For the exact
flow, physical observables evaluated at $k=0$ are, of course,
regulator independent by construction. However, truncations
generically induce spurious regulator dependencies, the amount of which
provides a measure for the importance of higher-order terms outside a
given truncation. Resulting regulator dependencies of physical
observables can thus be used for a quantitative error estimate in a
rather direct manner.

Moreover, the freedom to choose the regulator function within the mild
set of conditions provided by Eqs.~\eqref{1.13a}-\eqref{1.13c} can
also actively be used for an optimization of the flow. A truncated
flow is optimized if the results for physical observables lie as close
as possible to the true results; the influence of operators outside a
given truncation scheme on the physical results is then minimized. If
a truncated flow is optimized at each order in a systematic expansion,
the physical results converge most rapidly towards the true result,
also implying that the optimized flows exhibit enhanced numerical
stability. Optimization of RG flows has conceptually been advanced in
\cite{Ball:1994ji,Liao:1999sh,%
Litim:2000ci,Canet:2002gs,Pawlowski:2005xe}. In particular, an
optimization criterion for the IR regulated propagator at vanishing
field based on stability considerations has lead to the construction
of optimized regulators for general truncation schemes
\cite{Litim:2000ci}; we have used such a regulator already in
\Eqref{AO.3}. A full functional approach to optimization has been
presented in \cite{Pawlowski:2005xe}. Loosely speaking, the optimal RG
flow within a truncation is identified with the most direct, i.e.,
shortest trajectory in theory space.

Since optimization helps improving quantitative predictions, enhances
numerical stability and can at the same time be used to reduce
technical effort, it is of high practical relevance. Optimization is
therefore recommended for all modern applications of the functional
RG.

\section{Functional RG for Gauge Theories}

\subsection{RG flow equations and symmetries}

Before we embark on the complex machinery of quantum field theories
with non-abelian gauge symmetries, let us study the interplay between
flow equations and symmetries from a more general perspective with
emphasis on the structural aspects. 

Consider a QFT which is invariant under a continuous symmetry
transformation which can be realized linearly on the fields; let
$\mathcal G$ be the generator of an infinitesimal version of this
transformation, i.e., $\mathcal G \phi$ is linear in $\phi$. For
example, a global O($N$) symmetry in a QFT for $N$-component scalar
fields is generated by
\begin{equation}
\mathcal{G}^a=-f^{abc} \int \D^D x\,\phi^b(x)\, \frac{\delta}{\delta
  \phi^c(x)}. \label{2.1}
\end{equation}
Incidentally, a local symmetry would be generated by the analogue of
\Eqref{2.1} without the spacetime integral. Together with the
invariance of the measure under this symmetry, cf. \Eqref{1.2}, the
invariance of the QFT can be stated by
\begin{equation}
0=\frac{1}{Z} \int \mathcal D \varphi\, \mathcal{G}\, \E^{-S+\int
  J\varphi}. \label{2.2}
\end{equation}
In other words, a transformation of the action can be undone by a
transformation of the measure. 

What does this symmetry of the QFT imply for the effective action?
First, we observe that \Eqref{2.2} yields
\begin{eqnarray}
0&=&\frac{1}{Z} \int \mathcal D \varphi \left( -(\mathcal G S) + \int
  J (\mathcal G \varphi)\right)\, \E^{-S+\int J\varphi} \nonumber\\
&=& -\langle \mathcal G S\rangle_J + \E^{-W[J]}\int J(\mathcal G
  \varphi)|_{\varphi=\frac{\delta}{\delta J}} \E^{W[J]}. \label{2.3}
\end{eqnarray}
Performing the Legendre transform, we find at $J=J_{\sup}\equiv
J[\phi]$:
\begin{equation}
0=-\langle \mathcal G S\rangle_{J[\phi]} + \int
  \frac{\delta\Gamma}{\delta\phi} \mathcal G
  \phi. \label{2.4a}
\end{equation}
The last term is nothing but $\mathcal G \Gamma$, and we obtain the
important identity (Ward identity) 
\begin{equation}
\mathcal G\Gamma[\phi] = \langle \mathcal G S \rangle_{J[\phi]}.
\label{2.4} 
\end{equation}
It demonstrates that the
effective action is invariant under a symmetry if the bare action as
well as the measure are invariant. 

On this level, the statement sounds rather trivial, but it can readily
be generalized to the effective average action by keeping track of the
regulator term,
\begin{equation}
\mathcal G\Gamma_k[\phi] = \langle \mathcal G (S+\varDelta
S_k)\rangle_{J[\phi]} -\mathcal G\varDelta S_k[\phi].
\label{2.5} 
\end{equation}
Here, we learn that the whole RG trajectory $\Gamma_k$ is invariant
under the symmetry if $\mathcal G S=0$ and $\mathcal G \varDelta
S_k=0$. This is the case if the regulator preserves the symmetry. For
instance for the globally O($N$)-symmetric theory, a regulator of
the form
\begin{equation}
\varDelta S_k= \frac{1}{2} \int \frac{\D^D p}{(2\pi)^D}\,
\varphi^a(-p) \delta^{ab} R_k(p) \varphi^b(p) \label{2.6}
\end{equation}
is such an invariant regulator. 

Note the following crucial point: since the regulator vanishes,
$\varDelta S_k\to0$, for $k\to 0$, \Eqref{2.5} appears to tell us that
$\mathcal G \Gamma_{k=0}[\phi] = \langle \mathcal G S
\rangle_{J[\phi]}$ always holds, implying that the symmetry is always
restored at $k=0$ even for a non-symmetric regulator. This is indeed
true, but requires that the initial conditions at $k=\Lambda$ have to
be carefully chosen in a highly non-symmetric manner, since
\begin{equation}
\mathcal G \Gamma_\Lambda[\phi] = \langle \mathcal G S \rangle +
\langle \mathcal G \varDelta S_{k=\Lambda} \rangle -\mathcal G
\varDelta S_{k=\Lambda}. \label{2.7}
\end{equation}
Here, the last two terms do generally not cancel each other.
Therefore, even for non-symmetric regulators, the Ward identity tells
us how to choose initial conditions with non-symmetric UV counterterms,
such that the latter are exactly eaten up by non-symmetric flow
contributions, see Fig.~\ref{fig:WardFlow}.

{\unitlength=1mm
\begin{figure}[t]
\centering
\begin{picture}(100,60)
\put(0,0){
\includegraphics[height=6cm]{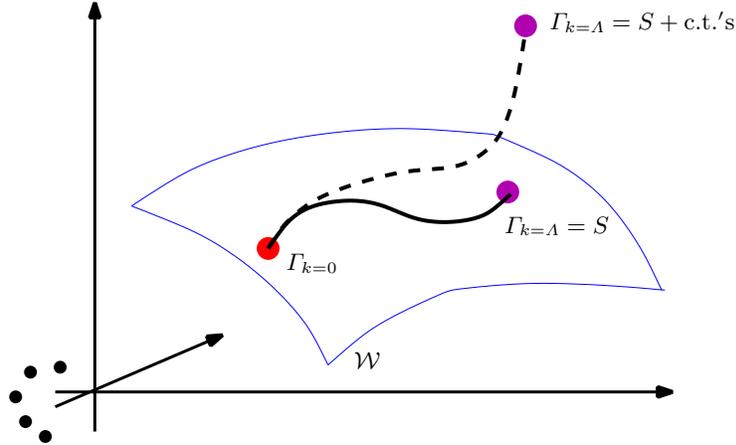}}
\put(73,55){$\Gamma_{k=\Lambda}=S + \text{c.t.'s}$}
\put(38,23){$\Gamma_{k=0}$} 
\put(67,28){$\Gamma_{k=\Lambda}=S$}
\put(47,10){$\mathcal{W}$} 
\end{picture}

\caption{Sketch of the RG flow in theory space with symmetries: the
  symmetry relation cuts out a hypersurface $\mathcal W$ where the
  action is invariant under the given symmetry. A symmetric regulator
  keeps the invariance explicitly, such that the RG trajectory always
  stays inside $\mathcal W$ (solid line). A non-symmetric regulator
  can still be used if non-symmetric counterterms (c.t.'s) are chosen
  such that they eat up the non-symmetric flow contributions (dashed
  line) by virtue of the Ward identity \Eqref{2.7}. Also in this case,
  the resulting effective action is invariant,
  $\Gamma_{k=0}\in\mathcal W$. }
\label{fig:WardFlow}
\end{figure}
}

In general, it is advisable to use a symmetric regulator, since the
space of symmetric actions is smaller, implying that fewer couplings
have to be studied in a truncation. However, if a symmetric regulator
is not available, the flow equation together with the Ward identity
can still be used. In fact, in gauge theories there is no simple
formalism with a symmetric regulator. 

\subsection{Basics of Gauge Theories}

Let us review a few basic elements of quantum gauge theories for
reasons of completeness and in order to introduce our notation. 

Consider the classical Yang-Mills action,
\begin{equation}
S_{\mathrm{YM}}= \int \D^D x\, \frac{1}{4} F_{\mu\nu}^a F_{\mu\nu}^a,
\label{2.8}
\end{equation}
with the field strength
\begin{equation}
F_{\mu\nu}^a =\partial_\mu A_\nu^a -\partial_\nu A_\mu^a + g f^{abc}
A_\mu^b A_\nu^c, \label{2.9}
\end{equation}
where the gauge field $A_\mu^a$ carries an internal-symmetry index
(color), and $f^{abc}$ are the structure constants of a compact
non-abelian Lie group. The hermitean generators $T^a$ of this group
form a Lie algebra and satisfy
\begin{equation}
[T^a,T^b]=\I f^{abc} T^c, \label{2.10}
\end{equation}
e.g., $a,b,c,=1,2,\dots,\Nc^2-1$ for the group SU($\Nc$). (In the
fundamental representation, the $T^a$ are hermitean $\Nc\times\Nc$
matrices which can be normalized to $\tr[T^a,T^b]=\frac{1}{2}
\delta^{ab}$.)

The naive attempt to define the corresponding quantum gauge theory by
\begin{equation}
Z[J] \stackrel{?}{=} \int\mathcal D A \, \E^{-S_{\mathrm{YM}}[A]+\int
  J_\mu^a A_\mu^a} \label{2.11}
\end{equation}
fails and generically leads to ill-defined quantities plagued by
infinities. The reason is that the measure $\mathcal D A_\mu^a$
contains a huge redundancy, since many gauge-field configurations
$A_\mu^a$ are physically equivalent. Namely, the action is invariant
under the local symmetry
\begin{equation}
A_\mu^a \to A_\mu^a -\partial_\mu \omega^a+gf^{abc} \omega^b
A_\mu^c\equiv A_\mu^a + \delta A_\mu^a,
\label{2.12}
\end{equation}
where $\omega^a(x)$ is considered to be infinitesimal and
differentiable, but otherwise arbitrary. The set of all possible
transformations forms the corresponding gauge group. The generator of
this symmetry is
\begin{equation}
\mathcal{G}_A^a(x) =D_\mu^{ab} \frac{\delta}{\delta A_\mu^b},
\label{2.13}
\end{equation}
where $D_\mu^{ab}=\partial_\mu\delta^{ab} + g f^{abc} A_\mu^c$ denotes
the covariant derivative in adjoint representation. It is a simple
exercise to show that $\int \D^D x\, \omega^b \mathcal{G}_A^b A_\mu^a
= \delta A_\mu^a$. The full symmetry transformation for
non-infinitesimal $\omega^a(x)$ can be written as  ($A_\mu\equiv A_\mu^a
T^a$)
\begin{equation}
A_\mu \to A_\mu^\omega = UA_\mu U^{-1} -\frac{\I}{g} (\partial_\mu U)
U^{-1}, \label{2.14}
\end{equation}
with $U=U[\omega]=\E^{-\I g \omega^a T^a}$ being an element of the Lie
group. Field configurations which are connected by \Eqref{2.14} are
\emph{gauge equivalent} and form the \emph{gauge orbit}:
\begin{equation}
[A_\mu^{\mathrm{orbit}}]=\{ A_\mu^\omega\, |\, A_\mu
=A_\mu^{\mathrm{ref}}, \quad U[\omega]\in
\text{SU}(\Nc)\}. \label{2.15} 
\end{equation}
Here, $A_\mu^{\mathrm{ref}}$ is a reference gauge field which is
representative for the orbit. 

In order to define the quantum theory, we would like to dispose of a
measure which picks one representative gauge-field configuration out
of each orbit. This is intended by choosing a gauge-fixing condition,
\begin{equation}
\mathsf{F}^a[A]=0.\label{2.16}
\end{equation}
For instance, $\mathsf{F}^a=\partial_\mu A_\mu^a$ is an example for a
Lorentz covariant gauge-fixing condition. Ideally, \Eqref{2.16} should
be satisfied by only one $A_\mu^a$ of each orbit. (As discussed below,
this is actually impossible for standard smooth gauge-fixing
conditions, owing to topological obstructions \cite{Singer:1978dk}).

Gauge fixing can be implemented in the generating functional by means
of the Faddeev-Popov trick which is usually derived from  
\begin{equation}
1=\int \mathcal D \mathsf{F}^a\, \delta[\mathsf{F}^a]=\int \D\mu(\omega)\,
\delta[\mathsf{F}^a(A^\omega)]\, \det \left( \frac{\delta
    \mathsf{F}^a[A^\omega]}{\delta \omega^b} \right), 
\label{2.17}
\end{equation}
where $\D\mu$ denotes the invariant Haar measure for an integration
over the gauge group manifold (at each spacetime point). This rule is
reminiscent to the corresponding rule for variable substitution in an
ordinary integral, $1=\int \D f\, \delta(f)= \int \D x\, \delta(f(x))
\, \left|\frac{\D f}{\D x} \right|$, for $f(x)$ having only one zero.
But as already this simple comparison shows, \Eqref{2.17} is
accompanied by the tacit assumption that $\mathsf{F}^a=0$ picks only
one representative and that the Faddeev-Popov determinant $\det(\delta
\mathsf{F}/\delta \omega)>0$.  Both assumptions are generally not
true; and both these properties, namely, that several gauge copies on
the same gauge orbit all satisfy a given standard gauge-fixing
condition and that the Faddeev-Popov determinant is not positive
definite, are characteristic of the famous Gribov problem
\cite{Gribov:1977wm}. A direct solution to this problem in the
functional integral formalism is by no means simple; nevertheless, let
us assume here that a solution exists which renders the gauge-fixed
functional integral well defined such that we can proceed with
deriving the flow equation. We will return to this problem later in
the discussion of the flow equation.

As discussed in standard textbooks, it is now possible to show that
the Faddeev-Popov determinant is gauge invariant,
$\Delta_{\text{FP}}[A^\omega]\equiv \det \frac{\delta
  \mathsf{F}^a[A^\omega]}{\delta \omega^b} =\Delta_{\text{FP}}[A]$. As
a consequence, $\delta[\mathsf{F}^a[A]] \Delta_{\text{FP}}[A]$
can be inserted into the functional integral, such that the redundancy
introduced by gauge symmetry is removed at least for perturbative
amplitudes; this renders the perturbative amplitudes well defined and
$S$ matrix elements are, in fact, independent of the gauge-fixing
condition. As a result, the Euclidean gauge-fixed generating
functional, replacing the naive attempt \Eqref{2.11}, becomes
\begin{equation}
Z[J]=\E^{W[J]}=\int \mathcal D A\, \Delta_{\text{FP}}[A]\,
\delta[\mathsf{F}^a[A]]\, \E^{-S_{\text{YM}}+\int JA}. \label{2.18}
\end{equation}
The additional terms can be brought into the exponent:
\begin{equation}
\delta[\mathsf{F}^a[A]]\to \E^{-\frac{1}{2\alpha} \int \D^D x
  \mathsf{F}^a\mathsf{F}^a} \big|_{\alpha\to 0} \equiv
  \E^{-S_{\text{gf}}[A]}, \label{2.19}
\end{equation}
where we have used a Gau\ss ian representation of the $\delta$
functional. The exponentiation of the Faddeev-Popov determinant can be
done with Grassmann-valued anti-commuting real \emph{ghost} fields $c,
\bar c$, yielding
\begin{equation}
\Delta_{\text{FP}}[A]=\int\mathcal D\bar c \mathcal D c \, \E^{-\int
  \D^D x\, \bar{c}^a \frac{\delta F^a}{\delta \omega^b} c^b} \equiv
  \int  \mathcal D\bar c \mathcal D c \,
  \E^{-S_{\text{gh}}}. \label{2.19b}
\end{equation}
The ghost fields transform homogeneously, 
\begin{equation}
c^a\to c^a+g f^{abc}\omega^b c^c, \quad
\bar{c}^a\to \bar{c}^a+g f^{abc}\omega^b \bar{c}^c, 
\label{2.20}
\end{equation}
as induced by a corresponding generator, 
\begin{equation}
\mathcal{G}_{\text{gh}}^a=-g f^{abc} \left( c^c \frac{\delta}{\delta
    c^b} + \bar{c}^c \frac{\delta}{\delta \bar{c}^b}
    \right). \label{2.20b}
\end{equation}
In perturbative $S$ matrix elements, the ghosts can be shown to cancel
unphysical redundant gauge degrees of freedom. 

Now, we generalize the generating functional by coupling sources also to
the ghosts, in order to treat them on the same footing as the gauge
field,
\begin{equation}
Z[J,\eta,\bar\eta]=\E^{W[J,\eta,\bar\eta]} 
=\int \mathcal D A\mathcal D c\mathcal D \bar{c}\, \E^{-S_{\text{YM}}
  -S_{\text{gh}} -S_{\text{gf}} +\int JA+\int \bar\eta c-\int
  \bar{c}\eta}. \label{2.21}
\end{equation}
The construction of the effective action $\Gamma$ now proceeds in a
standard fashion, 
\begin{equation}
\Gamma[A,\bar c, c] = \sup_{J,\eta,\bar\eta} \left( \int JA+\int
  \bar\eta c-\int \bar{c}\eta -W[J,\eta,\bar\eta]
  \right). \label{2.22}
\end{equation}
Since $\Gamma$ is the result of a gauge-fixed construction, it is not
manifestly gauge invariant. Gauge invariance is now encoded in a
constraint given by the Ward identity \Eqref{2.4} applied to the
present case with the generator
\begin{equation}
\mathcal{G}^a(x)=\mathcal{G}_A^a(x) + \mathcal{G}_{\text{gh}}^a(x)
=D_{\mu}^{ab} \frac{\delta}{\delta A_{\mu}^b} -gf^{abc} \left( c^c
  \frac{\delta}{\delta  c^b} + \bar{c}^c \frac{\delta}{\delta
    \bar{c}^b}  \right). \label{2.23}
\end{equation}
Since $\mathcal{G}^a S_{\text{YM}}=0$, the Ward identity boils down to
\begin{equation}
\mathcal W:= \mathcal{G}^a \Gamma[A,\bar c,c] -\langle \mathcal{G}^a
(S_{\text{gf}}+S_{\text{gh}})\rangle =0, \label{2.24}
\end{equation}
which in the context of nonabelian gauge theories is also commonly
referred to as Ward-Takahashi identity (WTI).  For instance in the
Landau gauge,
\begin{equation}
\mathsf{F}^a[A]=\partial_\mu A_{\mu}^a, \quad
\frac{\delta\mathsf{F}^a[A^\omega]}{\delta \omega^b} =-\partial_\mu
D_\mu^{ab}[A],\label{2.25}
\end{equation}
together with the gauge parameter $\alpha\to 0$, cf. \Eqref{2.19}, we
can work out this identity more explicitly by computing $\mathcal{G}^a
S_{\text{gf,gh}}$, e.g., in momentum space. This is done in
Subsect.~\ref{ssec:WTI}. As an example for how the Ward-Takahashi
identity constrains the effective action, let us consider a gluon mass
term, $\Gamma_{\text{mass}}=\frac{1}{2} \int m_{A}^2 A_\mu^a A_\mu^a$.
It can be shown order by order in perturbation theory that the
Ward-Takahashi identity enforces this gluon mass to vanish, $m_A^2=0$;
for more details, see Subsect.~\ref{ssec:WTI}. Hence, the gluon is
protected against acquiring a mass by perturbative quantum
fluctuations because of gauge invariance.

\subsection{RG flow equation for gauge theories}

From the gauge-fixed generating functional, the RG flow equation for
$\Gamma_k$ can straightforwardly be derived along the lines of
Subsect.~\ref{ssec:RGflow}. Using a regulator term,
\begin{eqnarray}
\varDelta S_k&=&\frac{1}{2} \int \frac{\D^D p}{(2\pi)^D}\, A_\mu^a(-p)\,
(R_{k,A})_{\mu\nu}^{ab}(p)\, A_\nu^b(p) \nonumber\\
&=& +\int \frac{\D^D p}{(2\pi)^D}\,
\bar{c}^a(p)\, (R_{k,\text{gh}})^{ab}(p)\, c^b(p), \label{2.29}
\end{eqnarray}
we obtain the flow equation
\begin{eqnarray}
\pat\Gamma_k[A,\bar c,c]&=&\frac{1}{2}\, \Tr\, \pat R_{k,A}
[(\Gamma_k^{(2)}+R_k)^{-1}]_A -\Tr \pat
R_{k,\text{gh}}[(\Gamma_k^{(2)}+R_k)^{-1}]_{\text{gh}} \nonumber\\
&\equiv& \frac{1}{2}\, \text{STr}\, \pat
R_k(\Gamma_k^{(2)}+R_k)^{-1}. \label{2.30}
\end{eqnarray}
The minus sign in front of the ghost term arises because of the
anti-commuting nature of these Grassmann-valued fields; the
super-trace in the second line of \Eqref{2.30} takes this sign into
account. In our notation in \Eqref{2.30}, $\Gamma_k^{(2)}$ is also
matrix-valued in field space, i.e., with respect to $(A,\bar c, c)$;
therefore, $[(\Gamma^{(2)}+R_k)^{-1}]_A$ denotes the gluon component
of the full inverse of $\Gamma^{(2)}+R_k$ (and not just the inverse of
$\delta^2\Gamma_k/\delta A\delta A +R_{k,A}$).

Is this a gauge-invariant flow? Manifest gauge invariance is certainly
lost, because the regulator is not gauge invariant; e.g., at small
$p$, the regulator -- here being similar to a mass term -- is
forbidden by gauge invariance, as discussed above.\footnote{One may
  wonder whether a gauge-invariant flow can be set up with a
  gauge-invariant regularization procedure. In fact, this is an active
  line of research, and various promising formalisms have been
  developed so far \cite{Morris:2000fs,Branchina:2003ek,%
    Pawlowski:2003sk}. However, the price to be paid for the resulting
  simple gauge constraints comes in the form of nontrivial Nielsen
  identities, non-localities or extensive algebraic constructions. For
  practical application, we thus consider the standard formulation
  described here as the most efficient approach so far.}  But manifest
gauge invariance is anyway lost, owing to the gauge-fixing procedure.
Gauge symmetry is encoded in the Ward-Takahashi identity.  From this
viewpoint, the regulator is merely another source of explicit
gauge-symmetry breaking, giving rise to further terms in the
Ward-Takahashi identity, the form of which we can directly read off
from \Eqref{2.5}:
\begin{equation}
\mathcal{W}_k:=\mathcal{G}\Gamma_k+\mathcal{G} \varDelta S_k 
-\langle \mathcal{G}(S_{\text{gf}}+S_{\text{gh}} +\varDelta
S_k)\rangle =0. \label{2.31}
\end{equation}
Owing to the additional regulator terms, this equation is called
\emph{modified Ward-Takahashi identity} (mWTI). With $\varDelta S_k$ being
quadratic in the field variables, these regulator-dependent
terms have a one-loop structure, since  $\langle \mathcal{G} \varDelta
S_k \rangle -\mathcal{G}\varDelta S_k$ corresponds to an integral over
the connected 2-point function with a regulator insertion $\sim
R_k$. Since the standard Ward-Takahashi identity already involves loop
terms (cf. Subsect. \ref{ssec:WTI}), the solution to \Eqref{2.31} is
no more difficult to find than that of the standard WTI. 

As before, we observe that
\begin{equation}
\lim_{k\to 0} \mathcal{W}_k \equiv \mathcal{W}, \label{2.32}
\end{equation}
such that a solution to the mWTI $\mathcal{W}_k=0$ satisfies the
standard WTI, $\mathcal{W}=0$, if the regulator is removed at $k=0$.
Such a solution is thus gauge invariant. Loosely speaking, the mWTI
$\mathcal{W}_k=0$ defines a modified gauge invariance that reduces to
the physical gauge invariance for $k\to0$. For a discussion of these
modified symmetry constraints from different perspectives, see e.g.,
\cite{Reuter:1993kw,Bonini:1993kt,Ellwanger:1994iz,%
D'Attanasio:1996jd,Bonini:1994kp,Litim:1998qi,Freire:2000bq,%
Igarashi:2001mf,Pawlowski:2005xe}, or the review \cite{Litim:1998nf}.

One further important observation is that the flow of the mWTI
satisfies 
\begin{equation}
\pat \mathcal{W}_k =-\frac{1}{2} \, G_k^{AB} \pat R_k^{AC} G_k^{CD}\,
\frac{\delta}{\delta \Phi^B} \frac{\delta}{\delta \Phi^D}\,
\mathcal{W}_k. \label{2.33}
\end{equation}
Here, we have used the collective field variable $\Phi=(A,\bar c, c)$.
The collective indices $A,B,C,\dots$ label these components, and
denote all discrete indices (color, Lorentz, etc.) as well as momenta;
e.g., the flow equation reads in this notation: $\pat
\Gamma_k=\frac{1}{2} \pat R_k^{AB} G_k^{BA}$.  The derivation of
\Eqref{2.33} from \Eqref{2.31} is indeed straightforward and a
worthwhile exercise.

Let us draw an important conclusion based on \Eqref{2.33}: if we
manage to find an effective action $\Gamma_k$ which solves the mWTI
$\mathcal{W}_k=0$ at some scale $k$, then also the flow of the mWTI
vanishes, $\pat \mathcal{W}_k=0$. In other words, the mWTI is a fixed
point under the RG flow. Now, if this $\Gamma_k$ is connected with
$\Gamma_{k'}$ at another scale $k'$ by the flow equation, also
$\Gamma_{k'}$ satisfies the mWTI at this new scale,
$\mathcal{W}_{k'}=0$. Gauge invariance at some scale therefore implies
gauge invariance at all other scales, if the corresponding
$\Gamma_k$'s solve the flow equation. The whole concept of the mWTI is
sketched and summarized in Fig.~\ref{fig:mWTI}.

{\unitlength=1mm
\begin{figure}[t]
\centering
\begin{picture}(100,60)
\put(0,0){
\includegraphics[height=6cm]{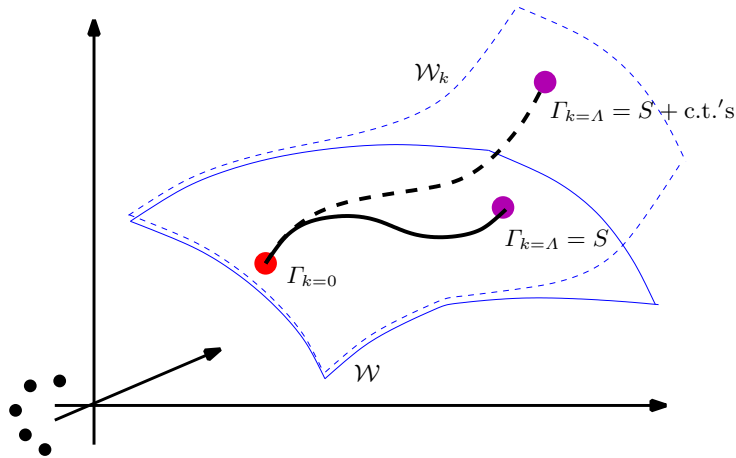}}
\put(73,45){$\Gamma_{k=\Lambda}=S + \text{c.t.'s}$}
\put(38,23){$\Gamma_{k=0}$} 
\put(67,28){$\Gamma_{k=\Lambda}=S$}
\put(47,10){$\mathcal{W}$} 
\put(55,50){$\mathcal{W}_k$} 
\end{picture}

\caption{Sketch of an RG flow with gauge symmetry in theory space: the
  standard Ward identity $\mathcal W$ cuts out a hypersurface of
  gauge-invariant action functionals. A gauge-invariant trajectory
  (solid line) would lie completely within this hypersurface. Instead,
  the presence of the regulator leads to the mWTI $\mathcal{W}_k$,
  cutting out a different hypersurface, which approaches $\mathcal W$
  in the limit $k\to 0$. A solution to the flow equation stays within
  the $\mathcal{W}_k$ hypersurface (dashed line), implying gauge
  invariance of the final full action $\Gamma\equiv \Gamma_{k=0}$.}
\label{fig:mWTI}
\end{figure}
}

Unfortunately, the picture is not as rosy as it seems for a simple
practical reason: in the general case, we will not be able to solve
the flow equation exactly. Hence, the identity \eqref{2.33} and thus
$\pat \mathcal{W}_k=0$ will be violated on the same level of accuracy.
This problem is severe if $\pat\mathcal{W}_k=0$ is violated by
RG-relevant operators, see Fig.~\ref{fig:mWTIrel}; the latter are
forbidden in the perturbative gauge-invariant theory.

{\unitlength=1mm
\begin{figure}[t]
\centering
\begin{picture}(100,60)
\put(0,0){
\includegraphics[height=6cm]{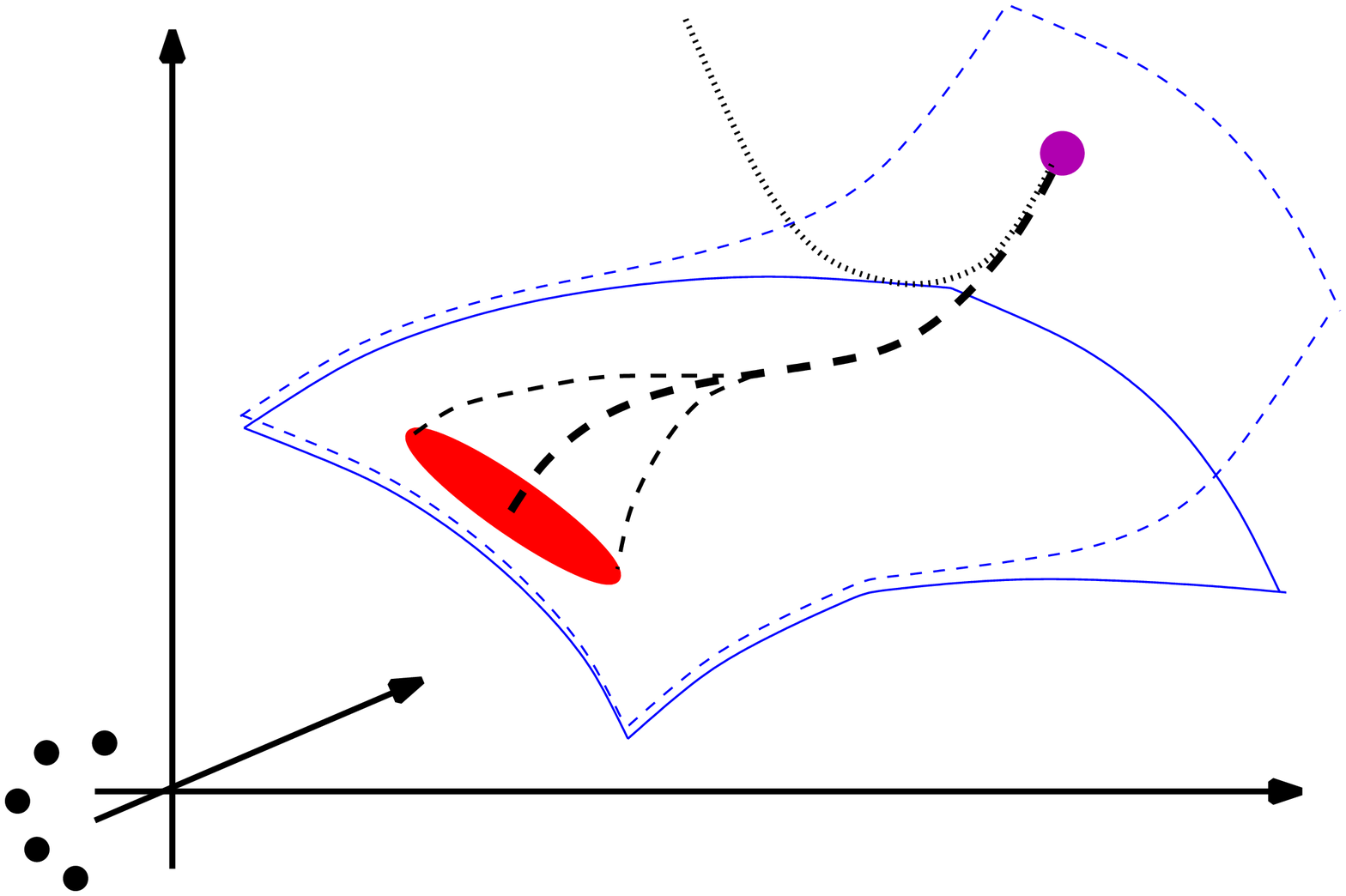}}
\put(73,45){$\Gamma_{k=\Lambda}=S + \text{c.t.'s}$}
\put(43,20){$\Gamma_{k=0}$}
\put(47,10){$\mathcal{W}$}
\put(55,50){$\mathcal{W}_k$}
\end{picture}

\caption{Sketch of an RG flow with gauge symmetry in theory space: a
  truncation of the effective action introduces an error, implying a
  possible range of estimates for $\Gamma_{k=0}$ as depicted by the
  extended ellipse. The error can also have a component orthogonal to
  the $\mathcal W$ or $\mathcal{W}_k$ hypersurface, if the Ward
  identities are solved on the same level of accuracy as the truncated
  flow.  A violation of the mWTI by RG relevant operators is
  particularly dangerous, since it can quickly drive the system away
  from the physical solution (dotted line) and thus must be avoided.}
\label{fig:mWTIrel}
\end{figure}
}

In the perturbative domain where naive power-counting holds, an RG
relevant operator is potentially given by the gluon mass term,
$\frac{1}{2} \int m_A^2 A_\mu^a A_\mu^a$. Gauge invariance in the form
of the standard WTI enforces $m_A^2=0$ as a consequence of $\mathcal
W=0$, as mentioned above. By contrast, such a bosonic mass term in a
system without gauge symmetry would receive large contributions from
fluctuations; perturbative diagrams are typically quadratically
divergent in such systems. In a naively truncated RG flow of a gauge
system, we can therefore expect the gluon mass to become large if the
gauge symmetry is not respected, $m_A^2\sim g^2\Lambda^2$.  Now, the
mWTI assists to control the situation: since gauge symmetry is not
manifestly present in our RG flow, we cannot expect the gluon mass to
vanish at all values of $k$. As discussed in more detail in
Subsect.~\ref{ssec:WTI}, the gluon mass becomes of order $m_A^2\sim
g^2 k^2$, as can be determined from $\mathcal{W}_k=0$.  As long as
perturbative power-counting holds, this implies that $m_A^2\to 0$ for
$k\to 0$, and the gauge constraint becomes satisfied in the limit when
the regulator is removed.
 
This consideration demonstrates that the mWTI can turn a potentially
dangerous relevant operator, which may appear in some truncation, into
an irrelevant harmless operator which dies out in the limit of
vanishing regulator. From another viewpoint, the mWTI tells us
precisely the right amount of gauge-symmetry breaking that we have to
put at the UV scale $\Lambda$ in the form of counterterms, such that
this explicit breaking is ultimately eaten up by the fluctuation-induced
breaking terms from the regulator, ending up with a perfectly
gauge-invariant effective action.

This simple gluon-mass example teaches an important lesson: for the
construction of a truncated gauge-invariant flow, the flow equation
and the mWTI should be solved simultaneously within the truncation. As
a standard recipe \cite{Ellwanger:1995qf,Ellwanger:1996wy}, the flow
equation can first be used to determine the flow of all
\emph{independent} operators (e.g., transverse propagators and
transverse vertex projections) which are not constrained by
$\mathcal{W}_k=0$. Then use the mWTI to compute the remaining
dependent operators (e.g., gluon mass, longitudinal terms).  In this
manner, the gauge constraint is explicitly solved on the truncation
and gauge-invariance of the truncation is guaranteed.  \footnote{An
  alternative option could be to use only the flow equation together
  with a regulator that does automatically suppress artificial
  relevant operators. In fact, this is conceivable in the framework of
  optimization \cite{Pawlowski:2005xe}.} It turns out that the use of
the mWTI (instead of the flow equation itself) for the computation of
a dependent operator generically corresponds to a resummation of a
larger class of diagrams \cite{Gies:2003dp}.

For the remainder of this subsection, let us return to the Gribov
problem discussed below \Eqref{2.17}. The fact that standard gauge
fixings do not uniquely pick exactly one representative of each gauge
orbit and that the Faddeev-Popov determinant hence is not positive
makes the nonperturbative definition of the functional integral
problematic. Any nonperturbative method which is related to the
functional integral such as the flow equation therefore appears to
face the same problem. As an example, let us concentrate on the Landau
gauge and consider the set of all gauge-field configurations that
satisfy the gauge-fixing condition $\mathsf{F}^a=\partial_\mu
A_\mu^a=0$; the Faddeev-Popov operator then is $\delta
\mathsf{F}^a[A^\omega]/\delta \omega^b=-\partial_\mu D_\mu^{ab}[A]$. A
perturbative expansion around $g\to0$ goes along with the
Faddeev-Popov operator at the origin of configuration space,
$-\partial_\mu D_\mu^{ab}[A]\to -\partial^2$; since this Laplacian is
a positive operator, the Gribov problem does not play a role in
perturbation theory to any finite order. 

Moving away from the origin, it is useful to consider the following
gauge-fixing functional, corresponding to the $L_2$ norm of the gauge
potential along the gauge orbit,
\begin{equation}
\mathcal{F}_A[\omega]\equiv ||A^\omega||^2 = ||A||^2 +2 \int_x
\omega^a \partial_\mu A_\mu^a + \int_x \omega^a (-\partial_\mu
D_\mu^{ab}) \omega^b + \mathcal{O}(\omega^3),
\label{gribov1}
\end{equation}
where we have expanded $A_\mu^\omega$ of \Eqref{2.14} to second order
in $\omega$. We can identify transverse gauge potentials that satisfy
the gauge condition as the stationary points of
$\mathcal{F}_A[\omega]$.  The Gribov problem implies that a gauge
orbit does not just contain one but many stationary points of
$\mathcal{F}_A[\omega]$. We observe that the subset of stationary
points given by (local) minima of $\mathcal{F}_A[\omega]$ corresponds
to a positive Faddeev-Popov operator $(-\partial_\mu D_\mu^{ab})>0$;
this subset constitutes the \emph{Gribov region} $\Omega_{\text{G}}$.
Restricting the gauge-field integration to the Gribov region,
$\int\mathcal D A \to \int_{\Omega_{\text{G}}} \mathcal D A$, cures
the most pressing problem of having a potentially ill-defined
generating functional owing to a non-positive Faddeev-Popov
determinant. Let us list some important properties of the Gribov
region, as detailed in \cite{Zwanziger:2003cf}: (i)
$\Omega_{\text{G}}$ contains the origin of configuration space and
thus all perturbatively relevant field configurations; (ii) the Gribov
region is convex and bounded by the \emph{(first) Gribov horizon}
$\partial\Omega_{\text{G}}$, consisting of those field configurations
for which the lowest eigenvalue of the Faddeev-Popov operator
vanishes; hence $\Delta_{\text{FP}}=0$ on $\partial\Omega_{\text{G}}$.

Does this restriction of the gauge-field integration to the Gribov
region modify the flow equation? In order to answer this question, let
us  go back to the functional-integral equation for the effective
action (without IR regulator) in \Eqref{1.8},
\begin{equation}
\E^{-\Gamma[\phi]}=\int \mathcal D \varphi\,
\exp\left(-S[\phi+\varphi] +\int \frac{\delta
    \Gamma[\phi]}{\delta\phi}\,\varphi \right), \label{gribov2}
\end{equation}
with the supplementary condition that $\langle \varphi\rangle=0$,
owing to the shifted integration variable,
cf. \Eqref{1.8}. Differentiating both sides with respect to $\phi$
yields
\begin{equation}
\frac{\delta \Gamma[\phi]}{\delta\phi} = \left\langle \frac{\delta
    S[\varphi+\phi]}{\delta \phi} \right\rangle_{J[\phi]},
    \label{gribov3}
\end{equation}
which is a compact representation of the Dyson-Schwinger
equations. Note that the same equation can be obtained from the
following identity:
\begin{equation}
0=\int\mathcal D \varphi \, \frac{\delta}{\delta\varphi}
\E^{-S[\phi+\varphi] + \int \frac{\delta\Gamma[\phi]}{\delta\phi}\,
  \varphi}, \label{gribov4}
\end{equation}
which holds, since the integrand is a total derivative. No boundary
terms appear here, because the action typically goes to infinity,
$S\to\infty$, for an unconstrained field $\varphi\to\infty$. 

Now, the crucial point for a quantum gauge theory with Faddeev-Popov
gauge fixing is that the identity corresponding to \Eqref{gribov4}
holds also for the constrained integration domain $\Omega_{\text{G}}$.
No boundary term arises, simply because the Faddeev-Popov operator and
thus the integrand vanishes on the boundary
$\partial\Omega_{\text{G}}$. We conclude that the Dyson-Schwinger
equations are not modified by the restriction to the Gribov region.
Finally, the same argument can be transfered to the flow-equation
formalism by noting that the effective average action has to satisfy
an identity similar to \Eqref{gribov2} including the regulator,
\begin{equation}
\E^{-\Gamma_k[\phi]-\varDelta S_k[\phi]}
=\!\int\! \mathcal D \varphi\,
\exp\left(\!-S[\phi+\!\varphi]-\!\varDelta S_k[\phi+\!\varphi] 
+\!\!\int\! \frac{\delta
    (\Gamma_k[\phi]+\!\varDelta S[\phi])}{\delta\phi}\,\varphi \right),
\label{gribov5} 
\end{equation}
with corresponding IR regulated Dyson-Schwinger equations,
\begin{equation}
\frac{\delta (\Gamma_k[\phi]+\varDelta S_k[\phi])}{\delta\phi} 
= \left\langle \frac{\delta
    (S[\varphi+\phi]+\varDelta S_k[\varphi+\phi])}{\delta \phi} 
  \right\rangle_{J[\phi]}.
    \label{gribov6}
\end{equation}
The latter can again be obtained from a functional integral over a
total derivative similar to \Eqref{gribov4}, indicating that a
restriction to the Gribov region does not modify \Eqref{gribov6}. The
final step of the argument consists in noting that the scale
derivative $\pat$ of \Eqref{gribov6} yields the flow equation (once
differentiated with respect to $\phi$). 

To summarize, solving quantum gauge theories by the construction of
correlation functions by means of functional methods (Dyson-Schwinger
equations or flow equations) precisely corresponds to an approach with
a build-in restriction to the Gribov region as an attempt to solve the
Gribov problem. From a flow-equation perspective, the argument can
even be turned around: taking the viewpoint that the quantum gauge
theory is defined by the flow equation, we can initiate the flow in
the perturbative deep UV where the Faddeev-Popov determinant is
guaranteed to be positive. Solving the flow, a resulting stable
trajectory necessarily stays within the Gribov region. 

Let us close this section with the remark that the picture developed
so far is not yet complete. The restriction to the Gribov region only
removes the problem of the non-positive Faddeev-Popov determinant. It
does not guarantee that we have integrated over the configuration
space by picking only one representative of each gauge orbit. In fact,
even within the Gribov region, there are still Gribov copies.
Therefore, the integration domain in gauge configuration space has to
be restricted even further by picking the global minimum of
$\mathcal{F}_A[\omega]$ in \Eqref{gribov1}. The resulting space is
known as the \emph{fundamental modular region} $\varLambda$. In
practice, the explicit construction of $\varLambda$ is difficult; for
instance, finding the global minimum of the gauge-fixing functional
$\mathcal{F}_A[\omega]$ on the lattice, corresponds to an extremely
involved spin-glass problem. However, it has been argued in
\cite{Zwanziger:2003cf} within a stochastic-quantization approach that
the problem of Gribov copies within the Gribov region does not affect
the correlation functions and their computation. This stresses even
further the potential of functional methods for nonperturbative
problems in gauge theories.

\subsection{Ward-Takahashi identity${}^*$}
\label{ssec:WTI}
 
The following subsection is devoted to a detailed discussion of the
gauge constraints in the form of the Ward-Takahashi identity (WTI) and
its modified counterpart in the presence of the regulator (mWTI). Let
us start with the standard WTI which we derived already in a compact
notation in \Eqref{2.24},
\begin{equation}
\mathcal W:= \mathcal{G}^a \Gamma[A,\bar c,c] -\langle \mathcal{G}^a
(S_{\text{gf}}+S_{\text{gh}})\rangle =0, \label{WTI1}
\end{equation}
which represents the gauge-symmetry encoding constraint that the
effective action $\Gamma$ has to satisfy in a gauge-fixed
formulation. In order to work out the single terms more explicitly,
it is useful to go to momentum space; we use the Fourier conventions
\begin{equation}
A_\mu^a(x)=\int_q \E^{\I qx} A_\mu^a(q), \quad c^a(x)=\int_q \E^{\I
  qx} c^a(q), \quad \bar{c}^a(x)=\int_q \E^{-\I qx}
  \bar{c}^a(q), \label{WTI2}
\end{equation}
where $\int_q\equiv \int \frac{\D^D q}{(2\pi)^D}$. This implies for
the functional derivatives, for instance,
\begin{equation}
\frac{\delta}{\delta A_\mu^a(x)}=\int_q \E^{-\I qx}
\frac{\delta}{\delta A_\mu^a(q)}, \quad \text{etc.} \label{WTI3}
\end{equation}
As a result, the generator of gauge transformations $\mathcal{G}^a$
reads in momentum space
\begin{eqnarray}
  \mathcal{G}^a(p)&=&
       \I p_\mu \frac{\delta}{\delta A_\mu^a(-p)}\label{WTI4}\\
&&    -g f^{abc}  \int_q \left[ A_\mu^c(q) \frac{\delta}{\delta
    A_\mu^b(q-p)} + c^c(q) \frac{\delta}{\delta c^b(q-p)}
      + \bar{c}^c(q) \frac{\delta}{\delta \bar{c}^b(q-p)} \right].
\nonumber
\end{eqnarray}
This allows us to compute the building blocks of the last two terms in
\Eqref{WTI1} in momentum space; we obtain the gauge transforms
\begin{eqnarray}
  \mathcal{G}^a(p)  S_{\text{gf}} &=&\!
  \frac{\I}{\alpha} p^2 p_\mu A_\mu^a(p) -\frac{1}{\alpha} g f^{abc}
  \!\!\int_q \!A_\mu^c(q) (p-q)_\mu (p-q)_\nu A_\nu^b(p-q),
  \label{WTI5}\\ 
  \mathcal{G}^a(p)  S_{\text{gh}} &=&-g f^{abc} \int_q \bar{c}^c(q)
  p\cdot(q+p) c^b(q+p) \nonumber\\
  &&-\I g^2 f^{feb} f^{abc}
  \int_{q_1,q_2} p_\mu \bar{c}^c (q_1) A_\mu^e (p+q_1+q_2) c^f(q_2).
  \label{WTI6} 
\end{eqnarray} 
Upon insertion into \Eqref{WTI1}, we arrive at a explicit
representation of the WTI in terms of full correlation functions in
the presence of a source $J$ which is field dependent by virtue of the
Legendre transform, $J=J_{\sup}=J[\phi]$,
\begin{eqnarray}
&&  \mathcal{G}^a(p)\left[ \Gamma_k -\int \left(\frac{1}{2\alpha}
      (\partial_\mu A_{\mu}^a)^2 + (\partial_\mu \bar{c}^a)  
      D_\mu^{ab}(A) c^b \right) \right] 
  \nonumber\\
  &&\quad = -\frac{1}{\alpha} g f^{abc} \int_q (p+q)_\mu (p+q)_\nu 
  \langle A_{\mu}^c(-q) A_{\nu}^b(p+q) \rangle_{\text{con}}\nonumber\\
  &&\qquad-g f^{abc} \int_{q_1,q_2} p_\mu \big( \delta(p+q_1-q_2)
  q_{2\mu} \delta^{bf} -\I g
  f^{bef}A_{\mu}^e(p+q_1-q_2)\big)\nonumber\\
  &&\qquad\qquad\qquad\times\langle
  \bar{c}^c(q_1) c^f(q_2)\rangle_{\text{con}} \nonumber\\
  &&\qquad -\I g^2 f^{abc}f^{feb} \int_{q_1,q_2} p_\mu\langle
  \bar{c}^c(q_1) A_{\mu}^e(p+q_1-q_2) c^f(q_2)\rangle_{\text{con}},
  \label{WTI7}
\end{eqnarray}
where $\langle \dots\rangle_{\text{con}}$ denotes only the connected
part of the correlation functions, e.g., $\langle \varphi
\varphi\rangle_{\text{con}} =\langle \varphi \varphi\rangle -\langle
\varphi \rangle \langle \varphi \rangle $. All terms on the right-hand
side are loop terms, the last term is even a two-loop term. It should
also be stressed that the WTI is expressed here in terms of
unrenormalized fields and couplings. 

As an example, let us see how the WTI imposes constraints on operators
in the effective action. For this, we discuss a gluon mass term in the
Landau gauge, $\alpha\to 0$. The gluon-mass operator reads,
\begin{equation}
\Gamma_{\text{mass}}=\frac{1}{2} \int m_A^2\, A_\mu^a
A_\mu^a. \label{WTI8}
\end{equation}
Its gauge transform yields
\begin{equation}
\mathcal{G}^a(p) \Gamma_{\text{mass}} = m_A^2\, \I p_\mu A_\mu^a(p). 
\label{WTI9}
\end{equation}
This implies that we have to project the remaining terms of the WTI
only onto the operator $\sim p_\mu A_\mu^a(p)$ in order to study the
gauge constraint on the gluon mass. Let us do so for the first term on
the right-hand side of \Eqref{WTI7} by way of example; shifting the
momentum $q$ by $q\to q-p$, this term reads
\begin{eqnarray}
&&-\frac{1}{\alpha} g f^{abc} \int_q q_\mu q_\nu 
  \langle A_{\mu}^c(p-q) A_{\nu}^b(q) \rangle_{\text{con}}|_{\alpha\to
  0} \nonumber\\
&&\qquad =-\frac{1}{\alpha} \int_q P_{\text{L},\mu\nu}(q)\, q^2\,
  G_{\mu\nu}^{cb}(p-q,q|A,c,\bar c)|_{\alpha\to 0}, \label{WTI10}
\end{eqnarray}
where we have introduced the longitudinal projector
$P_{\text{L},\mu\nu} =q_\mu q_\nu/q^2$ as well as the full gluon
propagator in the background of all fields
$G_{\mu\nu}^{cb}(p-q,q|A,c,\bar c)$. In view of \Eqref{WTI9}, this
propagator is needed only to linear order in the gauge field, 
\begin{eqnarray}
G_{\mu\nu}(p-q,q|A,c,\bar c)&=&G_{\mu\nu}(p-q,q) \nonumber\\
&&+ G_{\mu\kappa}(p\!-\!q,q\!-\!p)
V_{3A,\kappa\lambda\rho}(p\!-\!q,-p,q) G_{\rho\nu}(-q,q) A_\lambda(p)
\nonumber\\
&&+\mathcal{O}(A^2,\bar c c), \label{WTI11}
\end{eqnarray}
where $V_{3A}$ denotes the full three-gluon vertex, and all momenta
are counted as in-flowing. Inserting the order linear in $A$ of
\Eqref{WTI11} into \Eqref{WTI10}, we observe that both gluon
propagators are contracted with the longitudinal projector,
\begin{eqnarray}
&&-\frac{1}{\alpha} g f^{abc} \int_q q_\mu q_\nu 
  \langle A_{\mu}^c(p-q) A_{\nu}^b(q) \rangle_{\text{con}}|_{\alpha\to
  0} \nonumber\\
&&\qquad = -\frac{1}{\alpha} \int_q q^2 G_{\text{L},\mu\kappa} V_{3A,
  \kappa\lambda\rho} G_{\text{L}, \rho\mu} A_\lambda |_{\alpha\to 0}. 
\label{WTI12}
\end{eqnarray}
Now, the Landau gauge strictly enforces the gauge fields to be
transverse. Any longitudinal modes have to decouple in the
Landau-gauge limit; in particular, we have $G_{\text{L}}\sim \alpha\to
0$. As a consequence, the whole expression \eqref{WTI12} goes to zero
linearly with $\alpha$, at least order by order in perturbation
theory. We conclude that this first term on the right-hand side of the
WTI \eqref{WTI7} does not support a nonvanishing value for the gluon
mass. Let us mention without proof that the same property can be shown
also for all other terms on the right-hand side of \Eqref{WTI7}.
Therefore, the WTI enforces the gluon mass term to vanish to any order
in perturbation theory, $\Gamma_{\text{mass}}=0$, implying $m_A^2=0$.
The WTI protects the zero mass of the gluon against perturbative
quantum contributions, because of gauge invariance.

Let us now turn to the modifications of the gauge constraint in the
presence of a regulator. We already derived the regulator-modified WTI
(mWTI) in \Eqref{2.31}, 
\begin{equation}
\mathcal{W}_k:=\mathcal{G}\Gamma_k+\mathcal{G} \varDelta S_k 
-\langle \mathcal{G}(S_{\text{gf}}+S_{\text{gh}} +\varDelta
S_k)\rangle =0. \label{WTI13}
\end{equation}
For an explicit representation, we need the gauge transforms of the
regulator terms,
\begin{eqnarray}
  \mathcal{G}^a(p)  \Delta S_{k,A} &=&
  {i}  p_\mu (R_{k,A})_{\mu\nu}^{ab}(p) A_\nu^b(p) 
  -g f^{abc}\!\! \int_q\! A_\mu^c(q) (R_{k,A})_{\mu\nu}^{bd}(p\!-\!q)
  A_\nu^d(p\!-\!q), \nonumber\\
&&  \label{WTI14}\\ 
  \mathcal{G}^a(p)  \Delta S_{k,\text{gh}} &=&-g f^{abc} \int_q
  \bar{c}^c(q) [R_{k,\text{gh}}(q+p)-R_{k,\text{gh}}(q)]
  c^b(q+p). \label{WTI15} 
\end{eqnarray}
Using our previous result for the standard WTI \eqref{WTI7}, the mWTI
can now be displayed in the more explicit form, 
\begin{eqnarray}
  &&\mathcal{G}^a(p)\left[ \Gamma_k -\int \left(\frac{1}{2\alpha}
      (\partial_\mu A_{\mu}^a)^2 + (\partial_\mu \bar{c}^a)  
      D_\mu^{ab}(A) c^b \right) \right] 
  \nonumber\\
  &&\quad =-\frac{1}{\alpha} g f^{abc} \int_q (p+q)_\mu (p+q)_\nu 
  \langle A_{\mu}^c(-q) A_{\nu}^b(p+q) \rangle_{\text{con}} \nonumber\\
  &&\qquad-g f^{abc} \int_{q_1,q_2} p_\mu \big( \delta(p+q_1-q_2)
  q_{2\mu} \delta^{bf} -\I g
  f^{bef}A_{\mu}^e(p+q_1-q_2)\big)\nonumber\\
  &&\qquad\qquad\qquad\times\langle
  \bar{c}^c(q_1) c^f(q_2)\rangle_{\text{con}} \nonumber\\
  &&\qquad -\I g^2 f^{abc}f^{feb} \int_{q_1,q_2} p_\mu\langle
  \bar{c}^c(q_1) A_{\mu}^e(p+q_1-q_2) c^f(q_2)\rangle_{\text{con}}
  \nonumber\\ 
  &&\qquad
  -\frac{1}{2} g
  f^{abc}\int_q\big[(R_{k,A})_{\mu\nu}(p+q)-(R_{k,A})_{\mu\nu}(q)\big]  
  \langle A_{\mu}^c(-q) A_{\nu}^b(p+q) \rangle_{\text{con}}
  \nonumber\\
  &&\qquad
  -g f^{abc}\int_q\big[R_{k,\text{gh}}(p+q)-R_{k,\text{gh}}(q)\big] 
  \langle\bar{c}^c(q) c^b(q+p)\rangle_{\text{con}}  \quad
  \Big|_{\alpha\to 0}. \label{WTI16}
\end{eqnarray}
The last two terms denote the modification of the mWTI in comparison
to the standard WTI. These two terms are one-loop terms with a
structure similar to the flow equation itself. Both terms vanish in
the limit $k\to0$ and the standard WTI is recovered, as it
should. Again, we stress that the mWTI is expressed in terms of
unrenormalized fields and couplings. 

As an example, it would be straightforward to work out the precise
contribution of the regulator terms to the gluon mass which does not
vanish in contrast to the WTI result
\cite{Ellwanger:1996wy,Fischer:2004uk}. However, here it suffices to
estimate the order of magnitude of this contribution.  The structure
$R_k(p+q)-R_k(q)$, together with the fact that we need to project only
on the terms linear in $A$ and $p_\mu$ (cf.  \Eqref{WTI9}), implies
that the $q$ integral is peaked around $q^2\simeq k^2$. The
dimensionful scales on the right-hand side for the gluon mass are thus
set by $k^2$, resulting in $m_A^2\sim g^2 k^2$ with a proportionality
coefficient that depends on the regulator. This is a very unusual
bosonic-mass running which guarantees that the gluon mass is not a
relevant operator in the flow, but vanishes with $k\to 0$.

Let us close this subsection by briefly discussing the connection of
the present formalism with the BRST formalism. The latter involves one
further conceptual step, emphasizing BRST invariance as a residual
invariance of the gauge-fixed functional integral. The corresponding
symmetry constraints on the effective action, the \emph{Slavnov-Taylor
  identities}, have the advantage in the standard formulation that
they are bilinear in derivatives of the effective action. This allows
for an algebraic resolution of the gauge constraints in contrast to
the loop computations necessary for the WTI,
cf. \Eqref{WTI7}. If the regulator term is included, modified
Slavnov-Taylor identities can still be derived \cite{Reuter:1993kw,%
  Ellwanger:1994iz,D'Attanasio:1996jd,Bonini:1994kp}, but the result
no longer has a bilinear structure. We conclude that the BRST
formulation has no real advantage in the case of the RG flow equation,
such that the present formalism can fully be recommended also for
practical applications.

\subsection{Further reading: Landau-gauge IR propagators${}^*$}
\label{sec:FRLG}

The following paragraphs give a short introduction to recent lines of
research, and may serve as a guide to the literature. 

The functional RG techniques for gauge theories developed above can
now be used for computing the effective action in a vertex expansion,
cf. \Eqref{1.}. In momentum space, the expansion reads,
\begin{equation}
\Gamma_k[\phi]= \sum_{n}\frac{1}{n\!}\, \int_{p_1,\dots,p_n}\,
\delta(p_1+\dots+p_n) \, \Gamma_k^{(n)} (p_1,\dots,p_n)\, \phi(p_1) \dots
\phi(p_n), \label{V2.4}
\end{equation}
where $\int_p =\int d^D p/(2\pi)^{D}$, $\delta(p)=(2\pi)^{D}
\delta^{(D)}(p)$, and $\phi=(A_\mu^a,\bar{c}^a, c^a)$. Inserting
\Eqref{V2.4} into the flow equation \eqref{2.30}, we obtain an
infinite set of coupled first-oder differential equations for the
proper vertices $\Gamma_k^{(n)}$.  Truncating the expansion at order
$n_{\text{max}}$ leaves all equations for the vertices
$\Gamma_k^{(n\leq n_{\text{max}}-2)}$ unaffected. In order to close
this tower of equations, the vertices of order $n_{\text{max}}$ and
$n_{\text{max}}-1$ can either be derived from their truncated
equations or taken as bare or constructed by further considerations;
see, e.g., \cite{Blaizot:2006vr}.  This defines a consistent
approximation scheme that can in principle be iterated to arbitrarily
high orders in $n_{\text{max}}$.

Let us consider here the lowest nontrivial order, 
\begin{equation}
\Gamma_k=\frac{1}{2}\!\int_q \! A_\mu^a(-q)\! \left[ {Z_A(q^2)}{q^2}\,
  P_{\text{T}\mu\nu} + m_k^2 \delta_{\mu\nu} \right]\!  A_\nu^a(q)
        +\int_q \!\bar{c}^a(q)\, Z_{\text{gh}}(q^2){q^2}\, c^a(q) +
  \dots,\label{LG1}
\end{equation}
where $P_{\text{T}}$ is the transverse projector, and the ellipsis
denotes higher-order vertices and longitudinal gluonic terms. In the
following, we will confine ourselves to the Landau gauge $\alpha=0$
where longitudinal modes decouple completely; moreover, the Landau
gauge is known to be a fixed-point of the RG flow
\cite{Ellwanger:1995qf,Litim:1998qi}. The nontrivial ingredients
consist in the fully momentum-dependent wave function renormalizations
$Z_A(p^2)$ and $Z_{\text{gh}}(p^2)$, as well as a gluon mass term
which has been discussed in detail above. The flow equations for the
inverse of the transverse gluon and ghost propagators
($G_k=(\Gamma_k^{(2)}+R_k)^{-1}$),
\begin{equation}
\Gamma_{k,A\,\text{T}}^{(2)}(p^2) = Z_A(p^2)p^2 + m_k^2, \quad
\Gamma_{k,\text{gh}}^{(2)}(p^2) = Z_{\text{gh}}(p^2)p^2, \label{LG2}
\end{equation}
are shown in Fig.~\ref{fig:QCDprops}, except for diagrams involving
quartic vertices. The flow-equation diagrams are reminiscent to those
of Dyson-Schwinger equations. But there are two differences:
Dyson-Schwinger equations also involve two-loop diagrams, whereas the
flow equation imposes its one-loop structure also on the propagator
and vertex equations. Second, all internal propagators and vertices
are fully dressed quantities in the flow equation, whereas
Dyson-Schwinger equations always involve one bare vertex. Both vertex
expansions are slightly different infinite-tower expansions of the
same generating functional, with the flow equations being amended by
the differential RG structure.

{\unitlength=1pt
\begin{figure}[t]
  \centering
  \begin{eqnarray*}
    \pat\,\, 
    \begin{minipage}{1.9cm} 
      \begin{picture}(50,40)
        \Gluon(0,20)(25,20){4}{2.5}
        \Gluon(25,20)(50,20){4}{2.5}
        \GCirc(25,20){5}{0.5}
      \end{picture}
    \end{minipage}
    {}^{-1}  &=& -\,
    \begin{minipage}{3cm}
      \begin{picture}(60,40)
        \Gluon(0,20)(20,20){4}{2.5}
        \Gluon(60,20)(80,20){4}{2.5}
        \DashCArc(40,20)(20,180,360){2}
        \DashArrowArc(40,20)(20,0,90){2}
        \DashArrowArc(40,20)(20,90,180){2}
        \CBoxc(40,40)(10,10){Red}{Red} 
        \GCirc(40,0){5}{0.5}
        \GCirc(26,34){5}{0.5}
        \GCirc(54,34){5}{0.5} 
        \GCirc(20,20){5}{0.3} 
        \GCirc(60,20){5}{0.3}
      \end{picture}
    \end{minipage}
    \,+\,\frac{1}{2}\,
    \begin{minipage}{3cm}
      \begin{picture}(60,40)
        \Gluon(0,20)(20,20){4}{2.5}
        \Gluon(60,20)(80,20){4}{2.5}
        \GlueArc(40,20)(20,180,360){4}{5.5}
        \GlueArc(40,20)(20,0,90){4}{3.5}
        \GlueArc(40,20)(20,90,180){4}{3.5}
        \CBoxc(40,40)(10,10){Red}{Red} 
        \GCirc(40,0){5}{0.5}
        \GCirc(26,34){5}{0.5}
        \GCirc(54,34){5}{0.5} 
        \GCirc(18,20){5}{0.3}
        \GCirc(62,20){5}{0.3}
      \end{picture}
    \end{minipage}
    \,+\,\dots \\
    &&\\
    &&\\
    &&\\
    \pat\,\, 
    \begin{minipage}{1.9cm} 
      \begin{picture}(50,40)
        \DashArrowLine(0,20)(25,20){2}
        \DashArrowLine(25,20)(50,20){2}
        \GCirc(25,20){5}{0.5}
      \end{picture}
    \end{minipage}
    {}^{-1}  &=& -\,
    \begin{minipage}{3cm}
      \begin{picture}(60,40)
        \Gluon(0,20)(20,20){4}{2.5}
        \Gluon(60,20)(80,20){4}{2.5} 
        \DashCArc(40,20)(20,180,360){2}
        \GlueArc(40,20)(20,0,90){4}{3.5}
        \GlueArc(40,20)(20,90,180){4}{3.5} 
        \CBoxc(40,40)(10,10){Red}{Red} 
        \GCirc(40,0){5}{0.5}
        \GCirc(26,34){5}{0.5}
        \GCirc(54,34){5}{0.5} 
        \GCirc(20,20){5}{0.3} 
        \GCirc(60,20){5}{0.3}
      \end{picture}
    \end{minipage}
    \,+\,\dots
  \end{eqnarray*}
\caption{Flow equations for gluon and ghost propagators in a vertex
  expansion. All internal lines and vertices denote fully dressed
  quantities, indicated by filled circles. To each diagram, there is
  another one with identical topology but with the regulator insertion
  occurring at the opposite internal line. The ellipses denote
  diagrams involving quartic vertices which are not displayed.}
\label{fig:QCDprops}
\end{figure}
}

A truncation at this order requires information about the triple and
quartic vertices. In a minimalistic approach, they may be taken as
bare (possibly accompanied by $k$-dependent renormalization
constants). In general, this nonperturbative approximation is expected
to be reliable at weak coupling. For instance, the perturbative result
is rediscovered at high scales, $k=\Lambda_{\text{p}}$; for momenta
$p^2$ larger than this perturbative scale $\Lambda_{\text{p}}^2$,
the wave function renormalizations yield,
\begin{eqnarray}
Z_{A}(p^2)&\simeq& Z_{\Lambda_{\text{p}},A} \left( 1-\eta_A\,
  \frac{11\Nc \alpha_{\Lambda_{\text{p}}}}{12 \pi} 
  \ln \frac{p^2}{\Lambda_{\text{p}}^2}\right), \nonumber\\
Z_{\text{gh}}(p^2)&\simeq& Z_{\Lambda_{\text{p}},\text{gh}}
  \left( 1-\eta_{\text{gh}}\, \frac{11\Nc
\alpha_{\Lambda_{\text{p}}}}{12 \pi} \ln
  \frac{p^2}{\Lambda_{\text{p}}^2} \right),  
\label{LG3}
\end{eqnarray}
where $\eta_A=-13/22$ and $\eta_{\text{gh}}=-9/44$ denote the
anomalous dimensions for gluons and ghosts, respectively.
$Z_{\Lambda_{\text{p}},A},Z_{\Lambda_{\text{p}},\text{gh}}$ are the
normalizations of the fields, and $\alpha_{\Lambda_\text{p}}$ is the
value of the coupling constant at $\Lambda_{\text{p}}$.

At first glance, there is no reason why the higher-order vertex
structures which are dropped in the minimalistic truncation should not
become dominant at strong coupling.  However, as we learned from the
anharmonic oscillator example in Subsect.~\ref{ssec:AnhOsc}, the
quality of a low-order truncation does not depend on how large
higher-order terms may get but whether they exert a strong influence
on the low-order equations. Moreover, mechanisms may exist that
systematically suppress higher-order contributions, such as an IR
suppression of certain propagators; since higher-order vertex
equations involve more propagators, such a propagator suppression
would control large classes of diagrams.

In recent years, evidence has been provided that such a suppression is
indeed operative in the gluon sector in the Landau gauge: low-order
vertex expansions reveal an IR suppressed gluon propagator which
renders the contributions from higher gluonic vertices subdominant.
This solution has been pioneered in \cite{vonSmekal:1997is} using
truncated Dyson-Schwinger equations, see \cite{Alkofer:2000wg,%
Fischer:2006ub} for reviews; IR gluon suppression in the Landau gauge
has meanwhile been confirmed by many lattice simulations
\cite{Cucchieri:1997dx,Leinweber:1998uu,Alexandrou:2000ja,%
Langfeld:2001cz,Furui:2004cx,Sternbeck:2005tk,Silva:2005hb,%
Boucaud:2005ce}. In the continuum, this scenario goes along with an IR
enhanced ghost propagator, i.e., IR \emph{ghost dominance}. This IR
enhancement does not spoil the vertex expansion owing to a
nonrenormalization theorem for the ghost-gluon vertex
\cite{Taylor:1971ff}, the running of which is thus protected against
strong renormalization effects. Also ghost dominance has been observed
on the lattice \cite{Gattnar:2004bf,Sternbeck:2005tk}, even though
Gribov-copy and/or finite-volume/size effects appear to affect the IR
ghost sector more strongly.

The nonrenormalization theorem of the ghost-gluon vertex in the Landau
gauge gives rise to a nonperturbative definition of the running
coupling in terms of the wave function renormalizations,
\begin{equation}
\alpha(p^2)= \frac{g^2}{4\pi}\, \frac{1}{Z_A(p^2)\,
  Z_{\text{gh}}^2(p^2)}. \label{LG4}
\end{equation}
In the IR, the above-described scenario of ghost dominance and gluon
suppression is quantitatively observed in terms of a power-law
behavior of the wave function renormalizations,\footnote{In the
  Dyson-Schwinger literature, the gluon and ghost propagator behavior
  is often characterized by \emph{dressing functions}
  $Z_{\text{DSE}},G_{\text{DSE}}$ which are related to the wave
  function renormalizations by $Z_A(p^2)=Z_{\text{DSE}}^{-1}(p^2)$ and
  $Z_{\text{gh}}(p^2)=G_{\text{DSE}}^{-1}(p^2)$ for $k\to 0$.}
\begin{equation}
Z_A(p^2)\sim (p^2)^{-2\kappa}, \quad Z_{\text{gh}}(p^2)\sim
(p^2)^{\kappa}, 
\label{LG5}
\end{equation}
where $\kappa$ denotes a positive IR exponent. In the functional RG
framework, this solution can be shown to be a fixed point of the flow
equations for the propagators, cf. Fig.~\ref{fig:QCDprops}, in the
momentum regime $k^2\ll p^2\ll \Lambda_{\text{QCD}}^2$
\cite{Pawlowski:2003hq}. The interrelation of the ghost and gluon
propagators owing to the simultaneous occurrence of the exponent
$\kappa$ arises in all functional approaches from self-consistency
arguments; as a direct consequence, the running coupling \eqref{LG4}
approaches a fixed point in the IR, $\alpha(p^2\ll
\Lambda_{\text{QCD}}^2)\to \alpha_\ast$. For instance, a truncation
with a bare ghost-gluon vertex results in $\kappa\simeq 0.595$;
possibly induced momentum dependencies of the vertex can lead to
slightly lower values $0.5\leq\kappa\leq0.595$ \cite{Lerche:2002ep}.
Regulator dependencies arising in an RG calculation also lie in this
range \cite{Pawlowski:2004ip}.  The IR solutions \eqref{LG5} are IR
attractive fixed-point solutions for a wide class of initial
conditions and momentum-dependencies of the gluon vertices, once the
gluon propagator has developed a mass-like structure at intermediate
momenta at a few times $\Lambda_{\text{QCD}}$ \cite{Fischer:2004uk}.
Using suitable vertex \emph{ans\"atze}, full solutions connecting the
perturbative UV branch \Eqref{LG3} and the IR power-laws \eqref{LG5}
have been found with Dyson-Schwinger equations, see
\cite{Fischer:2002hn,Fischer:2006ub}. 

Most importantly, the IR power-law behavior featuring gluon
suppression and ghost dominance agrees with criteria which are
expected to be satisfied in two different scenarios of confinement:
the Kugo-Ojima \cite{Kugo:1979gm} and the Gribov-Zwanziger
\cite{Gribov:1977wm,Zwanziger:1991gz,Zwanziger:1993qr} confinement
scenario. In particular, a strongly IR divergent ghost propagator
represents a signature of confinement in these scenarios, which
describe the absence of color charged asymptotic states and
(indirectly) a linear rise of the potential between a static
quark-antiquark pair. The study of correlation functions in
connection with these scenarios of low-energy gauge theories clearly
demonstrates the potential of functional methods to access even the
strongly-coupled gauge sector by analytical means.

\section{Background-field flows}

Imagine for a second that we knew nothing about computing loops and
constructing amplitudes in some sort of expansion which involves a
perturbative or even a fully dressed propagator. If we knew only the
degrees of freedom of our gauge system and the symmetries we would be
trying to write down an effective action in terms of all possible
gauge-invariant gluon operators, such as $F_{\mu\nu}^a F_{\mu\nu}^a$,
$F_{\mu\nu}^a (D_\kappa D_\kappa)^{ab}F_{\mu\nu}^b$ etc. and determine
the coefficients, e.g., in a manner similar to chiral perturbation
theory.  The result would be gauge invariant by construction, and we
would never worry about Ward identities and how gauge invariance can
be encoded in a nontrivial manner in a gauge-fixed formulation.

\subsection{Background-field formalism}

The background-field formalism aims precisely at the construction of
such an effective action, nevertheless by computing loops and
integrating out fluctuations in a special gauge-fixed manner. Here is
a rough sketch of the idea:
\begin{itemize}
\item[(1)] Introduce an auxiliary, non-dynamical field $\bar A_\mu^a$
  (background field) with its own auxiliary symmetry transformation
  $\bG$. 
\item[(2)] Construct a gauge-fixed QFT which has broken invariance
  under the standard gauge transformation $\mG$ but a manifest
  invariance under $(\mG+\bG)$. 
\item[(3)] Let the full $\Gamma$ inherit the symmetry properties in
  the end by setting $A=\bar A$ after the gauge-fixed calculation. 
\end{itemize}
Let us start with (1): we introduce $\bar{A}_\mu^a$ and corresponding
covariant derivatives $\bar{D}_\mu^{ab} =\partial_\mu \delta^{ab} + g
f^{acb} \bar{A}_\mu^c$, and the generator of symmetry transformations
\begin{equation}
\bG^a(x)=\bar{D}_\mu^{ab} \frac{\delta}{\delta \bar{A}_\mu^b},
\label{3.1} 
\end{equation}
which we call the background transformation. Note that the ghosts are
not affected by $\bG$. Together with \Eqref{2.23}, it is obvious that
$(A-\bar{A})$ now transforms homogeneously under $\mG+\bG$, 
\begin{equation}
\int \D^D y\, \omega^b(y) (\mG+\bG)^b(y) (A-\bar{A})_\mu^a(x)
=gf^{abc} \omega^b(x) (A-\bar{A})_\mu^c(x).\label{3.2}
\end{equation}
As step (2), we choose a gauge fixing $\mathsf{F}^a$ which fixes the
$\mG$ symmetry but is invariant under $\mG+\bG$:
\begin{eqnarray}
\mathsf{F}^a&=&\bar{D}_\mu^{ab}(A_\mu^b-\bar{A}_\mu^b) \nonumber\\
&\Rightarrow&(\mG+\bG)S_{\text{gf}}=\frac{1}{2\alpha}\, (\mG+\bG) \int
\D^Dx\, [\bar{D}(A-\bar{A})]^2=0. \label{3.3}
\end{eqnarray}
In fact, the gauge-fixing term is invariant under the combined
transformation. With the Faddeev-Popov operator
\begin{equation}
\frac{\delta \mathsf{F}^a}{\delta \omega^b}=-\bar{D}_\mu^{ac}
D_\mu^{cb}, \label{3,4a}
\end{equation}
it is also straightforward to show that
\begin{equation}
(\mG+\bG)S_{\text{gh}}=-(\mG+\bG)\int \D^D x\, \bar{c}^a \bar{D}_\mu^{ac}
D_\mu^{cb} c^b =0. \label{3.4}
\end{equation}
Finally, let us merely sketch step (3): the price to be paid so far is
that $\Gamma$ now depends on $A$ and $\bar A$. But at the end of the
calculation, we can identify $A=\bar{A}$, such that 
\begin{equation}
0=(\mG+\bG)\Gamma[A,\bar{A}]\big|_{A=\bar{A}} =\mG \Gamma[A,A],
\label{3.5}
\end{equation}
where the first equality holds by construction and the second arises
from setting $A=\bar{A}$. Now, it is possible to prove that the
background effective action with $A=\bar{A}$ reduces precisely to the
standard effective action, $\Gamma[A,A]\equiv\Gamma[A]$
\cite{Abbott:1980hw,QFTtextbook2,Dittrich:1985tr}.  Therefore,
\Eqref{3.5} verifies the desired gauge-invariant construction of
$\Gamma[A]$. This $\Gamma[A]$ thus only consists of gauge-invariant
building blocks. Of course, there is still a nontrivial constraint
which becomes visible if we go away from the limit $A=\bar A$; namely,
$\mG\Gamma[A,\bar{A}]$ has to satisfy the standard WTI
\cite{Freire:2000bq}.

\subsection{Background-field flow equation}

The desired properties of the effective action expressed by
\Eqref{3.5} can be maintained in the construction of the corresponding
RG flow equation, if also the regulator satisfies
\begin{equation}
(\mG+\bG)\varDelta S_k=0. \label{3.6}
\end{equation}
This holds, e.g., for the choice 
\begin{equation}
\varDelta S_k=\frac{1}{2} \int (A-\bar{A})
R_{k,A}(\bar{\Delta}_A) (A- \bar{A}) + \int \bar c
R_{k,\text{gh}}(\bar{\Delta}_{\text{gh}}) c, \label{3.7}
\end{equation}
where $\bar{\Delta}_{A,\text{gh}}$ are operators that can depend
on $\bar{A}$ and transform homogeneously. For the gluon sector, a
suitable choice can, for instance, be given by
\begin{equation} 
(\bar{\Delta}_A)_{\mu\nu}^{ac}=\{ -\bar{D}_\kappa^{ab}
\bar{D}_\kappa^{bc} \delta_{\mu\nu}, \,
 -\bar{D}_\kappa^{ab} \bar{D}_\kappa^{bc} \delta_{\mu\nu} +2\I g
 (\bar{F}_{\mu\nu}^b T^b)^{ac}, \dots\}, \label{3.8}
\end{equation}
where the first form corresponds to the background-covariant
Laplacian, and the second also contains the spin-one coupling to the
background field. For the ghost sector, the Laplacian is also an
option, $\bar{\Delta}_{\text{gh}}=-\bar{D}_\kappa^{ab}
\bar{D}_\kappa^{bc}$. The resulting flow equation in the
background-field gauge reads \cite{Reuter:1997gx}
\begin{equation}
\pat \Gamma_k[A,\bar c,c,\bar{A}]=\frac{1}{2} \, \text{STr}\, \left\{
  \pat R_k(\bar{\Delta}) [ \Gamma_k^{(2)}[A,\bar{A}]
  +R_k(\bar{\Delta})]^{-1} \right\}. \label{3.9}
\end{equation}
Here, it is a temptation to set $A=\bar{A}$ in the flow equation;
however, the above construction tells us that this should be done only
at the end of the calculation at $k=0$. 

Let us parameterize (suppressing ghosts for a moment) the effective
action as \cite{Reuter:1997gx}
\begin{equation}
\Gamma_k[A,\bar{A}]
=\Gamma_k^{\text{inv}}[A]+\Gamma_k^{\text{gauge}}[A,\bar{A}],
\label{3.10}
\end{equation}
where $\Gamma_k^{\text{inv}}$ is a gauge-invariant functional, and
$\Gamma_k^{\text{gauge}}$ denotes the gauge-non-invariant remainder
that satisfies $\Gamma_k^{\text{gauge}}[A,\bar{A}=A]=0$,
cf. \Eqref{3.5}. Considering the second functional derivative with
respect to $A$, 
\begin{equation}
\Gamma_k^{(2)}[A,\bar{A}]
=\Gamma_k^{\text{inv}(2)}[A]+\Gamma_k^{\text{gauge}(2)}[A,\bar{A}],
\label{3.11}
\end{equation}
we observe that $\Gamma_k^{\text{inv}(2)}[A]$ must be singular. This
is because of the zero modes associated with gauge invariance: a
variation with respect to $A$ which points tangentially to the gauge
orbit has to leave $\Gamma_k^{\text{inv}}$ invariant, corresponding
to a flat direction. For the flow equation \eqref{3.9} to be well
defined, the contribution of $\Gamma_k^{\text{gauge}(2)}$ to the
denominator in \Eqref{3.9} has to lift these flat directions. This
demonstrates that $\Gamma_k^{\text{gauge}}$ must not be dropped from
a truncation, even though we may ultimately be interested only in
$\Gamma_k^{\text{inv}}$. 

On the other hand, the mWTI does not impose any constraint on
$\Gamma_k^{\text{inv}}$, 
\begin{equation}
0=\mathcal{W}_k[\Gamma_k]\equiv
\mathcal{W}_k[\Gamma_k^{\text{gauge}}], \label{3.12}
\end{equation}
which implies that we have the  full freedom to choose any
gauge-invariant functional as an ansatz for $\Gamma_k^{\text{inv}}$,
and its solution will solely be determined by the flow equation. 

To summarize, the background formalism facilitates the construction of
a gauge-invariant RG flow in which the manifestly gauge-invariant
parts of the effective action $\Gamma_k^{\text{inv}}$ can be separated
from the gauge-dependent parts $\Gamma_k^{\text{gauge}}$. In practice,
we can construct our ansatz for $\Gamma_k^{\text{inv}}$ by picking
gauge-invariant building blocks, in a similar manner as we would do
for other effective field theories. In addition, we have to construct
(a truncation for) $\Gamma_k^{\text{gauge}}$ with the aid of the mWTI
\eqref{3.12}, in order to lift the gauge zero modes in the flow
equation.

\subsection{Running coupling}

The background formalism also provides for a convenient
nonperturbative definition of the running coupling. As a general
remark, let us stress that there is no unique definition of the
running coupling in the nonperturbative domain. Universality of the
running coupling holds only near fixed points; e.g., only the one-loop
coefficient of the perturbative $\beta$ function is definition and
scheme independent (in a mass-independent regularization scheme, also
the two-loop coefficient is universal). Hence, any result for the
running coupling in the nonperturbative domain has to be understood
strictly in the context of its definition. 

The definition within the background formalism follows, for instance,
from the RG invariance of the background-covariant derivative, which
is a gauge-covariant building block of $\Gamma_k^{\text{inv}}$,
\begin{equation}
\bar{D}_\mu^{ab}[\bar{A}]=\partial_\mu \delta^{ab} + \bar{g} f^{abc}
\bar{A}_\mu^c. \label{3.13}
\end{equation}
Here, we have used the notation $\bar g$ for the bare
coupling. Obviously, the first term $\partial_\mu\delta^{ab}$ is RG
invariant. Hence, also the product of $\bar{g}$ and $\bar{A}$ must be
RG invariant \cite{Abbott:1980hw},
\begin{equation}
\pat (\bar{g} \bar{A}_\mu^c) =0, \quad \Rightarrow\quad
\bar{g}\bar{A}_\mu^c =g \bar{A}_{\text{R},\mu}^c, \label{3.13a}
\end{equation}
where $g$ now denotes the renormalized coupling and $\bar{A}_\text{R}$
the renormalized background field. Renormalization of the background
field is described by a wave function renormalization factor,
$\bar{A}_{\text{R}}=Z_k^{1/2} \bar{A}$. Consequently, also the running
of the coupling is tied to the same wave function renormalization,
\begin{equation}
g^2=Z_k^{-1} \bar{g}^2. \label{3.13b}
\end{equation}
We obtain for the $\beta$ function of the running coupling
\begin{equation}
\beta_{g^2}\equiv \pat g^2=\eta g^2, \quad \eta=-\pat \ln
Z_k,\label{3.14} 
\end{equation}
where $\eta$ denotes the anomalous dimension of the background field.
The wave function renormalization of the background field can be read
off from the kinetic term of the background gauge potential,
\begin{equation}
\Gamma_k^{\text{inv}}[A]=\int \frac{Z_k}{4} \, F_{\mu\nu}^a
F_{\mu\nu}^a +\dots, \label{Fqtrunc}
\end{equation}
where the dots represent further terms in the truncation. The running
coupling is thus linked to the lowest-order term of an operator
expansion of the effective action.

According to its definition, the running coupling can be viewed as the
response coefficient to excitations about the background field. If
the background field, as a natural choice, is associated with the
vacuum state the running coupling measures the coupling between the
vacuum and excitations, for instance, in the form of (effective) quark
and gluon fluctuations.  

\subsection{Truncated background flows}

Let us study the background flow in a simple approximation, following
\cite{Reuter:1996ub}. In particular, we will be satisfied by a minimal
truncation for $\Gamma_k^{\text{gauge}}$. For this, we expand
$\Gamma^{\text{gauge}}$ to leading order in $A-\bar A$,
\begin{equation}
\Gamma_k^{\text{gauge}}[A,\bar A] = \int (A-\bar A)_\mu^a\,
M_{\mu\nu}^{ab}\, (A-\bar{A})_\nu^b+ \mathcal O ((A-\bar
A)^3). \label{3.31}
\end{equation}
Then, we fix $M_{\mu\nu}^{ab}$ by the tree-level order of the mWTI:
\begin{equation}
\Gamma_k^{\text{gauge}}[A,\bar A]=\frac{1}{2\alpha} \int
(A-\bar{A})_\mu^a\, (-\bar{D}_\mu^{ac} \bar{D}_\nu^{cb})\,
(A-\bar{A})_\nu^b+\dots, \label{3.32}
\end{equation}
which is just the classical gauge-fixing term. This approximation has
an important consequence: to this order, 
\begin{equation}
\Gamma_k^{(2)}=\Gamma_k^{\text{inv}(2)}+\Gamma_k^{\text{gauge}(2)}
\label{3.33}
\end{equation}
is independent of $(A-\bar{A})$, and we can set $A=\bar{A}$ under the
flow for all $k$. The form of $\Gamma_k^{\text{gauge}}$ in
\Eqref{3.32} is just enough to lift the gauge zero modes. This gives
us an approximate flow for $\Gamma_k^{\text{inv}}$, 
\begin{eqnarray}
\pat \Gamma_k^{\text{inv}}[A]&=&\frac{1}{2} \, \text{Tr}\, \left\{
  \pat R_{k,A}(\Delta_A) \left[ \Gamma_k^{\text{inv}(2)} +
  \frac{1}{\alpha} (-DD) +R_k(\Delta) \right]^{-1}_A \right\}
  \nonumber\\
&&-\text{Tr}\, \left\{  \pat R_{k,\text{gh}}(\Delta_{\text{gh}}) \left[
  \Gamma_k^{\text{inv}(2)}  +R_k(\Delta) \right]^{-1}_{\text{gh}}
  \right\}, \label{3.34}
\end{eqnarray}
where we have dropped the bars on the right-hand side, since
$A=\bar{A}$. So far, we have suppressed the ghost fields
$c,\bar{c}$. If the dependence of $\Gamma_k^{\text{inv}}$ on the
ghosts is such that  ghosts are contracted with homogeneously
transforming color tensors, the mWTI is satisfied to the same level of
accuracy as by the ghost-independent part. The classical ghost action
in background-field gauge reads $S_{\text{gh}}=- \int \D^D x\,
\bar{c}^a \bar{D}^{ab} D^{bc} c^c$; hence, the lowest-order
approximation for our truncation is given by the classical term at
$A=\bar A$, 
\begin{equation}
\Gamma_{k,\text{gh}}^{\text{inv}}[A,\bar c,c]=-\int \D^Dx\, \bar{c}^a
{D}^{ab} {D}^{bc} c^c. \label{3.35}
\end{equation}
Let us now concentrate on the running coupling in the background
gauge. As discussed above, this can be read off from the kinetic terms
of the gauge field which we will therefore choose as the only
nontrivial part of our simplest nontrivial truncation, 
\begin{equation}
\Gamma_{k,\text{glue}}^{\text{inv}}[A]
=\int \frac{Z_k}{4} \, F_{\mu\nu}^a F_{\mu\nu}^a , \label{beta1}
\end{equation}
where the running of the wave function renormalization remains to be
calculated. This is done by projecting the right-hand side of the flow
equation onto this kinetic operator only; any other operator generated
by the flow, e.g., containing ghost fields, is dropped. Hence, we can
set $c,\bar c=0$ in $\Gamma^{(2)}$ (of course, \emph{after} functional
differentiation), implying that $\Gamma^{(2)}$ becomes block-diagonal
with respect to gluon and ghost sectors. In the gluon sector, we
obtain 
\begin{equation}
\delta^2 \Gamma^{\text{inv}}_k|_{\text{glue}} 
=  {Z_k}\int \D^D x\, \delta A_\mu^a\left[
  \mathcal{D}_{\text{T},\mu\nu}^{ab}+D_\mu^{ac}   D_\nu^{cb} 
-    \frac{1}{\alpha} D_\mu^{ac} D_\nu^{cb} \right] \delta A_\nu^b, 
\label{beta2}
\end{equation}
where we have introduced the notation
\begin{equation}
\mathcal{D}_{\text{T},\mu\nu}^{ab} = -D^{ac}_\kappa D^{cb}_{\kappa}
    \delta_{\mu\nu} +2 \bar{g} f^{abc} F_{\mu\nu}^{c}, 
\label{beta3}
\end{equation}
for the covariant spin-one Laplacian. Using the Feynman gauge
$\alpha=1$ here for simplicity, the gluon sector thus reduces to
\begin{equation}
( \Gamma^{\text{inv}(2)}_k )_{\mu\nu}^{ab}|_{\text{glue}}
=  {Z_k} \mathcal{D}_{\text{T},\mu\nu}^{ab}. 
\label{beta4}
\end{equation}
From \Eqref{3.35}, we can immediately read off the form of
$\Gamma^{(2)}$ in the ghost sector,\footnote{Be aware of footnote
  \ref{foot:grass} on page \pageref{foot:grass}.}
\begin{equation}
\delta^2 \Gamma^{\text{inv}}_k |_{\text{gh}}
=-\int \D^Dx\, \delta \bar{c}^a {D}^{ac} {D}^{cb} \delta c^b, \quad
\Rightarrow\quad (\Gamma^{\text{inv}(2)}_k)^{ab}|_{\text{gh}} 
=-D^{ac}_\kappa D^{cb}_{\kappa}. 
\label{beta5}
\end{equation}
Upon insertion into \Eqref{3.34}, the flow equation boils down to 
\begin{equation}
\pat \Gamma^{\text{inv}}_k[A]=\frac{1}{2} \,\Tr\, 
    \left[ \pat R_{k,A} ({Z_k}\mathcal{D}_{\text{T}}+R_{k,A})^{-1}
    \right]
    -\Tr\, 
    \left[ \pat R_{k,{\text{gh}}} (-D^2+R_{k,\text{gh}})^{-1}
    \right].
\label{beta6}
\end{equation}
A convenient choice for the regulator is given by (cf. \Eqref{1.13d})
\begin{equation}
  R_{k,A}={Z_k}\, \mathcal{D}_{\text{T}} \,
  r(\mathcal{D}_{\text{T}}/k^2) , \quad R_{k,\text{gh}} =
  (-D^2)\, r(-D^2/k^2), \quad \lim_{y\to 0} r(y) \to\frac{1}{y}.
\label{beta7}
\end{equation}
The insertion of the wave function renormalization $Z_k$ into the
regulator is useful, since it maintains the invariance of the flow
equation under RG rescalings of the fields (the ghost wave function
renormalization has been set to $Z_{k,\text{gh}}=1$ in our
truncation).  The choice $r(y) \to\frac{1}{y}$ implies that the IR
modes with $p^2\lesssim k^2$ are regulated by acquiring a mass term
$\sim k^2$; the identification of the one-loop running is more
straightforward with this choice. Then, the flow equation
reads
\begin{eqnarray}
\pat \Gamma^{\text{inv}}_k[A]&=&\frac{1}{2} \,\Tr\, 
    \left[ \frac{\pat (Z_{k} r(\mathcal{D}_{\text{T}}/k^2)) }
      {Z_{k}(1+ r(\mathcal{D}_{\text{T}}/k^2))} 
       \right]
    -\Tr\, 
    \left[ \frac{\pat r(-D^2/k^2) }
      {1+ r(-D^2/k^2)} 
    \right]\nonumber\\
&=:&\Tr\, \mathcal{H}_{Z_k}(\mathcal{D}_{\text{T}}/k^2) - 2 \,\Tr\,
    \mathcal{H} (-D^2/k^2).
\label{beta8}
\end{eqnarray}
The obvious definition of the $\mathcal H$ functions in the last line
expresses the fact that both terms on the right-hand side are traces
over a function of a single operator. It is useful to formally
introduce the Laplace transform of the $\mathcal H$ functions, 
\begin{equation}
\mathcal H (y) = \int_0^\infty ds\, \widetilde{\mathcal H}(s)\,
\E^{-ys}, \label{beta9}
\end{equation}
such that the flow equation can be written as
\begin{equation}
\pat \Gamma^{\text{inv}}_k[A]=
\int_0^\infty ds\,  \widetilde{\mathcal{H}}_{Z_k}(s)\,
\Tr\,\E^{-s(\mathcal{D}_{\text{T}}/k^2)}
 - 2 \int_0^\infty ds\, \widetilde{\mathcal H}(s)\,\Tr\,
    \E^{-s(-D^2/k^2)}.
\label{beta10}
\end{equation} 
As a result, we have brought the flow equation into \emph{propertime}
form. Note that this was possible because of the particular operator
structure resulting from our truncation: we were able to choose the
regulators such that they depend on the operators which appear as
entries of the block-diagonal $\Gamma^{(2)}$. Such a propertime form
of the flow can, for instance, always be established to leading order
in a derivative expansion of the RG flow \cite{Litim:2001hk}. In fact,
a wide class of truncations can be mapped onto a propertime form
\cite{Litim:2002xm,Gies:2002af} which is computationally advantageous;
moreover, propertime flows have extensively and successfully been used
in the literature \cite{Liao:1994fp,Floreanini:1995aj,Schaefer:em,%
Papp:2000he, Bonanno:2000yp,Zappala:2002nx}. In the present case, we
are finally dealing with traces over operator exponentials, so-called
\emph{heat kernels}, which can conveniently be dealt with by standard
methods.  The heat kernels of the covariant Laplacians occurring above
have frequently been studied in the literature; see, e.g.,
\cite{Gies:2002af}; here we merely need the term quadratic in the
field strength,
\begin{eqnarray}
\Tr\,\E^{-s(\mathcal{D}_{\text{T}}/k^2)} |_{F^2} &=&
\frac{\Nc(24-\!D)}{3} \frac{\bar{g}^2}{(4\pi)^{D/2}}
\left(\frac{s}{k^2}\right)^{2-(D/2)}\!\!
 \int\! \D^D x \frac{1}{4} F_{\mu\nu}^a F_{\mu\nu}^a, \nonumber\\
&&\label{beta11}\\
\Tr\,\E^{-s(-D^2/k^2)} |_{F^2} &=&
-\frac{\Nc}{3} \frac{\bar{g}^2}{(4\pi)^{D/2}}
\left(\frac{s}{k^2}\right)^{2-(D/2)} \int \D^D x \frac{1}{4} F_{\mu\nu}^a
F_{\mu\nu}^a, \label{beta12}
\end{eqnarray}
Let us from now on concentrate on four dimensional spacetime,
$D=4$. From \Eqref{beta10}, we can extract the flow of the wave
function renormalization
\begin{eqnarray}
\pat Z_k&=& \frac{20\Nc}{3} \frac{\bar{g}^2}{(4\pi)^{2}} \int_0^\infty
ds\, \widetilde{\mathcal H}_{Z_k}(s) + \frac{2\Nc}{3}
\frac{\bar{g}^2}{(4\pi)^{2}} \int_0^\infty 
ds\, \widetilde{\mathcal H}(s)\nonumber\\
&=& \frac{20\Nc}{3} \frac{\bar{g}^2}{(4\pi)^{2}} {\mathcal H}_{Z_k}(0)
 + \frac{2\Nc}{3} \frac{\bar{g}^2}{(4\pi)^{2}} \mathcal
 H(0),\label{beta13} 
\end{eqnarray}
where we have used \Eqref{beta9} for $y=0$. Together with $\pat(Z_k
r(\mathcal{D}_{\text{T}}/k^2)) = -Z_k[2\mathcal{D}_{\text{T}}/k^2
r'(\mathcal{D}_{\text{T}}/k^2) +\eta\,
r(\mathcal{D}_{\text{T}}/k^2)]$, and the anomalous dimension
$\eta=-\pat \ln Z_k$ as defined in \Eqref{3.14}, we observe that the
$\mathcal H$ functions for zero argument uniquely yield
\begin{equation}
\mathcal{H}_{Z_k}(0)= 1-\frac{\eta}{2}, \quad \mathcal{H}(0)=1,
\label{beta14} 
\end{equation}
independently of the precise form of the regulator shape function
$r(y)$ introduced in \Eqref{beta7}. This is a direct manifestation of
the regularization-scheme independence of the one-loop $\beta$
function coefficient, as will become clear soon.  

Introducing the renormalized coupling $g^2=Z_k^{-1} \bar{g}^2$, cf. 
\Eqref{3.13b}, we obtain
\begin{equation}
-\eta=\frac{22\Nc}{3} \frac{g^2}{(4\pi)^2} -\frac{10\Nc}{3}
 \frac{g^2}{(4\pi)^2}\, \eta.\label{beta15}
\end{equation}
This brings us to our final result for the Yang-Mills $\beta$ function
\begin{eqnarray}
\beta_{g^2}\equiv\pat g^2 &=& \eta g^2 = -\frac{22\Nc}{3}
\frac{g^4}{(4\pi)^2} \left( 1- \frac{10\Nc }{3} \frac{g^2}{(4\pi)^2}
\right)^{-1}\label{beta16}\\
&=& -\frac{22\Nc}{3}
\frac{g^4}{(4\pi)^2} -\frac{220\Nc^2}{9} \frac{g^6}{(4\pi)^4} -\dots
\label{beta17}
\end{eqnarray}
It is instructive to compare our result to the full perturbative
two-loop $\beta$ function:
\begin{equation}
\beta_{g^2}^{\text{2-loop}}= -\frac{22\Nc}{3}
\frac{g^4}{(4\pi)^2} -\frac{204\Nc^2}{9} \frac{g^6}{(4\pi)^4} -\dots
\label{beta18}
\end{equation}
Obviously, we have rediscovered the one-loop result exactly, as we
should. Moreover, the two-loop coefficient comes out remarkably well
with an error of only $8\%$. This is astonishing in two ways: first,
we have dropped a number of operators that contribute to the two-loop
coefficient coupling, e.g., $(F_{\mu\nu}^aF_{\mu\nu}^a)^2$ or
$(F_{\mu\nu}^a\widetilde{F}_{\mu\nu}^a)^2$, and also the ghost wave
function renormalization, which hence appear to be less relevant.
Second, whereas \Eqref{beta18} is a universal result obtained within a
mass-independent regularization scheme such as $\overline{\text{MS}}$,
our result arises from a mass-dependent regularization scheme with a
mass scale $k$; for such a scheme, the two-loop coefficient generally
is not universal. However, in the present truncation, our $\beta$
function indeed is universal to all orders computed above, owing to
the fact that the dependence of the regulator shape function drops out
in \Eqref{beta14}. We conclude that this simple truncation contains
already much relevant information about the universal part of this
two-loop coefficient. Larger truncations indeed contribute further
terms which are partly non-universal, as expected. An exact two-loop
calculation based on the functional RG can be found in
\cite{Pawlowski:2001df}.

In view of the quality of this simple truncation, it is tempting to
speculate whether the result also contains reliable nonperturbative
information. However, our result for the $\beta$ function in
\Eqref{beta16} develops a pole at $g^2=3(4\pi)^2/(10\Nc)$, clearly
signaling the breakdown of the truncation. This could already have
been expected from the physics content of the truncation:
independently of the coupling strength, the resulting quantum
equations of motion are identical to classical Yang-Mills theory. This
equations allow for plain-wave solutions, describing freely
propagating gluons. The truncation misses the important IR
phenomenon of confinement and thus cannot be reliable in the deeply
nonperturbative domain. For a stable flow from the UV down to the IR,
a larger truncation is required that in addition has the potential to
describe the relevant degrees of freedom for confinement. 

\subsection{Further reading: IR running coupling${}^*$}

The nonperturbative estimate of the $\beta$ function for the running
coupling derived above has a remarkable property. All higher-loop order
corrections can be summed into a geometric series, resulting in the
structure of \Eqref{beta16}. The origin of this structure lies in the
fact that we have included the wave function renormalization $Z_k$ in
the regulator, cf. \Eqref{beta7}. The apparent formal reason was that
the RG invariance of the flow equations is thus maintained; but in
addition, we thereby obtain a result which contains a resummation of a
larger class of perturbative diagrams. 

From a more physical viewpoint, the wave function renormalization
describes the deformation of the perturbative spectrum of
fluctuations, $S^{(2)}\sim p^2$, as it is induced by quantum
fluctuations; at lower scales, we find the spectrum
$\Gamma_k^{(2)}\sim Z_k p^2$. Therefore, the inclusion of $Z_k$ in the
regulator leads to a better adjustment of the regulator to the
deformation of the spectrum, i.e., to the \emph{spectral flow} of
$\Gamma_k^{(2)}$.

If $Z_k$ had not been included in the regulator, we would have found
only the one-loop $\beta$ function in the truncation \eqref{beta1}
without any higher-loop orders. The latter would have been encoded in
the flow of higher-order operators outside this simple truncation. For
instance, the inclusion of the higher-order operator $(F_{\mu\nu}^a
F_{\mu\nu}^a)^2$ would have resulted in a $\beta$-function estimate of
the form $\beta_{g^2}=\eta g^2$, with
\begin{equation}
\eta=-b_0 \frac{g^2}{(4\pi)^2} -b_1\, \frac{g^2}{(4\pi)^2}\, w_2,
\label{in5.1} 
\end{equation}
with some coefficient $b_1$, and $b_0$ being the correct one-loop
result, and $w_2$ denoting the generalized coupling of this
higher-order operator.  In this truncation, all nonperturbative
information is contained in the flow of $w_2$, which in turn can
reliably be computed only by including even higher-order operators.  A
good estimate therefore probably requires a very large truncation.
Even if the precise infrared values of the higher couplings $w_2, w_3,
\dots$ may not be very important, their flow exerts a strong influence
on the running coupling in this approximation.

Together with the inclusion of $Z_k$ in the regulator, our estimate
for the $\beta$ function would be of the form $\beta_{g^2}=\eta g^2$
with
\begin{equation}
\eta=-\frac{b_0 \frac{g^2}{(4\pi)^2} +b_1\, \frac{g^2}{(4\pi)^2} w_2}
{1+d_1\,\frac{g^2}{(4\pi)^2}+d_2\,\frac{g^2}{(4\pi)^2} w_2},
\label{in5.2}
\end{equation}
with a further coefficient $d_2$, and $d_1<0$ can be read off from
\Eqref{beta16}. Particularly this $d_1$ makes an important
contribution to the two-loop $\beta$ function coefficient, as
discussed above.  Contrary to \Eqref{in5.1}, this equation contains
information to all orders in $g^2$, even for the strict truncation
$w_2=0$, solely due to the spectral adjustment of the regulator.

Of course, for more complicated truncations, the deformation of the
spectrum of $\Gamma_k^{(2)}$ becomes much more involved. A natural
generalization would thus be the inclusion of the full
$\Gamma_k^{(2)}$ in the regulator. However, any dependence of $R_k$ on
the fluctuation would invalidate the derivation of the flow equation
in \Eqref{1.14}, especially spoiling the one-loop structure. But
within the background formalism, we can at least include
$\bar{\Gamma}_k^{(2)}$ where we have set all field dependence of
$\Gamma$ equal to the background field $\bar{A}$, 
\begin{equation}
R_k(\bar{\Gamma}_k^{(2)})=\bar{\Gamma}_k^{(2)}\,
r(\bar{\Gamma}_k^{(2)}/(Z_k k^2), \label{in5.2a}
\end{equation}
such that the regulator is fully adjusted to the spectral flow of the
fluctuation operator evaluated at the background field. The resulting
flow equation then contains also $\bar{\Gamma}_k^{(2)}$ derivatives;
for instance, at $A=\bar{A}$ where
$\Gamma_k^{(2)}=\bar{\Gamma}_k^{(2)}$, we find
\begin{eqnarray}
\pat\Gamma_k&=&\frac{1}{2}\, \text{STr} \left[ (2-\eta)\, \frac{- y\,
    r'(y)}{1+r(y)} 
+ \frac{\pat \Gamma_k^{(2)}}{\Gamma_k^{(2)}} \,
 \frac{r(y)+yr'(y)}{1+r(y)} \right]_{y=\frac{\Gamma_k^{(2)}}{Z_k k^2}}.
    \label{in3.5}
\end{eqnarray}
Despite the additional terms, this form of the flow equation (together
with the approximation of setting $A=\bar{A}$ already for finite $k$)
is technically advantageous, since it can be written in generalized
propertime form by means of a Laplace transformation, see the
discussion in the preceding section. 

The additional $\pat \Gamma_k^{(2)}$ terms are the generalization of
the $\eta$ term in \Eqref{beta14} and thus can support a further
resummation of a large class of diagrams. In \cite{Gies:2002af}, this
has been used together with an operator-expansion truncation involving
arbitrary powers of the field strength invariant
$(F_{\mu\nu}^aF_{\mu\nu}^a)^n$, $n=1,2,\dots$, to get a
nonperturbative estimate of the $\beta$ function. The anomalous
dimension has the structure,
\begin{equation}
 \eta=-\frac{b_0 \frac{g^2}{(4\pi)^2} +b_1\, \frac{g^4}{(4\pi)^4}+b_2
   \, \frac{g^6}{(4\pi)^6} +\dots}
{1+d_1\,\frac{g^2}{(4\pi)^2}+d_2\,\frac{g^4}{(4\pi)^4}+d_3
   \, \frac{g^6}{(4\pi)^6} +\dots },
\label{in5.3}
\end{equation}
the form of which can approximately be resummed in a closed-form
integral expression, see \cite{Gies:2002af}. The resulting $\beta$
function exhibits a second zero at $g^2\to g_\ast^2>0$, corresponding
to an IR fixed point, see Fig.~\ref{fig:alpha}. Hence the
flow-equation results for the IR running coupling in background gauge
based on an operator expansion show strong similarities to those in
the Landau gauge based on a vertex expansion mentioned in
Subsect.~\ref{sec:FRLG}. Within low-order perturbation theory, the
universality of the running coupling is a well-known property.
Independently of the different definitions of the coupling, the
one-loop (and in mass-independent schemes also the two-loop)
$\beta$-function coefficient is always the same. Here, we also observe
a qualitative agreement between the Landau gauge and the background
gauge in the nonperturbative IR in the form of an attractive fixed
point. This points to a deeper connection between the two gauges which
deserves further study and may be traced back to certain
non-renormalization properties in the two gauges.

{\unitlength=0.75mm
\begin{figure}[t]
\begin{center}
\begin{picture}(120,70) 
\put(-8,-10){
\includegraphics[width=9.75cm,height=6cm]{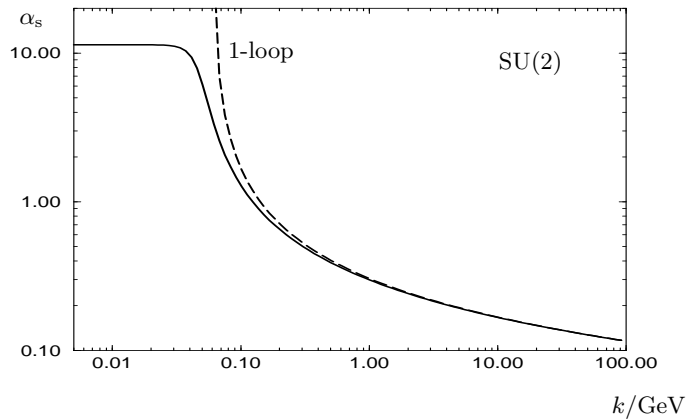}} 
\put(5,61){$\alpha_{\text{s}}$}
\put(90,53){SU(2)} 
\put(110,-8){$k/$GeV} 
\put(42,55){1-loop}
\end{picture} 
\end{center} 
\medskip

\caption{Running coupling $\alpha_{\text{s}}$ versus momentum scale
  $k$ in GeV for gauge group SU(2), using the initial value
  $\alpha_{\text{s}}(M_Z)\simeq0.117$ for illustration. The solid line
  represents the result of an infinite-order resummation of
  \Eqref{in5.3} as taken from \cite{Gies:2002af} in comparison with
  one-loop perturbation theory (dashed line).}
\label{fig:alpha}
\end{figure}}

The background formalism has also been used for finite-temperature
studies of Yang-Mills theory and the approach to chiral symmetry
breaking, \cite{Braun:2006jd,Braun:ECT}, and for a study of
nonperturbative renormalizability of gauge theories with extra
dimensions \cite{Gies:2003ic}. The background-field formalism lies
also at the heart of a series of RG flow-equation studies of quantum
gravity \cite{Reuter:1996cp}; it is, of course, also useful for the
study of abelian gauge theories such as the abelian Higgs model
\cite{Litim:1994jd} or strong-coupling QED \cite{Gies:2004hy}. An
alternative strategy to exploit a background field in RG flow
equations has been proposed in \cite{Litim:2002ce}.  Finally, the
spectral adjustment of the regulator described here is a special case
of the general strategy of functional optimization
\cite{Pawlowski:2005xe}, cf. Subsect.~\ref{ssec:FRO}.

\section{From Microscopic to Macroscopic Degrees of Freedom}

A typical feature of strongly interacting field theories is given by
the fact that macroscopic degrees of freedom can be very different
from microscopic degrees of freedom. For instance in QCD, quarks and
gluons represent the microscopic degrees of freedom, whereas
macroscopic degrees of freedom are mesons and baryons. The latter are
bound states of quarks and gluons. Prominent representatives are the
light pseudo-scalar mesons (pions, kaons, \dots) which carry
bi-fermionic quantum numbers, $\phi\sim \yb\psi$. This type of
fermionic pairing occurs in many systems, also in condensed-matter
physics with strongly correlated electrons.  Generically, a strong
fermionic (self-)interaction is required for this pairing. In QCD,
this is, of course, induced via the interactions with gluons.
For an efficient description of the physics, it is advisable to take
this transition from microscopic to macroscopic degrees of freedom
into account \cite{Polonyi:2000fv,Gies:2001nw,Schwenk:2004hm,%
Schutz:2004rn, Salmhofer1,Harada:2005tw}.

\subsection{Partial bosonization}
\label{ssec:partbos}

An explicit example for a fermion-to-boson transition is provided by
the Hubbard-Stratonovich transformation, or \emph{partial
  bosonization}. Let us discuss this transformation with the aid of a
specific system: the Nambu--Jona-Lasinio (NJL) model
\cite{NambuJonaLasinio}. We consider a version with one Dirac fermion,
defined by the action
\begin{equation}
S_{\text{F}}=\int \D^4x \left\{ \yb \I \fss{\partial} \psi +
  \frac{1}{2}\, \lb [(\yb\psi)^2 -(\yb\gamma_5 \psi)^2]\right\}.
  \label{4.1}
\end{equation}
This model has a U(1)$\times$U(1) symmetry which here plays a similar
role as chiral symmetry in QCD; in particular, it protects the
fermions against acquiring a mass due to fluctuations. 

Partial bosonization is obtained with the aid of the following mixed
fermionic-bosonic theory,
\begin{equation}
S_{\text{FB}}=\int \D^4x \left\{ \yb \I \fss{\partial} \psi + \mb^2
  \phi^\ast \phi + \hb [\yb P_{\text{L}} \phi \psi -\yb P_{\text{R}}
  \phi^\ast \psi ]\right\}, \label{4.2}
\end{equation}
with $P_{\text{L,R}}=(1/2)(1\pm \gamma_5)$ being the projectors onto
left- and right-handed components of the Dirac fermion. In fact, the
model \eqref{4.2} is equivalent to that of \eqref{4.1} also on the
quantum level if 
\begin{equation}
\mb^2=\frac{\hb^2}{2\lb}. \label{4.3}
\end{equation}
For a proof, it suffices to realize that 
\begin{eqnarray}
\int \mathcal D \phi\, \E^{-S_{\text{FB}}}
&=& \E^{-\int \yb \I\fss{\partial}\psi} \int \mathcal D\phi\, 
\E^{-\int \left(\phi^\ast + \frac{\hb}{\mb^2} \yb P_{\text{L}} \psi
  \right) \mb^2\left(\phi - \frac{\hb}{\mb^2} \yb P_{\text{R}} \psi
  \right)} \nonumber\\
&&\qquad\qquad\qquad \times\E^{-\int \frac{\hb^2}{\mb^2} (\yb
  P_{\text{L}}  \psi)(\yb P_{\text{R}} \psi)} \nonumber\\
&=& \mathcal N \, \E^{-\int \left(\yb \I\fss{\partial}\psi  +
    \frac{1}{4} \frac{\hb^2}{\mb^2} [(\yb\psi)^2-(\yb\gamma_5
    \psi)^2]\right)} \stackrel{\eqref{4.3}}{\equiv} \mathcal N\,
\E^{-S_{\text{F}}}, \label{4.4} 
\end{eqnarray}
where we have used that $(\yb P_{\text{L}} \psi)(\yb P_{\text{R}}
\psi)=(\yb\psi)^2-(\yb\gamma_5 \psi)^2$, and $\mathcal N$ abbreviates
the Gau\ss ian integral over $\phi$ which is a pure number and can be
absorbed into the normalization of the remaining fermionic integral
$\int \mathcal D\yb\mathcal D\psi\, \E^{-S_{\text{F}}}$. Also on the
classical level, the equations of motion display the fermionic
pairing, i.e., bosonization,
\begin{equation}
\phi= \frac{\hb}{\mb^2}\, \yb P_{\text{R}} \psi, \quad \phi^\ast =
-\frac{\hb}{\mb^2}\, \yb P_{\text{L}}\psi. \label{4.5}
\end{equation}
Equation \eqref{4.2} is the starting point for mean-field theory. The
fermionic integral is Gau\ss ian now, 
\begin{equation}
\int \mathcal D\phi \mathcal D\yb \mathcal D\psi\, \E^{-S_{\text{FB}}}
=\int \mathcal D\phi\, \E^{-S_{\text{B}}}, \label{4.6a}
\end{equation}
resulting in the purely bosonic action
\begin{equation}
S_{\text{B}}=\int \D^4x\left\{ \mb^2 \phi^\ast\phi -\ln \det [
  \I\fss{\partial} +\hb (P_{\text{L}}\phi -P_{\text{R}}\phi^\ast) ]
  \right\}. \label{4.6}
\end{equation}
Mean-field theory now neglects bosonic fluctuations and assumes that
the bosonic ground state corresponds to that of the classical bosonic
action. Of course, the $\ln\det$ is still a complicated nonlinear and
nonlocal expression; nevertheless, assuming that the ground state is
homogeneous in space and time, $\phi=$const., the determinant can be
computed by standard means \cite{QFTtextbook1}. For our purposes, it
suffices to know that for
\begin{equation}
\lb>\frac{8\pi^2}{\Lambda^2} \equiv \lb{}_{\text{cr}}, \label{4.6b}
\end{equation}
with $\Lambda$ being the UV cutoff, the resulting effective potential
$V_{\text{B}}(\phi^\ast\phi)$ has a nonzero minimum, implying a
nonzero vacuum expectation value $\langle\phi\rangle \neq0$. In the
fermionic language, this vacuum expectation value corresponds to a
bi-fermionic condensate $\langle\phi\rangle\sim \langle \yb\psi\rangle$
(a chiral condensate in the QCD context). The expectation value
generates fermion mass terms $\sim m_{\text{f}} \yb\gamma_5\psi$
with\footnote{The occurrence of $\gamma_5$ in the fermion mass term
arises from our fermion conventions \cite{Wetterich:1990an}; these are
related to more standard conventions by a discrete chiral rotation.}
\begin{equation}
m_{\text{f}} \sim \hb\langle\phi\rangle, \label{4.7}
\end{equation}
and the U(1)$\times$U(1) symmetry is spontaneously broken to U(1)
(fermion number conservation); this implies the existence of one
Goldstone boson corresponding to excitations of the phase of the
$\phi$ field. In the QCD context, this scenario corresponds to
the spontaneous break-down of chiral symmetry with the pseudo-scalar
mesons as Goldstone bosons. For $\lb<\lb{}_{\text{cr}}$, the vacuum
expectation value is zero, $\langle\phi\rangle=0$ and the system
remains in the symmetric phase. 

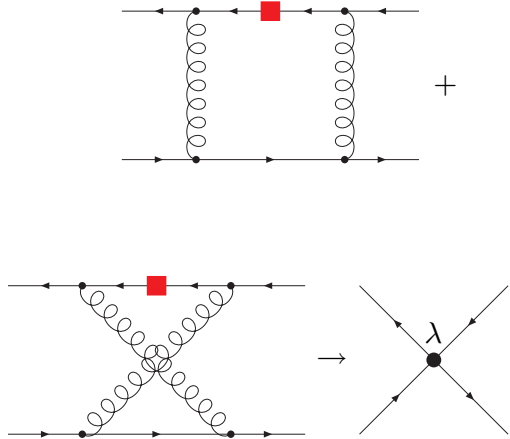
\begin{figure}[t]
\begin{center}
\subfigure{}{\scalebox{0.7}[0.7]{
\begin{picture}(190,140)(-10,0)
  \SetOffset(3,10)
  \ArrowLine(40,100)(0,100)
  \ArrowLine(120,100)(80,100)
  \ArrowLine(80,100)(40,100)
  \CBoxc(80,100)(10,10){Red}{Red} 
  \ArrowLine(160,100)(120,100)
  \ArrowLine(0,20)(40,20)
  \ArrowLine(40,20)(120,20)
  \ArrowLine(120,20)(160,20)
  \Vertex(40,100){2}
  \Vertex(120,100){2}
  \Vertex(40,20){2}
  \Vertex(120,20){2}
  \Gluon(40,100)(40,20){-5}{7.5}
  \Gluon(120,100)(120,20){5}{7.5}
  \Text(175,60)[c]{\scalebox{1.8}[1.8]{$+$}}
\end{picture}}}
%\hspace{1cm}
\subfigure{}{\scalebox{0.7}[0.7]{%\fbox{
\begin{picture}(190,140)(-10,0)
  \SetOffset(3,10)
  \ArrowLine(40,100)(0,100)
  \ArrowLine(120,100)(80,100)
  \ArrowLine(80,100)(40,100)
  \CBoxc(80,100)(10,10){Red}{Red} 
  \ArrowLine(160,100)(120,100)
  \ArrowLine(0,20)(40,20)
  \ArrowLine(40,20)(120,20)
  \ArrowLine(120,20)(160,20)
  \Vertex(40,100){2}
  \Vertex(120,100){2}
  \Vertex(40,20){2}
  \Vertex(120,20){2}
  \Gluon(40,100)(120,20){-5}{10.5}
  \Gluon(120,100)(40,20){5}{10.5}
  \Text(175,60)[c]{\scalebox{1.8}[1.8]{$\rightarrow$}}
\end{picture}}}%}
\subfigure{}{\scalebox{0.7}[0.7]{%\fbox{
\begin{picture}(120,140)(-10,0)
  \SetOffset(3,10)
  \ArrowLine(40,60)(0,100)
  \ArrowLine(0,20)(40,60)
  \ArrowLine(80,100)(40,60)
  \ArrowLine(40,60)(80,20)
  \Vertex(40,60){4}
  \Text(40,75)[c]{\scalebox{1.8}[1.8]{$\lb$}}
%  \Text(102,60)[c]{\scalebox{1.8}[1.8]{$\rightarrow$}}
\end{picture}}}%}
%\subfigure{}{\scalebox{0.7}[0.7]{%\fbox{
%\begin{picture}(190,140)(-10,0)
%  \SetOffset(3,10)
%  \ArrowLine(40,60)(0,100)
%  \ArrowLine(0,20)(40,60)
%  \DashLine(40,60)(80,60){2}
%  \ArrowLine(120,100)(80,60)
%  \ArrowLine(80,60)(120,20)
%  \Vertex(40,60){2}
%  \Vertex(80,60){2}
%\end{picture}}}%}
\end{center}
\vspace{-1.0cm}
\caption{Box diagrams with fundamental QCD interactions generate
  effective four-fermion self-interactions interactions $\lb$. (Only
  one diagram per topology is shown; further diagrams with the
  regulator insertion (filled box) attached to other internal lines,
  of course, also exist.) The resulting flow-equation contribution to
  the running of $\lb$ is given in \Eqref{4.8}.}
\label{fig:box}
\end{figure}

These models of NJL type with broken symmetry show many similarities
with QCD phenomenology, but there are important caveats from a
microscopic viewpoint.  First, there is no microscopic four-quark (or
higher) self-interaction beyond criticality in QCD, $\lb|_\Lambda\to
0$. In other words, $\lb [(\yb\psi)^2-(\yb\gamma_5\psi)^2]$ and other
four-fermion operators are RG irrelevant; the Euclidean microscopic
action for vanishing current quark masses is given by
\begin{equation}
S_{\text{QCD}}=\int \D^D x\left( \frac{1}{4} F_{\mu\nu}^a F_{\mu\nu}^a
  +\I \yb \fsl{D} \psi \right), \label{4.7a} 
\end{equation}
where $D_\mu=\partial_\mu + \I \bar{g} \tau^c A_\mu^c$ denotes the gauge
covariant derivative in the fundamental representation, which is
generated by the $\tau^a$ with $\tr[\tau^a\tau^b]=(1/2) \delta^{ab}$.
Of course, four-quark operators in the effective action are generated
by gluon exchange from fluctuations described by box diagrams; see
Fig.  \ref{fig:box}. The resulting $\beta$ function for the four-quark
coupling reads to lowest order
\begin{equation}
\pat \lb\equiv \beta_{\lb}=-c_\lambda\, \frac{1}{k^2} \, g^4,
\label{4.8}
\end{equation}
where $c_\lambda$ is a coefficient which depends on the algebraic
structure of the theory and the details of the IR regularization. For
definiteness, let us consider QCD with an SU(3) gauge sector but only
one massless quark flavor $\Nf=1$, the classical action of which has
the same ``chiral'' symmetry properties as the NJL model used above.
For this system, the coefficient $c_\lambda$ obtains $c_\lambda=5/(12
\pi^2)>0$ for the regulator \eqref{AO.3} and a Fierz decomposition as
chosen in \cite{Gies:2002hq}. Obviously, the four-quark
self-interaction $\lb$ is asymptotically free as it should be for a
QCD-like theory. Of course, for increasing gauge coupling $g$ towards
the IR, a naive extrapolation of \Eqref{4.8} predicts that the
fermionic self-interaction can become critical $\lb>\lb{}_{\text{cr}}$
for some IR scale $k_{\text{cr}}$. If this holds also in the full
theory, the quark self-interaction $\lb
[(\yb\psi)^2-(\yb\gamma_5\psi)^2]$ becomes strongly RG relevant at
scales below $k_{\text{cr}}$ and we expect the system to end up in the
symmetry-broken phase \cite{Aoki:1996fh}.

This scenario appears to match our expectations for QCD. But in order
to arrive at a quantitative description we do not only face the
problem of computational control in the nonperturbative domain;
moreover, we have to deal with the conceptual problem of how to switch
from the description in terms of quarks and gluons to that involving
boson fields as well. In view of the Hubbard-Stratonovich
transformation, we may be tempted to apply partial bosonization at
some scale $k_{\text{B}}<k_{\text{cr}}$. However, it turns out that
this leads to a strong spurious dependence on the precise choice of
$k_{\text{B}}$ in generic truncations. This has to do with the
following observation. Let us naively partially bosonize at
$k_{\text{B}}<k_{\text{cr}}$ . Diagrammatically,
\begin{eqnarray}
  k_{\text{B}}:\qquad \lb\!\!\!\!\!\!
  \begin{minipage}{1.6cm} 
    \scalebox{0.4}[0.4]{ 
      \begin{picture}(120,140)(-10,0)
        \SetOffset(3,10)
        \ArrowLine(40,60)(0,100)
        \ArrowLine(0,20)(40,60)
        \ArrowLine(80,100)(40,60)
        \ArrowLine(40,60)(80,20)
        \Vertex(40,60){4}
                                %  \Text(40,75)[c]{\scalebox{1.8}[1.8]{$\lb$}}
                                %  \Text(102,60)[c]{\scalebox{1.8}[1.8]{$\rightarrow$}}
      \end{picture}
    }
  \end{minipage}
&\Longrightarrow&\qquad \hb \!\!\!\!\!\!\!
\begin{minipage}{2cm}
  \scalebox{0.4}[0.4]{ 
    \begin{picture}(190,140)(-10,0)
      \SetOffset(3,10)
      \ArrowLine(40,60)(0,100)
      \ArrowLine(0,20)(40,60)
      \DashLine(40,60)(80,60){2}
      \ArrowLine(120,100)(80,60)
      \ArrowLine(80,60)(120,20)
      \Vertex(40,60){2}
      \Vertex(80,60){2}
    \end{picture}
  }
\end{minipage}
 \!\!\!\!\! \hb
\nonumber\\
S_{\text{F}}\quad&\Longrightarrow&\quad S_{\text{FB}}, \label{4.8a}
\end{eqnarray}
where $\lb=0$ in $S_{\text{FB}}$ after bosonization. Now, let us
perform another RG step  and integrate out another momentum shell
$\varDelta k$. Owing to the box diagrams, new quark self-interactions
are generated again in this RG step, 
\begin{equation}
k-\varDelta k: \quad \pat\lb =
\begin{minipage}{2cm}
  \scalebox{0.4}[0.4]{ 
    \begin{picture}(190,140)(-10,0)
      \SetOffset(3,10)
      \ArrowLine(40,100)(0,100)
      \ArrowLine(120,100)(80,100)
      \ArrowLine(80,100)(40,100)
      \CBoxc(80,100)(10,10){Red}{Red} 
      \ArrowLine(160,100)(120,100)
      \ArrowLine(0,20)(40,20)
      \ArrowLine(40,20)(120,20)
      \ArrowLine(120,20)(160,20)
      \Vertex(40,100){2}
      \Vertex(120,100){2}
      \Vertex(40,20){2}
      \Vertex(120,20){2}
      \Gluon(40,100)(40,20){-5}{7.5}
      \Gluon(120,100)(120,20){5}{7.5}
    \end{picture}
  }
\end{minipage}
\quad\sim g^4 \neq 0.\label{4.8b}
\end{equation}
In other words, the bosonizing field $\phi$ which did a perfect job at
$k_{\text{B}}$ is no longer perfect at $k-\varDelta k$; it does no
longer bosonize all quark self-interactions which it was introduced
for. Incidentally, this problem does not only occur if gauge
interactions are present, as \Eqref{4.8b} seems to suggest. The same
problem arises, e.g., in purely fermionic systems where fermion
self-interactions are generated by the flow in many different channels
also with nonlocal structure; since partial bosonization with a local
bosonic interaction can never account for all four-fermion vertices,
the remaining four-fermion structure will again generate the full
structure in the RG flow. 

In a regime where many couplings run fast, neglecting the newly
generated terms \eqref{4.8b} can introduce large errors. But keeping
these terms seems to make partial bosonization redundant. The solution
of this dilemma is the subject of the following subsection.

\subsection{Scale-dependent field transformations}

In Eqs. \eqref{4.8a},\eqref{4.8b}, we have observed that different
bosonizing fields are needed to compensate the quark self-interactions
at different scales. This points already to a solution of the problem
\cite{Gies:2001nw}: we promote the bosonizing field $\phi$ to be scale
dependent, $\phi\to\phi_k$, the flow of which can be written as
\begin{equation}
\pat \phi_k = \mathcal{C}_k [\phi,\psi,\yb,\dots]. \label{4.9}
\end{equation}
Here, $\mathcal{C}_k$ is an a priori arbitrary functional of possibly
all fields in the system. For the present problem, the idea is to
choose $\mathcal{C}_k$ such that the resulting effective action
$\Gamma_k[\phi_k]$ does not possess fermionic self-interactions. For
more general cases, the functional $\mathcal{C}_k$ can be chosen such
that the effective action $\Gamma_k[\phi_k]$ becomes simple, since its
simplicity is a strong criterion for the proper choice of the relevant
degrees of freedom.

We can formulate this idea in a differential fashion: we are looking
for a functional $\mathcal{C}_k$ which yields a flow of
$\Gamma_k[\phi_k]$ taken at fixed $\phi_k$, (suppressing further field
dependencies on $\yb,\psi,\dots$),
\begin{equation}
\pat \Gamma_k[\phi_k]|_{\phi_k}=\pat \Gamma_k[\phi_k] -\int
\frac{\delta\Gamma_k[\phi_k]}{\delta \phi_k}\, \pat\phi_k,
\label{4.10}
\end{equation}
such that the flow of the fermion self-interaction vanishes for all
$k$, $\pat \lb|_{\phi_k}=0$. Therefore, if $\lb=0$ holds at one scale
$k$, $\lb$ stays zero at all scales, and $\phi_k$ becomes the
``perfect'' boson an all scales. The right-hand side of \Eqref{4.10}
can be read as follows: the first term denotes the full RG flow given
in terms of a flow equation, whereas the second term with
$\pat\phi_k=\mathcal{C}_k$ characterizes how $\phi_k$ has to be
modified scale by scale in order to partially bosonize fermion
self-interactions on all scales. This fixes the functional form of
$\mathcal{C}_k$. 

Now, we need a flow equation for the effective action
$\Gamma_k[\phi_k]$ with scale-dependent field variables. This can
indeed be formulated in various ways. Here, we follow a general and
flexible exact construction given in \cite{Pawlowski:2005xe}.
Consider the modified generating functional
\begin{equation}
Z_k[J]=\E^{W_k[J]}
= \int\mathcal D\varphi\, \E^{-S[\varphi]-\frac{1}{2} \int \varphi_k
  R_k\varphi_k +\int J\varphi_k}, \label{4.11}
\end{equation}
where we have coupled a scale-dependent $\varphi_k$ to the source and
the regulator. This combination guarantees that the resulting flow
equation has a one-loop structure \cite{Litim:2002xm}. The scale
dependence of $\varphi$ is given by
\begin{equation}
\pat\varphi_k=\tilde{\mathcal{C}}_k[\varphi],\label{4.11a}
\end{equation}
similar to \Eqref{4.9} with the difference that \Eqref{4.11a} is
formulated under the functional integral, whereas \Eqref{4.9} holds for
the fields conjugate to the source $J$, 
\begin{equation}
\phi_k= \langle\phi_k\rangle\equiv \frac{\delta W_k[J]}{\delta
  J}. \label{4.11b} 
\end{equation}
Hence, $\mathcal{C}_k$ and $\tilde{\mathcal{C}}_k$ generally are not
identical.

The derivation of the flow of $W_k[J]$ is straightforward:
\begin{eqnarray}
\pat W_k[J]&=&\frac{1}{Z_k[J]} \int \mathcal D\varphi\, \left(
  J\pat\varphi_k -\frac{1}{2} \int \varphi_k \pat R_k\varphi_k -\int
  \varphi_k R_k \pat \varphi_k\right) \nonumber \\
&& \qquad\qquad  \times \E^{-S[\varphi] -\varDelta
  S_k[\varphi_k] +\int J\varphi_k} \nonumber\\
&=& J \langle \pat \varphi_k\rangle -\frac{1}{2} \text{Tr} \pat R_k
  G_k -\int\frac{\delta}{\delta J} R_k \langle
  \pat \varphi_k \rangle \nonumber\\
&&-\int \phi_k R_k \langle \pat \varphi_k \rangle -\frac{1}{2} \int
  \phi_k \pat R_k \phi_k. \label{4.12}
\end{eqnarray}
Here, we have defined the propagator similar to \Eqref{1.15}
\begin{equation}
G_k(p)= \left(\frac{\delta^2 W_k}{\delta J \delta J}\right)(p) =
\langle \varphi_k(-p) \varphi_k(p) \rangle -\phi_k(-p) 
\phi_k(p) . \label{4.12a}
\end{equation}
We have also used the relation $\langle \varphi_k \pat
\varphi_k\rangle = (\frac{\delta}{\delta J}+\phi_k) \langle \pat
\varphi_k \rangle$. As usual, we define the effective action by means
of a Legendre transformation, this time involving the scale-dependent
field variables (cf. \Eqref{1.16}) ,
\begin{equation}
\Gamma_k[\phi_k]= \sup_{J} \left(\int J \phi_k -W_k[J]\right)
-\frac{1}{2} \int \phi_k R_k\phi_k. \label{4.13}
\end{equation}
The resulting flow of this effective action is
\begin{equation}
\pat \Gamma_k[\phi_k]=\frac{1}{2} \text{Tr}\, \pat R_k G_k + \int
\left( G_k \frac{\delta}{\delta \phi_k} \right) R_k \langle \pat
\varphi_k \rangle + \int \frac{\delta \Gamma_k}{\delta\phi_k} ( \pat
\phi_k - \langle \pat \varphi_k \rangle), \label{4.14}
\end{equation}
where, in the course of the Legendre transformation, all $J$
dependence turns into a $\phi_k$ dependence by virtue of
$J=J_{\sup}=J[\phi_k]$; this also implies $\frac{\delta}{\delta J} =
G_k\frac{\delta}{\delta \phi_k}$. 

For a given scale-dependent field transformation $\pat
\varphi_k=\tilde{\mathcal{C}}_k[\varphi]$, we can successively work
out $\langle\pat\varphi_k\rangle$, $\phi_k=\langle\varphi_k\rangle$
and $\pat\phi_k=\pat \langle \varphi_k\rangle$; in general, the latter
is not identical to $\langle \pat \varphi_k\rangle$. Following this
strategy, we would only then be able to compute the flow of
$\Gamma_k[\phi_k]$. Also $\mathcal{C}_k$ would be a derived quantity,
fixed implicitly by the choice of $\tilde{\mathcal{C}}_k$.

By contrast, we can supplement the flow equation \eqref{4.14} with a
bootstrap argument: since all we want to choose in the end is $\pat
\phi_k= \mathcal{C}_k[\phi]$, the precise form of $\pat \varphi
=\tilde{\mathcal{C}}_k$ need not be known; in fact,
$\tilde{\mathcal{C}}_k$ does not occur directly in \Eqref{4.14}, but
only in expectation values. Therefore, we simply assume that a
suitable $\tilde{\mathcal{C}}_k$ exists for a desired $\mathcal{C}_k$
such that 
\begin{equation}
\langle \pat \varphi_k \rangle \stackrel{!}{=} \pat
\phi_k. \label{4.15pre}
\end{equation}
Of course, this is a highly implicit construction, and in view of the
complicated structure of the mapping $\tilde{\mathcal C}_{k} \to
\mathcal{C}_k$, the existence of a suitable $\tilde{\mathcal{C}}_k$
for an arbitrary $\mathcal{C}_k$ is generally not guaranteed or, at
least, difficult to prove. Nevertheless, since the resulting flow
equation will, in practice, be used together with a truncation, it is
reasonable to assume that \Eqref{4.15} can at least be satisfied to
the order of the truncation. As a consequence of \Eqref{4.15}, the flow
equation simplifies, 
\begin{eqnarray}
\pat \Gamma_k[\phi_k]|_{\phi_k}&\stackrel{\eqref{4.10}}{=}&
\pat \Gamma_k[\phi_k] -\int
\frac{\delta\Gamma_k[\phi_k]}{\delta \phi_k}\, \pat\phi_k
\label{4.15}\\
&=&
\frac{1}{2}\, \text{Tr}\, \pat R_k G_k + \int \left(
  G_k\frac{\delta}{\delta \phi_k} \right) R_k \pat \phi_k
-\int \frac{\delta\Gamma_k[\phi_k]}{\delta \phi_k}\, \pat\phi_k,
\nonumber 
\end{eqnarray}
which is the desired flow equation for scale-dependent field
variables. Apart from the standard first term $\sim \text{Tr}\pat R_k
G_k$ and the third term which carries the explicit scale dependence of
$\phi_k$, we encounter the second term which takes care of fluctuation
contributions to the renormalization flow of the operator insertion
$\pat \varphi_k$ in the functional integral. Actually, this second
term will generally be subdominant for not too large coupling: first,
it is of higher order in the coupling, and second, $R_k$ insertions
lead to weaker numerical coefficients than $\pat R_k$ insertions for
standard regulators. We expect that this term does not induce strong
modifications for couplings up to $\mathcal{O}(1)$. 

\subsection{Scale-dependent field transformations for QCD:
  Rebosonization}  

Let us now turn back to our original problem of QCD-like systems,
where quark self-interactions are generated by gluon exchange. In
order to arrive at a mesonic description in the infrared, we now want
to apply scale-dependent field transformations that translate the
quark self-interactions into the bosonic sector an all scales -- a
process that may be termed partial \emph{rebosonization}. 

In order to illustrate the formalism, let us study one-flavor QCD in a
simple truncation. Apart from the standard kinetic terms for the quark
and gluons, supplemented by wave function renormalization factors, we
include a point-like four-quark self-interaction in the
scalar--pseudo-scalar sector\footnote{Of course, in order to avoid any
  ambiguity with respect to possible Fierz rearrangements of the
  four-fermion interactions in the point-like limit, all possible
  linearly-independent four-fermion interactions, in principle, have
  to be included in the truncation. For simplicity, we confine
  ourselves here just to the scalar--pseudo-scalar channel, where chiral
  condensation is expected to occur. For the four-fermion interactions
  that will be generated by the flow, we use the Fierz decomposition
  as proposed in \cite{Gies:2002hq}.}
\begin{equation}
\Gamma_{\text{F},k}=\int \D^4 x\left( \frac{Z_k}{4} F_{\mu\nu}^a
  F_{\mu\nu}^a 
  +\I Z_\psi \yb \fsl{D} \psi + \frac{1}{2}\, \lb [(\yb\psi)^2
  -(\yb\gamma_5 \psi)^2] \right). \label{4.15a} 
\end{equation}
The initial condition of the four-quark operator in the UV is
obviously given by $\lb|_{k\to\Lambda}\to 0$.  As a first step towards
rebosonization, we include a complex mesonic scalar field in the
truncation on equal footing,
\begin{eqnarray}
\Gamma_k=\Gamma_{\text{F},k}+\int\D^4 x\left( Z_\phi\partial_\mu
  \phi^\ast \partial_\mu \phi + V(\phi^\ast\phi) 
+ \hb [\yb P_{\text{L}} \phi \psi -\yb P_{\text{R}}  \phi^\ast \psi ]
  \right), \label{4.15b}
\end{eqnarray}
with a scalar potential $V(\phi^\ast\phi)=\mb^2 \phi^\ast\phi +
\mathcal{O}((\phi^\ast\phi)^2)$.  The initial conditions for the scalar
field at $k\to\Lambda$ need to be chosen such that the scalar has no
observable effect on the QCD sector whatsoever. This is easily done by
demanding that the Yukawa interaction with the quarks vanishes
$\hb|_{k\to\Lambda}\to 0$. We also choose a large scalar mass
$\mb^2|_{k\to\Lambda}= \mathcal{O}(\Lambda^2)$, ensuring a fast
decoupling of the scalar; finally, we set $Z_\phi|_{k\to\Lambda}\to 0$
which makes the scalar non-dynamical at the UV scale. Solving the flow
with these initial conditions, the scalars rapidly decouple and only
the standard QCD flow remains, revealing the purely formal character
of this first step towards rebosonization. 

As a second step, we now use the freedom to perform scale-dependent
field transformations, as suggested in \Eqref{4.9}. We promote the
field $\phi$ to be scale dependent, and choose the functional
$\mathcal{C}_k$ characterizing this scale dependence to be of the
form 
\begin{equation}
\pat \phi_k= \yb P_{\text{R}}\psi\, \pat \alpha_k, \quad \pat
\phi_k^\ast=-\yb P_{\text{L}}\psi\, \pat \alpha_k, \label{4.16}
\end{equation}
with some function $\alpha_k$ to be determined below. At this stage,
let us study the consequences of the last term in \Eqref{4.15} on the
resulting flow; note that we are now dealing with a complex field,
such that this term goes over into, $\int
\frac{\delta\Gamma_k}{\delta\phi_k} \pat \phi_k\to \int
\frac{\delta\Gamma_k}{\delta\phi_k} \pat \phi_k +\int
\frac{\delta\Gamma_k}{\delta\phi^\ast_k} \pat \phi^\ast_k$. From the
Yukawa interaction, we get from this term, together with \Eqref{4.16},
\begin{equation}
\hb [\yb P_{\text{L}} \phi\psi -\yb P_{\text{R}} \phi^\ast \psi]
\to \frac{1}{2} \hb \pat\alpha_k
[(\yb\psi)^2-(\yb\gamma_5\psi)^2], \label{4.17}
\end{equation}
which is a contribution to the flow of the quark
self-interaction. Together with contributions from the first term of
\Eqref{4.15}, and neglecting the second term of \Eqref{4.15} here and
in the following as discussed above, we obtain the flow of the
four-quark coupling at fixed $\phi_k$,
\begin{equation}
\pat\lb|_{\phi_k}=-c_\lambda\, \frac{1}{k^2}\, g^4 -\hb\, \pat \alpha_k,
\label{4.18a}
\end{equation}
where we used the result given in \Eqref{4.8}. Choosing the
transformation function 
\begin{equation}
\pat \alpha_k =- c_\lambda\, \frac{1}{k^2}\, \frac{g^4}{\hb},
\label{4.18b}
\end{equation}
we obtain
\begin{equation}
\pat \lb|_{\phi_k} =0, \label{4.18c}
\end{equation}
which, together with the initial condition $\lb|_{k\to\Lambda}\to 0$
implies that $\lb=0$ holds for all scales $k$. The scale-dependent
transformation \Eqref{4.16} thus has removed the point-like four-quark
interaction in the scalar--pseudo-scalar sector completely by partial
rebosonization. The information about this interaction is transformed
into the scalar sector; for instance, the scalar mass term is also
subject to the last term of \Eqref{4.15}:
\begin{equation}
\mb^2\phi^\ast\phi \to -\mb^2\pat\alpha_k[\yb P_{\text{L}} \phi\psi
-\yb P_{\text{R}} \phi^\ast \psi]. \label{4.19}
\end{equation}

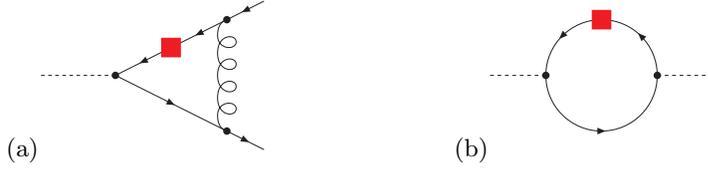
\begin{figure}[t]
\begin{center}
\subfigure{}{\scalebox{0.7}[0.7]{
\begin{picture}(190,140)(20,0)
  \SetOffset(3,10)
  \Text(30,20)[c]{\scalebox{1.5}[1.5]{(a)}}
  \DashLine(40,60)(80,60){2}
  \ArrowLine(160,100)(140,90)
  \ArrowLine(140,90)(110,75)
  \ArrowLine(110,75)(80,60)
  \ArrowLine(80,60)(140,30)
  \ArrowLine(140,30)(160,20)
  \Gluon(140,90)(140,30){-5}{4.5} 
  \Vertex(80,60){2} 
  \Vertex(140,90){2}
  \Vertex(140,30){2} 
  \CBoxc(110,75)(10,10){Red}{Red}
\end{picture}}}
\hspace{1cm}
\subfigure{}{\scalebox{0.7}[0.7]{
\begin{picture}(190,140)(20,0)
  \SetOffset(3,10)
  \Text(30,20)[c]{\scalebox{1.5}[1.5]{(b)}} 
  \DashLine(40,60)(70,60){2}
  \DashLine(130,60)(160,60){2}
  \ArrowArc(100,60)(30,0,90)
  \ArrowArc(100,60)(30,90,180)
  \ArrowArc(100,60)(30,180,0)
  \Vertex(70,60){2}
  \Vertex(130,60){2}
  \CBoxc(100,90)(10,10){Red}{Red} 
\end{picture}}}
\end{center}
\vspace{-1.0cm}
\caption{Diagrams contributing to the RG flow of the scalar sector:
  (a) flow of the Yukawa coupling, see \Eqref{4.20}; (b) flow of the
  scalar mass, see \Eqref{4.20a}. Only one diagram per topology is
  shown; further diagrams exhibit the regulator insertion (filled box)
  attached to other internal lines.}
\label{fig:yukawa}
\end{figure}

This yields a contribution to the flow of the Yukawa
coupling. Together with the contribution from the first term of
\Eqref{4.15}, i.e., the standard flow term, we obtain
\begin{eqnarray}
\pat \hb|_{\phi_k}&=&-\frac{1}{2}\, c_h\, g^2 \hb +\mb^2\,
\pat\alpha_k \nonumber\\
&\stackrel{\eqref{4.18b}}{=}&-\frac{1}{2}\, c_h\, g^2 \hb 
-c_\lambda \, \frac{\mb^2}{k^2}\,
\frac{g^4}{\hb}, 
\label{4.20}
\end{eqnarray}
where the coefficient $c_h$ is a result of the diagram shown in
Fig.~\ref{fig:yukawa}(a). For the linear regulator and in the Landau
gauge, the result is $c_h=1/\pi^2$ for SU(3). The second term of
\Eqref{4.20} accounts for the fact that the fermions couple to a
scale-dependent boson. The flow of the scalar mass term is not
transformed by the choice of \Eqref{4.16}; only the standard
flow-equation term contributes,
\begin{equation}
\pat \mb^2=c_m\, k^2\, \hb^2, \label{4.20a}
\end{equation}
where the coefficient $c_m$ yields for the regulator \eqref{AO.3}
$c_m=\Nc/(8\pi^2)$, resulting from the diagram in
Fig.~\ref{fig:yukawa}(b). The physical properties of the resulting
boson field can best be illustrated with the convenient dimensionless
composite coupling
\begin{equation}
\te:= \frac{\mb^2}{k^2 \hb^2}, \label{4.21}
\end{equation}
and its $\beta$ function
\begin{equation}
\pat \te=-2\te + c_m + c_h g^2 \te + 2c_\lambda g^4 \te^2,
\label{4.22}
\end{equation}
resulting from Eqs.~\eqref{4.8},\eqref{4.20},\eqref{4.20a}. The last
term comes directly from rebosonization. Since all $c_i>0$, this
$\beta$ function looks like a parabola; see Fig.~\ref{fig:parabola},
solid line. Without rebosonization, this $\beta$ function would have
corresponded to a straight line, see Fig.~\ref{fig:parabola}, dashed
line. 

{\unitlength=1pt
\begin{figure}[t]
  \centering
    \scalebox{1}[1]{
      \begin{picture}(300,170)(0,-20)
        \put(0,-20){ 
          \includegraphics[height=170pt]{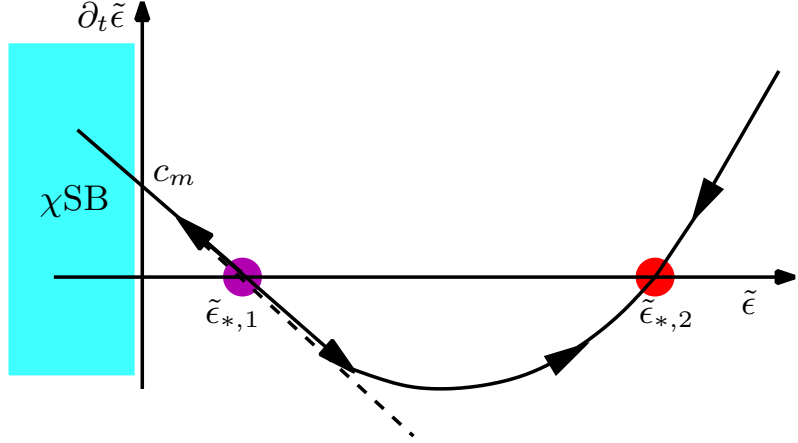} 
        }
        \Text(30,70)[c]{\scb{$\chi$SB}} 
        \Text(40,140)[c]{\scb{$\pat\te$}}
        \Text(68,80)[c]{\scb{$c_m$}}
        \Text(90,27)[c]{\scb{$\te_{\ast,1}$}} 
        \Text(254,27)[c]{\scb{$\te_{\ast,2}$}}
        \Text(285,32)[c]{\scb{$\te$}}
      \end{picture}
    }
\caption{Schematic plot of the $\beta$ function \eqref{4.22} for the composite
  coupling $\te=\frac{\mb^2}{k^2 \hb^2}$ with arrows pointing along
  the flow towards the IR. The fixed point $\te_{\ast,1}$ is IR
  repulsive; in its vicinity, the scalar field behaves as a
  fundamental scalar (dashed line). If the flow is initiated with
  $\te|_{k=\Lambda}<\te_{\ast,1}$, $\te$ drops quickly below zero and
  the system runs into the regime with chiral symmetry breaking
  ($\chi$SB). For $\te|_{k=\Lambda}>\te_{\ast,1}$, the system rapidly
  approaches the bound-state IR fixed point $\te_{\ast,2}$, where the
  scalar exhibits bound-state behavior. QCD initial conditions
  correspond to $\te|_{k\to\Lambda}\to \infty$.}
\label{fig:parabola}
\end{figure}
}

The $\beta$ function \Eqref{4.22} exhibits two fixed points: $\te_{\ast,1}$
is IR repulsive and $\te_{\ast,2}$ is IR attractive. In a
small-gauge-coupling expansion, the positions of the fixed points are
given by
\begin{equation}
\te_{\ast,1}\simeq \frac{c_m}{2} + \mathcal O(g^2), \quad \te_{\ast,2}\simeq
\frac{1}{c_\lambda g^4} + \mathcal{O}(1/g^2). \label{4.22a}
\end{equation}
Without rebosonization, only $\te_{\ast,1}$ is present. 

If we start with initial conditions such that
$\te|_{k\to\Lambda}<\te_{\ast,1}$, $\te$ quickly becomes negative,
corresponding to the bosonic mass term dropping below zero, $\mb^2<0$.
This indicates that the potential develops a nonzero minimum, giving
rise to chiral symmetry breaking and quark mass generation. However,
we obtain this initial condition $\te_{k\to\Lambda}<\te_{\ast,1}$ only
if either the scalar mass is small or the Yukawa coupling is large or
both; but this is in conflict with our QCD initial conditions,
specified below \Eqref{4.15b}. With this initial condition, the system
is not in the QCD universality class. Near $\te_{\ast,1}$, the slope
of the $\beta$ function \eqref{4.22} is $-2$, which is nothing but a
typical quadratic renormalization of a bosonic mass term; the scalar
behaves like an ordinary fundamental scalar here. In fact, there is a
correspondence between $\te_{\ast,1}$ and the critical coupling of
NJL-like systems,
\begin{equation}
\te_{\ast,1} \simeq \frac{Nc}{2 k^2 \lb{}_{\text{cr}}}, \label{4.22b}
\end{equation}
cf. Eqs. \eqref{4.3},\eqref{4.6b},\eqref{4.21}; the factor of $\Nc$
takes care of the additional color degree of freedom of the quarks
which was not present in the NJL system of
Subsect.~\ref{ssec:partbos}. The initial condition
$\te|_{k\to\Lambda}<\te_{\ast,1}$ agrees with
$\lb|_{k\to\Lambda}>\lb{}_{\text{cr}}$, and the system is in the
broken phase of the NJL model.

For $\te|_{k\to\Lambda}>\te_{\ast,1}$,  the system quickly approaches $\te_{\ast,2}$
either from above or below rather independently of the initial  values
of $\mb,\hb|_{k\to\Lambda}$. QCD initial conditions with large initial
$\mb$ and small initial $\hb$ correspond to $\te|_{k\to\Lambda}\to
\infty$. But also for much smaller initial $\te$, the system rapidly
flows to $\te_{\ast,2}$, and the memory of the precise initial values gets
lost. There, the system is solely determined by the gauge coupling
$g^2$ which governs the fixed-point position. This is precisely how it
should be in QCD. 

Near $\te_{\ast,2}$, the boson is not really a fully developed degree of
freedom. The flow does not at all remind us of the flow of a
fundamental scalar, but points to the composite nature of the
scalar. This justifies to call $\te_{\ast,2}$ the bound-state fixed
point; for instance, in weakly coupled systems such as QED, the boson
at the bound-state fixed point describes a positronium-like bound
state. 

It is a particular strength of the RG approach with scale-dependent
field transformations that one and the same field can describe bound
state formation on the one hand and condensate formation as well as meson
excitations on the other hand; whether the field behaves as a bound
state or as a fundamental scalar is solely governed by the dynamics
and the coupling strength of the system. 

So far, our analysis of the system was essentially based on
weak-coupling arguments, revealing that QCD at initial stages of the
flow approaches the bound-state fixed point. But, we still have to
answer a crucial question: how does QCD leave the bound-state fixed
point and ultimately approach the chiral-symmetry broken regime? The
answer is again given by the gauge coupling which controls the whole
flow. For increasing gauge coupling, the parabola characterizing the
$\beta$ function \eqref{4.22} for $\te$ is lifted, as depicted in
Fig.~\ref{fig:parlift}. At a critical coupling value,
$g^2=g_{\text{cr}}^2$, the fixed points $\te_{\ast,1}$ and
$\te_{\ast,2}$ annihilate each other and the system runs towards the
chiral-symmetry broken regime. This transition is unambiguously
triggered by gluonic interactions. For instance in the present case of
one-flavor QCD with gauge group SU(3), the critical coupling is given
by $\alpha_{\text{cr}}\equiv \frac{g^2_{\text{cr}}}{4\pi} \simeq 0.74$
for the regulator \Eqref{AO.3}. Since this coupling value is not a
universal quantity, one should not overemphasize its meaning; however,
it is interesting to observe that this coupling strength is in the
nonperturbative domain, as expected, but not very deeply. In
particular, since loop expansions go along with the expansion
parameter $\alpha/\pi$, the critical coupling and thus the approach to
chiral symmetry breaking appears still to be in reach of weak-coupling
methods (not to be confused with perturbation theory).

{\unitlength=1pt
\begin{figure}[t]
  \centering
    \scalebox{1}[1]{
      \begin{picture}(300,190)(0,-20) 
        \put(-10,-20){ 
          \includegraphics[height=170pt]{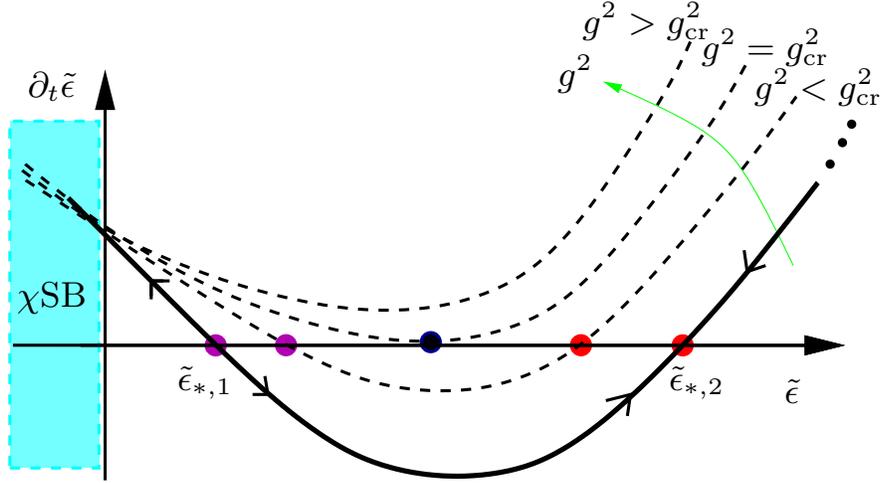} 
        }
        \Text(10,50)[c]{\scb{$\chi$SB}} 
        \Text(10,130)[c]{\scb{$\pat\te$}}
        \Text(208,135)[c]{\scb{$g^2$}}
        \Text(300,130)[c]{\scb{$g^2<g_{\text{cr}}^2$}} 
        \Text(280,145)[c]{\scb{$g^2=g_{\text{cr}}^2$}}
        \Text(235,155)[c]{\scb{$g^2>g_{\text{cr}}^2$}}
        \Text(68,18)[c]{\scb{$\te_{\ast,1}$}}
        \Text(254,18)[c]{\scb{$\te_{\ast,2}$}}
        \Text(290,15)[c]{\scb{$\te$}} 
      \end{picture}
    } 
\caption{Schematic plot of the $\beta$ function \eqref{4.22} for the
  composite coupling $\te=\frac{\mb^2}{k^2 \hb^2}$ with arrows
  pointing along the flow towards the IR. At weak gauge coupling,
  QCD-like systems first flow to the bound-state fixed point
  $\te_{\ast,2}$ where they remain over a wide range of scales. For
  increasing gauge coupling $g^2$, the $\beta$ function is lifted
  (dashed lines). At the critical coupling $g^2=g_{\text{cr}}^2$, the
  fixed points are destabilized and the system rapidly runs into the
  chiral symmetry broken regime ($\chi$SB).}
\label{fig:parlift}
\end{figure}
}

We conclude that the continuous scale-dependent translation allows for
a controllable transition between microscopic to macroscopic degrees
of freedom and between different dynamical regimes of a system. From a
quantitative viewpoint, no spurious dependence on a bosonization
scale, i.e., a scale at which degrees of freedom are discretely
changed, is introduced, because field transformations are continuously
performed on all scales. This helps maintaining the predictive power
of truncated RG flows. As a result, macroscopic parameters can
quantitatively be related to microscopic input.

\subsection{Further reading: aspects of field transformations${}^*$}

The scale-dependent field transformations introduced above were
illustrated with the aid of $n$-point interactions in the point-like
limit, e.g., a four-fermion interaction in the zero-momentum limit. Of
course, the formalism can also be used if momentum dependencies of the
vertices are taken into account. This is particularly important in the
context of rebosonization, since composite bosonic fluctuations and
bound states manifest themselves by a characteristic momentum
dependence in the fermionic $n$-point correlators; for instance, a
bosonic bound state corresponds to a pole in the $s$ channel of the
fermionic Minkowskian $4$-point vertex. 

A generic momentum dependence of an $n$-point vertex is nonlocal in
coordinate space. By means of a field transformation, certain
nonlocalities can be mapped onto a local description. The
Hubbard-Stratonovich transformation is an example for such a
transformation which maps a specific nonlocal four-fermion vertex onto
a local bosonic theory with a local Yukawa coupling to the fermion. 

With the following scale-dependent field transformation, this
nonlocal-to-local mapping can be performed on all scales for a
four-fermion vertex with $s$ channel momentum dependencies,
$\lb=\lb(s=q^2)$, 
\begin{eqnarray}
&&\pat\phi_k(q)=(\yb P_{\text{R}}\psi)(q)\, \pat\alpha_k(q),\quad
\pat\phi^\ast_k(q)=-(\yb P_{\text{L}}\psi)(-q)\,
\pat\alpha_k(q).\label{eq:d13b} 
\end{eqnarray}
This results in a flow of the 4-point interaction for the transformed
fields which reads
\begin{equation}
\pat\lb|_{\phi_k}= \pat\lb-\hb\, \pat \alpha_k(q).
\label{eq:4.18a}
\end{equation}
Obviously, the $q$ dependence of $\alpha_k$ can be chosen such the
fluctuation-induced $s$-channel $q$ dependence of $\lb$ is eaten up
for all scales and all values of $q$. Via the momentum-dependent
analog of \Eqref{4.20}, this induces a momentum dependence of the
Yukawa interaction $hb\to\hb(q)$; this can be taken care of by a
further generalization of the field transformation, 
\begin{eqnarray}
\pat\phi_k(q)&=&(\yb P_{\text{R}}\psi)(q)\, \pat\alpha_k(q) - \beta_k(q)
\phi_k(q), \nonumber\\
\pat\phi^\ast_k(q)&=&-(\yb P_{\text{L}}\psi)(-q)\,
\pat\alpha_k(q)- \beta_k(q)
\phi^\ast_k(q).\label{eq:d13c} 
\end{eqnarray}
The resulting flows for the Yukawa coupling and the inverse scalar
propagator for the transformed fields are then given by 
\begin{eqnarray}
\pat\hb(q)|_{\phi_k}&=&\pat\hb(q)+ \frac{Z_\phi q^2+\mb^2}{\hb}\,
\pat\lb(q^2) +\hb\,\pat\beta_k(q), \nonumber\\
(\pat Z_\phi(q) q^2+\pat \mb^2)|_{\phi_k}&=&\pat\mb^2
+2\pat\beta_k(q)\,(Z_\phi q^2+\mb^2). \label{eq:30a}
\end{eqnarray}
For a given momentum dependence of $\lb$, the first equation can be
used to determine the transformation function $\beta_k(q)$.\footnote{
  The momentum-independent part can, for instance, be fixed such that
  $\pat Z_\phi(q=k)=0$, ensuring that the approximation of a
  momentum-independent $Z_\phi$ is self-consistent.} The second
equation then fixes the running of the mass and of the wave function
renormalization of the transformed field.  In this manner, specific
nonlocal momentum structures can be transformed from the fermionic
interactions into a local boson sector.

This strategy has been used for the abelian gauged NJL model
\cite{Gies:2001nw}, QED \cite{Gies:2004hy} and QCD-like systems in
\cite{Gies:2002hq}. At weak coupling, one typically finds that the
bound-state fixed point discussed above of the composite coupling
$\te$ has similar counterparts in many other couplings, such as the
scalar dimensionless mass $\mb^2/k^2$ and the Yukawa coupling $\hb$.
This is a manifestation of their RG irrelevance at weak
coupling, with the dynamics of the scalar sector being fully
controlled by the fermions and their gauge interactions. If the gauge
coupling becomes large, the bound-state fixed points in all these
couplings is destabilized and interactions involving composite bosons
can become relevant or even dominant close to a transition into a
broken-symmetry regime. Since this switching from irrelevant to
relevant is a smooth process under the flow being controlled by the
dynamics itself, the predictive power of the computation is
maintained.

In QCD, the resulting effective action at IR scales naturally exhibits
a sector which is similar to a chiral quark-meson model
\cite{Jungnickel:1996fp,Schaefer:em,Schaefer:2004en} but with all
parameters fixed by the outcome of the RG flow with field
transformations. Also the gluonic sector can still be dynamically
active and contribute further to the running of the chiral sector.

A further application of the scale-dependent field transformation can
be found in \cite{Jaeckel:2002rm} where it is shown that the Fierz
ambiguities mentioned in the preceding section can be overcome by
treating all possible interaction channels and the corresponding
scale-dependent bosonic composites on equal footing.

\section*{Acknowledgment}

It is a great pleasure to thank A.~Schwenk and J.~Polonyi for
organizing the ECT$\ast$ school and for creating such a stimulating
atmosphere. I am particularly grateful to the students for their
active participation, critical questions and detailed discussions
which have left their traces in these lecture notes.  I would like to
thank J.~Braun, C.S.~Fischer, J.~Jaeckel, J.M.~Pawlowski, and
C.~Wetterich for pleasant and fruitful collaborations on some of the
topics presented here, and for numerous intense discussions, some
essence of which has condensed into these lecture notes. Critical
remarks on the manuscript by J.~Braun and J.M.~Pawlowski are
gratefully acknowledged. This work was supported by the DFG Gi 328/1-3
(Emmy-Noether program).

%%%%%%%%%%%%%%%%%%%%%%%%%%%%%%%%%%%%%%%%%%%%%%%%%%%%%%%%%%%%%%%%%%%%%%  }

%%%%%%%%%%%%%%%%%%%%%%%%%%%%%%%%%%%%%%%%%%%%%%%%%%%%%%%%%%%%%%%%%%%%%%

\printindex
\end{document}